\documentclass[draft,tightenlines,nofootinbib,preprint,aps,eqsecnum,amsmath,amssymb]{revtex4}

\newcommand{\beq}{\begin{equation}}
\newcommand{\eeq}{\end{equation}}
\newcommand{\bea}{\begin{eqnarray}}
\newcommand{\eea}{\end{eqnarray}}
\newcommand{\cir}{{\buildrel \circ \over =}}

\begin{document}

\title{Towards Relativistic Atomic Physics. I. The Rest-Frame
Instant Form of Dynamics and a Canonical Transformation for
a System of Charged Particles plus the Electro-Magnetic Field}

\author{David Alba}
\affiliation{Dipartimento di Fisica\\
Universita' di Firenze\\
Polo Scientifico, via Sansone 1\\
50019 Sesto Fiorentino, Italy\\
E-mail ALBA@FI.INFN.IT}
\author{Horace W. Crater}
\affiliation{The University of Tennessee Space Institute \\
Tullahoma, TN 37388 USA \\
E-mail: hcrater@utsi.edu}
\author{Luca Lusanna}
\affiliation{Sezione INFN di Firenze\\
Polo Scientifico\\
Via Sansone 1\\
50019 Sesto Fiorentino (FI), Italy\\
E-mail: lusanna@fi.infn.it}

\begin{abstract}

A complete exposition of the rest-frame instant form of dynamics for
arbitrary isolated systems (particles, fields, strings, fluids)
admitting a Lagrangian description is given. The starting point is
the parametrized Minkowski theory describing the system in arbitrary
admissible non-inertial frames in Minkowski space-time, which allows
one to define the energy-momentum tensor of the system and to show
the independence of the description from the clock synchronization
convention and from the choice of the 3-coordinates. The restriction
to the inertial rest frame, centered on the inertial observer having
the Fokker-Pryce center-of-inertia world-line, and the study of
relativistic collective variables replacing the non-relativistic
center of mass lead to the description of the isolated system as a
decoupled globally-defined non-covariant canonical external center
of mass carrying a pole-dipole structure (the invariant mass $M$ and
the rest spin ${\vec {\bar S}}$ of the system) and an external
realization of the Poincare' group. $Mc$ and ${\vec {\bar S}}$ are
the energy and angular momentum of a unfaithful internal realization
of the Poincare' group built with the energy-momentum tensor of the
system and acting inside the instantaneous Wigner 3-spaces where all
the 3-vectors are Wigner covariant. The vanishing of the internal
3-momentum and of the internal Lorentz boosts eliminate the internal
3-center of mass inside the Wigner 3-spaces, so that at the end the
isolated system is described only by Wigner-covariant canonical
internal relative variables.\medskip

Then an isolated system of positive-energy charged scalar articles
with mutual Coulomb interaction plus a transverse electro-magnetic
field in the radiation gauge is investigated as a classical
background for defining relativistic atomic physics. The electric
charges of the particles are Grassmann-valued to regularize the
self-energies. The external and internal realizations of the
Poincare' algebra in the rest-frame instant form of dynamics are
found. This allows one to define explicitly the rest-frame
conditions and their gauge-fixings (needed for the elimination of
the internal 3-center of mass) for this isolated system.\medskip

It is shown that there is a canonical transformation which allows
one to describe the isolated system as a set  of Coulomb-dressed
charged particles interacting through a Coulomb plus Darwin
potential plus a free transverse radiation field: these two
subsystems are not mutually interacting (the internal Poincare'
generators are a direct sum of the two components) and are
interconnected only by the rest-frame conditions and the elimination
of the internal 3-center of mass. Therefore in this framework with a
fixed number of particles there is a way out from the Haag theorem,
at least at the classical level.

\end{abstract}

\maketitle

\medskip

\today

\vfill\eject

\section{Introduction}

As shown in Refs.\cite{1b,2b,3b}, standard atomic physics is a
semi-relativistic treatment of a sector of QED in which:

\noindent a) the matter fields are approximated by scalar or
spinning particles at the classical level and by their first
quantization at the quantum level;

\noindent b) the relevant energies are below the threshold $2\, m\,
c^2$ of pair production so that there is a fixed number of
particles;

\noindent c) the electro-magnetic field is described in the Coulomb
gauge: it is often approximated with an external radiation field and
often one works in the long wavelength approximation
($\lambda_{em}\,\, >>\,\, size\, of\, atom$);

\noindent d) the atom is described in the electric dipole
representation (G$\o$ppert - Mayer or Power - Zieman - Woolley
unitary transformations \cite{1b,3b}) as a monopole
(non-relativistic center of mass) carrying an electric dipole and a
magnetic spin dipole; often the atom is approximated with a
two-level system carried by the monopole (Jaynes - Cummings - Paul
model \cite{3b}).
\bigskip

As a consequence, even if the theory is formulated in Minkowski
space-time, there is no consistent realization of the Poincare'
group available. As shown in Ref.\cite{4b}, due to the relation
$\epsilon_o\, \mu_o = 1/c^2$ ($\epsilon_o$ electric permittivity,
$\mu_o$ magnetic permeability), there are two non-relativistic
limits with two different realizations of the Galilei group: a) the
{\it electric} one without Faraday's law of induction (a
time-varying magnetic field does not produce an electric field); b)
the {\it magnetic} one without the displacement current (a
time-varying electric field does not produce a magnetic field).
\bigskip

The open problems are whether

\noindent a) it is possible to develop a consistent relativistic
version of atomic physics starting from a classical system of
relativistic charged particles plus the electro-magnetic field
followed by a quantization in which the number of particles is
fixed;

\noindent b) there is a clock synchronization convention defining
instantaneous 3-spaces in Minkowski space-time (possible Cauchy
surfaces for Maxwell equations);

\noindent c) the Poincare' algebra is correctly implemented.\bigskip

The need of this formulation comes from at least two directions:

\noindent A) The emergence of a new generation of extremely precise
and stable atomic clocks, to be put in space  and synchronized with
similar clocks on Earth so to be able to measure the gravitational
redshift of the Earth in a post-Newtonian approximation of the
geo-potential \cite{5b}, requires the consideration of  general
relativistic corrections in a special relativistic setting (i.e. the
post-Newtonian modification of null geodesics starting from the
Minkowski ones; these notions do not exist in non-relativistic
physics).

\noindent B) The absence of a relativistic theory of the
entanglement (see Ref.\cite{6b} for preliminary steps in this
direction) to be used, for instance, to formulate teleportation from
the Earth to the Space Station.
\bigskip

In this paper and subsequent ones we bring to bear intertwined tools
on the problem of relativistic atomic physics that do not appear in
the semi-relativistic treatments of the past. They are Dirac's
constraint theory, the maintenance of covariance by construction of
Poincare' generators, and the use of the three existing notions of
relativistic center-of-mass variables. These tools have been
developed over the past fifty years as a systematic response to
problems connected with the foundations of relativistic mechanics as
old as one hundred years. The result has been a full clarification
and solution of all these problems, see Refs.\cite{7b,8b,9b,10b} and
Subsection F of Section I. As shown in Ref.\cite{7b}, containing a
review of previous work \cite{8b,9b}, we have now a description of
every isolated system (particles, strings, fields, fluids) admitting
a Lagrangian formulation in arbitrary global inertial or
non-inertial frames in Minkowski space-time by means of {\it
parametrized Minkowski theories}. They allow one to get a
Hamiltonian description of the relativistic isolated systems, in
which the transition from a non-inertial (or inertial) frame to
another one is a gauge transformation generated by suitable
first-class Dirac constraints. Therefore, all the admissible
conventions for clock synchronization (identifying the instantaneous
3-spaces containing the system) turn out to be {\it gauge
equivalent}.\medskip

The basic motivation of parametrized Minkowski theories has been the
absence of an intrinsic notion of instantaneous 3-space (and of
spatial distance and 1-way velocity of light) in Minkowski
space-time, where to visualize the dynamics, due to the Lorentz
signature, only the light-cone, i.e. the conformal structure of
space-time identifying the allowed trajectories for incoming and
outgoing rays of light, is intrinsically given. Usually physics is
described in inertial frames centered on inertial observers, where
the instantaneous Euclidean 3-spaces are identified by means of
Einstein's ${1\over 2}$ clock synchronization convention
\footnote{An inertial observer  sends rays of light to another
arbitrary time-like observer, who reflects them back towards the
inertial observer. Given the emission ($x^o_i$) and adsorption
($x^o_f$) times on the inertial world-line, the point $P$ of
reflection on the other world-line is assumed to be simultaneous
with the mid-point $Q$ between emission and reabsorption  where
$x^o_P\, {\buildrel {def}\over =}\, x^o_Q = x^o_i + {1\over 2}\,
(x^o_f - x^o_i) = {1\over 2}\, (x^o_i + x^o_f)$. Only with this
convention the 1-way velocity of light is constant and isotropic and
coincides with the 2-way (or round-trip) velocity $c$.}.\medskip

However, very little is known about non-inertial frames \cite{9b}.
The only known way to have a global description of them is to choose
an arbitrary time-like observer and a 3+1 splitting of Minkowski
space-time with space-like hyper-surfaces (namely an arbitrary clock
synchronization convention) with a set of 4-coordinates adapted to
the foliation and having the observer as origin of the 3-coordinates
on each instantaneous 3-space. Each such foliation defines a {\it
global non-inertial frame} centered on the given observer if it
satisfies the M$\o$ller admissibility conditions \cite{11b},
\cite{9b}, and if the instantaneous (in general non-Euclidean)
3-spaces, described by the functions giving their embedding in
Minkowski space-time, tend to space-like hyper-planes at spatial
infinity \cite{9b}. The 4-metric in the non-inertial frame is a
function of the embedding functions, obtained from the flat metric
with a general coordinate transformation from the inertial Cartesian
4-coordinates to curvilinear 4-coordinates adapted to the
M$\o$ller-admissible foliation.\medskip

If we couple the Lagrangian of an arbitrary isolated system to an
external gravitational field and we replace the external
gravitational metric with the embedding-dependent 4-metric of a
non-inertial frame we get the Lagrangian of parametrized Minkowski
theories. It is a function of the matter of the isolated system (now
described in a non-inertial frame with variables {\it knowing} the
instantaneous 3-spaces) and of the embedding of the instantaneous
3-spaces of the non-inertial frame in Minkowski space-time. However,
the associated action is invariant \cite{8b}, \cite{7b,9b}, under
frame-preserving diffeomorphisms \footnote{Schmutzer and Plebanski
\cite{12b} were the only ones emphasizing the relevance of this
subgroup of diffeomorphisms in their attempt to obtain the theory of
non-inertial frames in Minkowski space-time as a limit from
Einstein's general relativity.} : this implies that the embeddings
are {\it gauge variables}, so that all M$\o$ller-admissible clock
synchronization conventions are gauge equivalent, As expected
special relativistic physics does not depend on how we
conventionally define the instantaneous 3-spaces!

\bigskip
As a particular case, it is now possible to get the {\it rest-frame
instant form of dynamics} of such isolated systems: every
configuration of the system is defined in an inertial frame whose
instantaneous 3-spaces are the Wigner hyper-planes orthogonal to the
conserved 4-momentum of the configuration (intrinsic definition of
the rest frame).
\medskip

In this instant form there are two realizations of the Poincare'
algebra:

\noindent 1) an {\it external} one in which the extended isolated
system is described as a point particle (by means of the canonical
non-covariant external 3-center of mass of the system \footnote{It
is the classical counterpart of the Newton-Wigner position operator:
its non-covariance is the way out from the no-interaction theorem in
relativistic mechanics \cite{10b}. This breaking of Lorentz
covariance is universal (independent from the isolated system), but
has no dynamical effect being associated with the decoupled
relativistic canonical center of mass.}) carrying an internal space
of Wigner-covariant relative variables determining the invariant
mass $M$ of the system and its overall spin $\vec S$;

\noindent 2) an unfaithful {\it internal} one inside the Wigner
hyper-planes, whose only non-vanishing generators are $M$ and $\vec
S$; the internal space is defined by the vanishing of the internal
3-momentum, i.e. by the rest-frame condition, and by the vanishing
of the internal Lorentz boosts, implying the elimination of the
internal 3-center of mass inside the Wigner hyper-planes which
avoids a double counting of the center of mass.\medskip

As said in Refs.\cite{7b,8b}, we have now a complete
characterization of the relativistic  collective variables (all of
them tend to the Newtonian center of mass in the non-relativistic
limit), which can be built {\it only in terms of the Poincare'
generators} \footnote{Besides the canonical non-covariant center of
mass there are only the covariant non-canonical Fokker-Pryce center
of inertia and the non-covariant non-canonical M$\o$ller center of
energy. See Refs.\cite{7b,8b,9b,10b} based on Pauri-Prosperi
formulation \cite{13b} of theory of canonical realizations of
(rotation, Galilei and Poincare') Lie groups in phase space and on
Dirac's theory of Hamiltonian constraints.} and, as shown in
Ref.\cite{10b}, it is now possible to reconstruct the relativistic
orbits of interacting particles.
\bigskip

In Ref.\cite{14b} there is the rest-frame description of N
relativistic positive-energy charged scalar particles plus the
electro-magnetic field (described in the radiation gauge, a special
case of Lorentz gauge). The particles have Grassmann-valued electric
charges to regularize the Coulomb self-energies. It was shown that
the use of the Lienard-Wiechert solution with {\it no incoming
radiation field} allows one to arrive at a description of N charged
particles dressed with a Coulomb cloud and mutually interacting
through the Coulomb plus the full relativistic Darwin potential.
This happens independently from the choice of the Green function
(retarded, advanced, symmetric,..) due to the Grassmann
regularization. In this way one can build the potential in the rest
frame describing all the static and non-static effects of the
one-photon exchange in QED. The quantization allows one to recover
the standard instantaneous approximation for {\it relativistic bound
states}, which till now had only been obtained starting from QFT
(either in the instantaneous approximations of the Bethe-Salpeter
equation or in the quasi-potential approach). In Ref.\cite{15b} the
same scheme was applied to spinning particles (with a
pseudo-classical Grassmann-valued spin generating the Dirac matrices
after quantization) and the Salpeter potential was identified after
a pseudo-classical Foldy-Wouthuysen transformation.
\bigskip

In this paper we will consider the isolated system of N relativistic
positive-energy charged scalar particles plus the electro-magnetic
field as a candidate for a description of relativistic atomic
physics, in which the atoms will turn out to be relativistic bound
states of subsets of the scalar (or spinning) particles. After
defining its rest-frame conditions, we will identify the external
and internal Poincare' generators of relativistic atomic physics.
Then we study whether it is meaningful to put the arbitrary
transverse electro-magnetic field as the sum of a transverse
radiation field plus the particle-dependent Lienard-Wiechert field
found in  Ref.\cite{14b}.\medskip

The main result will be to show the existence of a canonical
transformation in the rest-frame instant form, which sends the N
charged scalar particles, interacting only through Coulomb
potentials, plus an arbitrary transverse electro-magnetic field into
a set of N charged Coulomb-dressed particles mutually interacting
with the Coulomb + Darwin potential plus a decoupled transverse
radiation field.
\bigskip

Our surprising result shows that, in the rest-frame framework for
the description of relativistic atomic physics, at least at the
classical level, at every finite time (and not only asymptotically)
there is a canonical transformation from the radiation field to the
interpolating electro-magnetic one appearing in the phenomena
described by atomic physics (laser beams in a cavity interacting
with beams of atoms comes to mind).
\medskip

This has to be contrasted with QED, where we must consider only the
S matrix between IN and OUT free fields and not the interpolating
fields due to the Haag theorem, saying that the interaction picture
does not exist \cite{16b,a} and  that there is no unitary
transformation from the asymptotic IN and OUT states of the
radiation field to interpolating states of the general (non
radiation) electro-magnetic field. See the end of Section III.

\bigskip

After the canonical transformation each internal Poincare' generator
is the sum of the one of the radiation field plus the one of the
dressed particles interacting with the Coulomb + Darwin potential.
However the two subsystems are connected by the rest-frame condition
(vanishing of the internal 3-momentum) and by the condition
eliminating the overall internal center of mass (vanishing of the
internal Lorentz boosts).

\bigskip
In a second paper we will study the problems connected with the
collective variables of the isolated system "atoms plus
electro-magnetic field" and with the multipolar expansions of the
open subsystem formed by the atoms. We will show how to get
relativistic orbits for the atoms, how to define a relativistic
electric dipole representation and how to define a pseudo-classical
relativistic two-level atom.
\bigskip

Then, in a third paper, we will delineate how to make a canonical
quantization taking into account the relativistic need of clock
synchronization implying that the instantaneous 3-spaces are the
Wigner hyper-planes. If the canonical transformation of this paper
will turn out to be unitarily implementable, there will be a sound
definition of interpolating electro-magnetic fields starting from
asymptotic free radiation ones.\medskip

Moreover the absence of {\it relative times} among the particles
imposed by the clock synchronization, the non-locality of the
Poincare' generators and the non-covariance of the relativistic
canonical center of mass will be shown to introduce a {\it spatial
non-separability} of composite systems, whose description will be
the basic problem in the development of a theory about relativistic
entanglement independently from the adopted point of view about
quantum non-locality.

\bigskip

In Section II we give a concise updated review of the rest-frame
description of the isolated system "N positive-energy charged scalar
particles plus the electro-magnetic field"  and of its external and
internal Poincare' generators (Subsections A, B, C and D). In
Subsection F there is a comparison with other approaches to
relativistic mechanics, while in Subsection G there is the study of
the non-relativistic limit. We give also the rest-frame description
of the radiation field (Subsections E and H) and of the
particle-dependent Lienard-Wiechert fields of Ref.\cite{14b}
(Subsection L) and a discussion of the relation between the Fourier
transform of an arbitrary electro-magnetic field versus a radiation
field (Subsection I).

In Section III there is the definition of the canonical
transformation. In Section IV there is the form of the internal
Poincare' generators (Subsection A) and of the Hamilton equations
(Subsection B) after the canonical transformation. In Section V
there are some concluding remarks.\medskip

In Appendix A there is the definition of the transverse polarization
vectors for the radiation field in the radiation gauge and their
transformation properties under Poincare' transformations. In
Appendix B there are some properties of the Lienard-Wiechert
electro-magnetic fields and their Fourier transform. In Appendix C
there is the explicit evaluation of the internal Poincare'
generators after the canonical transformation. Finally in Appendix D
there are the conventions for the dimensions of the various
quantities adopted in this paper.

\vfill\eject

\section{N Charged Particles, with Grassmann-valued Charges, plus
the Electromagnetic Field in the Rest-Frame Instant Form of
Dynamics}

In this Section we shall review the results of Ref.\cite{14b}, which
make use of the radiation gauge for the electro-magnetic field,
after recalling the main steps of relativistic kinematics
\cite{8b,10b} \footnote{For a review on relativistic mechanics see
Refs.\cite{7b,10b}. For the details concerning parametrized
Minkowski theories and the definition of the rest-frame instant form
of dynamics see the first paper in Ref.\cite{8b}.} leading to the
rest-frame instant form of dynamics and emphasizing the differences
with previous approaches to relativistic mechanics.\bigskip

As said in the Introduction, the main tool is the description of
isolated relativistic systems (particles, fluids, strings, fields)
admitting a Lagrangian formulation by means of parametrized
Minkowski theories. After having done a (M$\o$ller- admissible
\cite{9b,11b}) 3+1 splitting of Minkowski space-time \footnote{We
use the metric $\eta_{\mu\nu} = (+---)$. See Appendix D for the
dimensions adopted in this paper.} (i.e. the choice of a global
non-inertial frame centered on an arbitrary time-like observer) and
defined observer-dependent radar 4-coordinates $\sigma^A = (\tau ,
\sigma^r)$ \footnote{To simplify the notation we will use $\vec
\sigma$ to denote the curvilinear 3-coordinates $\{ \sigma^r \}$.},
we use the embeddings $z^{\mu}(\tau ,\vec \sigma )$, identifying the
instantaneous 3-spaces $\Sigma_{\tau}$ as space-like surfaces in
Minkowski space-time, to find the metric $g_{AB}[z(\tau ,\vec \sigma
)]$ induced in the non-inertial frame. The action of parametrized
Minkowski theories is obtained by coupling the Lagrangian of the
isolated system to an external gravitational field and by replacing
the gravitational metric tensor $g_{\mu\nu}$ with $g_{AB}[z]$.
Therefore the resulting Lagrangian depends on both the matter
variables and on the embeddings $z^{\mu}(\tau ,\vec \sigma )$.
However the invariance of this action under frame-preserving
diffeomorphisms implies that the embeddings $z^{\mu}(\tau ,\vec
\sigma )$ are gauge variables. As a consequence the description of
the isolated system does not depend on the choice of the
non-inertial frame with its clock synchronization convention (the
choice of the proper time $\tau$) and its choice of 3-coordinates
$\sigma^r$ in the (in general non-Euclidean) instantaneous 3-space.
\medskip

An aspect of this approach that is distinctly new and not seen in
other approaches is the use of this embedding and it conjugate
momentum as an additional canonical pair of fields playing an equal
role with canonical particle and field variables. In essence, the
constraint formalism \footnote{For an elementary description see the
Appendix of Ref.\cite{7b}. The basic steps specified there which we
repeatedly use are a) identifying local symmetries of a singular
Lagrangian and associated first class constraints; b) this also
allows us to identify related gauge variables; c) one can then
explicitly break the gauge freedom by the introduction of gauge
fixing conditions on these gauge variable pairing up the first class
constraints of b): this is equivalent to have pairs of second class
constraints for the elimination of redundant pairs of canonical
variables.} is applied not only to the particle and field degrees of
freedom, but also to the foliation of space-time identifying the
instantaneous 3-spaces.
\bigskip

Subsections A-G of this Section will emphasize the particle aspect
of the rest-frame instant form with only Subsection E devoted to the
electro-magnetic field and to its reduction to the radiation gauge.
\medskip

In contrast, in Subsections I,H and G we will give the rest-frame
description of the radiation field in the radiation gauge, the
connection of it with a generic electro-magnetic field in presence
of charges (adapting the Coulomb gauge formulation of ch.1 of
Ref.\cite{1b} to the radiation gauge) and the expression of the
rest-frame radiation-gauge Lienard-Wiechert solutions of
Ref.\cite{14b}.

\subsection{Parametrized Minkowski Theories}

As shown in Ref.\cite{14b} the description of N scalar positive
energy particles with Grassmann-valued electric charges plus the
electro-magnetic field is done in parametrized Minkowski theories
with the action

\bea
  S &=&\int d\tau d^{3}\sigma \,{\cal L}(\tau ,\vec{\sigma})=\int d\tau L(\tau
),  \nonumber \\
 {\cal L}(\tau ,\vec{\sigma}) &=&{\frac{i}{2}}\sum_{i=1}^{N}\,
\delta ^{3}(\vec{ \sigma}-{\vec{\eta}}_{i}(\tau ))\, \Big[\theta
_{i}^{\ast }(\tau ){\dot{\theta}}_{i}(\tau ) -
{\dot{\theta}}_{i}^{\ast }(\tau )\theta _{i}(\tau )\Big]-  \nonumber \\
 &-&\sum_{i=1}^{N}\, \delta ^{3}(\vec{\sigma} - {\vec{\eta}}_{i}(\tau
))\, \Big[m_{i}\, c\, \sqrt{g_{\tau \tau }(\tau ,\vec{\sigma}) + 2\,
g_{\tau r}(\tau ,\vec{ \sigma})\, {\dot{\eta}}_{i}^{r}(\tau ) +
g_{rs }(\tau ,\vec{\sigma})\, {\dot{\eta}}_{i}^{r}(\tau )\,
{\dot{\eta}}_{i}^{s}(\tau )}+  \nonumber \\
&+&{{Q_i(\tau )}\over c}\, \Big(A_{\tau }(\tau ,\vec{\sigma}) +
A_{r}(\tau ,\vec{\sigma})\, {\dot{\eta}}_{i}^{r}(\tau )\Big)\Big]-  \nonumber \\
&-&{\frac{1}{4\, c}}\,\sqrt{g(\tau ,\vec{\sigma})}g^{AC }(\tau
,\vec{\sigma})g^{BD}(\tau ,\vec{\sigma})F_{AB}(\tau
,\vec{\sigma})F_{CD}(\tau ,
\vec{\sigma}),  \nonumber \\
&&{}  \nonumber \\
 Q_i(\tau ) &=& e\theta _{i}^{\ast }(\tau )\theta
_{i}(\tau ).
 \label{2.1}
 \eea

In this action the configuration variables are $z^{\mu}(\tau ,\vec
\sigma )$, ${\vec \eta}_i(\tau )$, $\theta_i(\tau )$ and $A_A(\tau
,\vec \sigma )$.\medskip

Here $z^{\mu}(\tau ,\vec \sigma )$ are the embeddings of the
instantaneous 3-spaces $\Sigma_{\tau}$, leaves of an arbitrary 3+1
splitting of Minkowski space-time. Instead of the standard Cartesian
4-coordinates, observer-dependent  Lorentz-scalar radar
4-coordinates $\sigma^A = (\tau ; \vec \sigma )$ are used, where
$\tau$ is a monotonically increasing function of the proper time of
an arbitrary time-like observer and $\sigma^r$ are 3-coordinates on
each $\Sigma_{\tau}$ having the observer as origin. The induced
metric is $g_{AB}(\tau ,\vec \sigma ) = z^{\mu}_A(\tau ,\vec \sigma
)\, \eta_{\mu\nu}\, z^{\nu}_B(\tau ,\vec \sigma )$ with $z^{\mu}_A =
{{\partial z^{\mu}}\over {\partial\, \sigma^A}}$: $g_{AB}$ is a
functional $g_{AB}[z]$ of $z^{\mu}(\tau ,\vec \sigma )$.\medskip

The scalar positive-energy particles are described by the
Lorentz-scalar 3-coordinates ${\vec \eta}_i(\tau )$ defined by
$x^{\mu}_i(\tau ) = z^{\mu}(\tau ,{\vec \eta}_i(\tau ))$, where
$x^{\mu}_i(\tau )$ are their world-lines \footnote{Since we identify
the instantaneous 3-space $\Sigma_{\tau}$ with a global clock
synchronization convention, the particles are identified by the
3-coordinates ${\vec \eta}_i(\tau )$ giving the intersection of
their world-lines with $\Sigma_{\tau}$. There are no relative times.
This implies an independent description for the positive-energy and
negative energy sectors of the particle mass shell $p^2_i(\tau ) =
m_i^2\, c^2$. Like the world-lines $x^{\mu}_i(\tau )$, also the
particle 4-momenta $p^{\mu}_i(\tau )$ are {\it derived} quantities
functions of the canonical variables of parametrized Minkowski
theories.}. $Q_i$ are the Grassmann-valued electric charges
satisfying $Q^2_i = 0$, $Q_i\, Q_j \not= 0$ for $i \not= j$: they
are described in terms of the complex Grassmann variables
$\theta_i(\tau )$, $\theta_i^*(\tau )$.\medskip

We also use Lorentz-scalar electro-magnetic potentials $A_A(\tau
,\vec \sigma ) = {\ \frac{{\partial\, z^{\mu}(\tau ,\vec \sigma
)}}{{\partial\, \sigma^A}}}\, A_{\mu}(z(\tau ,\vec \sigma ))$
adapted to the foliation, i.e. knowing the clock synchronization
convention. For the field strength we have $F_{AB}(\tau ,\vec \sigma
) = \Big(\partial_A\, A_B -
\partial_B\, A_A\Big)(\tau ,\vec \sigma ) = z^{\mu}_A(\tau ,\vec
\sigma )\, z^{\nu}_B(\tau ,\vec \sigma )\, F_{\mu\nu}(z(\tau ,\vec
\sigma ))$ with $F_{\mu\nu}(x) =
\partial_{\mu}\, A_{\nu}(x) - \partial_{\nu}\, A_{\mu}(x)$.\medskip

From Eqs.(\ref{2.1}) we get the canonical momenta \cite{11b}: $
\rho_{\mu}(\tau ,\vec \sigma ) = {{\delta\, S}\over {\delta\,
\partial_{\tau}\, z^{\mu}(\tau ,\vec \sigma )}}$, ${\vec
\kappa}_i(\tau ) = {{\delta\, S}\over {\delta\, {{d\,{\vec
\eta}_i(\tau )}\over {d\tau}}}}$, $\pi^A(\tau ,\vec \sigma ) = c\,
{{\delta\, S}\over {\delta\,\partial_{\tau}\, A_A(\tau ,\vec \sigma
)}}$ (with $\vec \pi = \vec E$ in inertial frames). The fundamental
Poisson brackets are $\{ z^{\mu}(\tau ,\vec \sigma ),
\rho_{\nu}(\tau ,{\vec \sigma}^{'})\} = - \eta^{\mu}_{\nu}\,
\delta^3(\vec \sigma - {\vec \sigma}^{'})$, $\{ A_A(\tau ,\vec
\sigma ), \pi^B(\tau ,{\vec \sigma}^{'})\} = c\, \eta^B_A\,
\delta^3(\vec \sigma - {\vec \sigma}^{'})$, $\{ \eta^r_i(\tau ),
\kappa_{js}(\tau )\} = \delta_{ij}\, \delta^r_s$ \footnote{Due to
this Poisson bracket we will use the following vector notation:
${\vec \eta}_i(\tau) = \Big(\eta^r_i(\tau)\Big)$, ${\vec
\kappa}_i(\tau) = \Big(\kappa_{ir}(\tau ) = -
\kappa^r_i(\tau)\Big)$.}.\medskip

The canonically conjugate variables $z^{\mu}(\tau ,\vec \sigma )$,
$\rho_{\mu}(\tau ,\vec \sigma )$; ${\vec \eta}_i(\tau )$, ${\vec
\kappa}_i(\tau )$; $A_A(\tau ,\vec \sigma )$, $\pi^A(\tau ,\vec
\sigma )$; are restricted by the first class constraints
($l^{\mu}(\tau ,\vec \sigma )$ is the unit normal to $\Sigma_{\tau
}$ while $z^{\mu}_r(\tau ,\vec \sigma )$ are tangent to it)

\beq
 {\cal H}^{\mu}(\tau ,\vec \sigma ) = \rho^{\mu}(\tau ,\vec \sigma ) -
 l^{\mu}(\tau ,\vec \sigma )\, T^{\tau\tau}(\tau ,\vec \sigma ) -
 z^{\mu}_r(\tau ,\vec \sigma )\, T^{\tau r}(\tau ,\vec \sigma )
 \approx 0,
 \label{2.2}
 \eeq

\noindent and by two first-class constraints for the
electro-magnetic field (see Ref.\cite{14b})\medskip

\bea
 \pi^{\tau}(\tau ,\vec \sigma ) &\approx & 0,\nonumber \\
 \Gamma(\tau ,\vec \sigma ) &=& \partial_r\, \pi^r(\tau ,\vec \sigma
 ) - \sum_{i=1}^N\, Q_i\, \delta^3(\vec \sigma - {\vec \eta}_i(\tau
 )) \approx 0.
 \label{2.3}
 \eea
\medskip

In Eqs.(\ref{2.2}) $T^{AB}(\tau ,\vec \sigma ) = - \Big({1\over
{\sqrt{g}}}\, {{\delta\, S}\over {\delta\, g_{AB}(\tau ,\vec \sigma
)}} \Big)(\tau ,\vec \sigma )$ is the energy-momentum of the
isolated system of particles plus the electro-magnetic field. See
Ref.\cite{14b} for the second class constraints eliminating the
momenta conjugate to the Grassmann variables $\theta_i(\tau )$: here
we will only use the Grassmann character of the electric charges
$Q_i$, which are constants of the motion.\medskip

The constraints  (\ref{2.3}) arise due to the invariance of the
action (\ref{2.1}) under frame-preserving diffeomorphisms \cite{7b}
and electro-magnetic gauge transformations.\medskip

As a consequence of the constraints (\ref{2.2}), $z^{\mu}(\tau ,\vec
\sigma )$ are {\it gauge variables} and this implies the gauge
equivalence of the clock synchronization conventions defining the
instantaneous 3-spaces $\Sigma_{\tau}$ and of the choice of the
3-coordinates on each $\Sigma_{\tau}$.\bigskip

The Dirac Hamiltonian is

\beq
 H_D = \int d^3\sigma\, \Big[\lambda^{\mu}\, {\cal H}_{\mu} + \lambda_{\tau}\,
 \pi^{\tau} - A_{\tau}\, \Gamma\Big](\tau ,\vec \sigma ),
 \label{2.4}
 \eeq

\noindent where the $\lambda$'s are the arbitrary Dirac multipliers
associated with the primary first-class constraints. In
Eq.(\ref{2.4}) $H_c = - \int d^3\sigma\, A_{\tau}(\tau ,\vec \sigma
)\, \Gamma (\tau ,\vec \sigma )$ is the canonical Hamiltonian
determined by the Legendre transformation: it is weakly zero due to
the secondary first-class constraint $\Gamma (\tau ,\vec \sigma )
\approx 0$ (the Gauss law).\bigskip

Since only the embedding variables have Minkowski indices, the
Poincare' generators are

\bea
 P^{\mu} &=& \int d^3\sigma \, \rho^{\mu}(\tau ,\vec \sigma
 ),\nonumber \\
 J^{\mu\nu} &=& \int d^3\sigma\, \Big(z^{\mu}\, \rho^{\nu} - z^{\nu}\,
 \rho^{\mu}\Big)(\tau ,\vec \sigma ).
 \label{2.5}
 \eea

\noindent they satisfy the Poincare' algebra: $\{ P^{\mu}, P^{\nu}
\} = 0$, $\{ P^{\mu}, J^{\alpha\beta} \} = \eta^{\mu\alpha}\,
P^{\beta} - \eta^{\mu\beta}\, P^{\alpha}$, $\{ J^{\mu\nu},
J^{\alpha\beta} \} = C^{\mu\nu\alpha\beta}_{\gamma\delta}\,
J^{\gamma\delta}$, $C^{\mu\nu\alpha\beta}_{\gamma\delta} =
\delta^{\nu}_{\gamma}\, \delta^{\alpha}_{\delta}\, \eta^{\mu\beta} +
\delta^{\mu}_{\gamma}\, \delta^{\beta}_{\delta}\, \eta^{\nu\alpha} -
\delta^{\nu}_{\gamma}\, \delta^{\beta}_{\delta}\, \eta^{\mu\alpha} -
\delta^{\mu}_{\gamma}\, \delta^{\alpha}_{\delta}\, \eta^{\nu\beta}
$.\bigskip

See Ref.\cite{7b} for the properties of the isolated system in
non-inertial frames. Here we will only describe the properties of
the  inertial foliation corresponding to its intrinsic rest frame.
\medskip

To this end we have to restrict the 3+1 splittings of Minkowski
space-time to inertial frames, whose instantaneous 3-spaces
$\Sigma_{\tau}$ are described by the following embeddings
($b^{\mu}_r(\tau )$ are three orthonormal space-like vectors)

\beq
 z^{\mu}_F(\tau ,\vec \sigma ) = x^{\mu}(\tau ) + b^{\mu}_r(\tau )\,
 \sigma^r.
 \label{2.6}
 \eeq

\noindent These space-like hyper-planes still depend on 10 residual
gauge degrees of freedom:

a) the world-line $x^{\mu}(\tau )$ of the inertial observer chosen
as origin of the 3-coordinates $\sigma^r$;

b) 6 variables parametrizing an orthonormal tetrad $b^{\mu}_A(\tau
)$ such that the constant (future-pointing) unit normal to the
hyper-planes is $l^{\mu} = b^{\mu}_{\tau } =
\epsilon^{\mu}{}_{\alpha\beta\gamma}\, b^{\alpha}_1(\tau )\,
b^{\beta}_2(\tau )\, b^{\gamma}_3(\tau )$.
\bigskip

If we impose the gauge fixings $z^{\mu}(\tau ,\vec \sigma ) -
z^{\mu}_F(\tau ,\vec \sigma ) \approx 0$ to the first-class
constraints (\ref{2.2}), only 10 degrees of freedom associated with
the momenta $\rho_{\mu}(\tau ,\vec \sigma )$ survive:

a) the total 4-momentum $P^{\mu}$ canonically conjugate to
$x^{\mu}(\tau )$, $\{ x^{\mu}, P^{\nu} \} = - \eta^{\mu\nu}$;

b) 6 momentum variables canonically conjugate to the tetrads
$b^{\mu}_A(\tau )$.\medskip

After this gauge fixing the Dirac Hamiltonian depends only on the 10
surviving first class constraints

\bea
 H_D &=& {\tilde \lambda}^{\mu}(\tau )\, {\tilde {\cal
 H}}_{\mu}(\tau ) - {1\over 2}\, {\tilde \lambda}^{\mu\nu}(\tau )\,
 {\tilde {\cal H}}_{\mu\nu}(\tau ) + (electromagnetic\,\, constraints),\nonumber \\
 &&{}\nonumber \\
 {\tilde {\cal H}}^{\mu}(\tau ) &=& \int d^3\sigma\, {\cal
 H}^{\mu}(\tau ,\vec \sigma ) \approx 0,\nonumber \\
 {\tilde {\cal H}}^{\mu\nu}(\tau ) &=& \int d^3\sigma\, \sigma^r\,
 \Big[b^{\mu}_r(\tau )\, {\cal H}^{\nu}(\tau ,\vec \sigma ) - b^{\nu}_r(\tau )\,
 {\cal H}^{\mu}(\tau ,\vec \sigma )\Big] \approx 0.
 \label{2.7}
 \eea

The Lorentz generators become

\bea
 J^{\mu\nu} &=& x^{\mu}\, P^{\nu} - x^{\nu}\, P^{\mu} +
 S^{\mu\nu},\nonumber \\
 &&{}\nonumber \\
 &&S^{\mu\nu} = \int d^3\sigma\, \sigma^r\, \Big[b^{\mu}_r(\tau )\,
 \rho^{\nu}(\tau ,\vec \sigma ) - b^{\nu}_r(\tau )\,
 \rho^{\mu}(\tau ,\vec \sigma )\Big].
 \label{2.8}
 \eea
\medskip

From now on we will use the notation $P^{\mu} = (P^o = E/c; \vec P)
= M\, c\, u^{\mu}(P)  = Mc\, (\sqrt{1 + {\vec h}^2}; \vec h)\,
{\buildrel {def}\over =}\, Mc\, h^{\mu}$, where $\vec h = \vec v/c$
is an a-dimensional 3-velocity and with $M\, c  = \sqrt{\,P^{2}}$.

\subsection{The Rest-Frame Instant Form}

To get the rest-frame instant form we add the  gauge fixings (only 6
of them are independent)

\beq
 b^{\mu}_A(\tau ) - \epsilon^{\mu}_A(u(P)) \approx 0,
 \label{2.9}
 \eeq

\noindent where $\epsilon^{\mu}_A(u(P)) = \epsilon^{\mu}_A(\vec h) =
L^{\mu}{}_A(P, {\buildrel \circ\over {P}})$ are the column of the
standard Wigner boost sending the 4-momentum $P^{\mu}$ (assumed
time-like) to its rest-frame form $P^{\mu} = L^{\mu}{}_{\nu}(P,
{\buildrel \circ\over {P}} )\, {\buildrel \circ\over {P}}^{\nu}$,
${\buildrel \circ\over {P}}^{\mu} = M\, c\, (1; \vec 0)$. As a
consequence of the $P^{\mu}$-dependence of the gauge fixings
(\ref{2.9}), as shown in Refs.\cite{8b}, the Lorentz-scalar
3-vectors ${\vec \eta}_i$, ${\vec \kappa}_i$ become Wigner spin-1
3-vectors (they transform under Wigner rotations if a Lorentz
transformation is done), so that the scalar product of two such
vectors is a Lorentz scalar. The same happens for all the 3-vectors
living inside these instantaneous 3-spaces, like the
electro-magnetic vector potential $\vec A(\tau ,\vec \sigma )$
($A_{\tau}(\tau ,\vec \sigma )$ remains a Lorentz-scalar).\medskip

From now on $i,j ..$ will denote Euclidean indices, while $r,s..$
will denote Wigner spin-1 indices. We have
$\epsilon^{\mu}_{\tau}(u(P)) = u^{\mu}(P) =
\epsilon^{\mu}_{\tau}(\vec h) = h^{\mu}$ and $
\epsilon^{\mu}_r(u(P)) = \Big(- u_r(P); \delta^i_r -
{\frac{{u^i(P)\, u_r(P)} }{{1 + u^o(P)}}}\Big) = \Big( - h_r;
\delta^i_r - {{h^i\, h_r}\over {1 + \sqrt{1 + {\vec h}^2}}}\Big) =
\epsilon^{\mu}_r(\vec h)$.\bigskip

\bigskip

In this way the 3+1 splittings of Minkowski space-time are
restricted to the inertial frames centered on the inertial observer
$x^{\mu}(\tau )$ and whose instantaneous 3-spaces $\Sigma_{\tau}$
are orthogonal to $P^{\mu}$: they are named {\it Wigner
hyper-planes}, because by construction the 3-vectors inside them are
Wigner spin-1 3-vectors.\bigskip

At this stage only 4 gauge degrees of freedom, $x^{\mu}(\tau )$,  of
the original embedding $z^{\mu}(\tau ,\vec \sigma )$ survive.
However, due to the dependence of the gauge fixings (\ref{2.9}) upon
$P^{\mu}$, the final canonical gauge variables are not $x^{\mu}$,
$P^{\mu}$, but a non-covariant ${\tilde x}^{\mu}(\tau )$ and
$P^{\mu}$ (see the first paper in Ref.\cite{8b})

\bea
 {\tilde x}^{\mu} &=& x^{\mu} -
{\frac{1}{{M\, c\, (P^o + M\, c) }}}\, \Big[P_{\nu}\, S^{\nu\mu} +
M\, c\, \Big(S^{o\mu} - S^{o\nu}\, {\frac{{ P_{\nu}\,
P^{\mu}}}{{M^2\, c^2}}}\Big)\Big],\nonumber \\
 &&{}\nonumber \\
 && u(p) \cdot \tilde x = u(P) \cdot x,\qquad \{ {\tilde
 x}^{\mu}(\tau ), P^{\nu} \} = - \eta^{\mu\nu}.
 \label{2.10}
 \eea
\medskip

After the gauge fixing (\ref{2.9}) the Lorentz generators become

\bea
 J^{\mu\nu} &=& {\tilde x}^{\mu}\, P^{\nu} - {\tilde x}^{\nu}\,
 P^{\mu} + {\tilde S}^{\mu\nu},\nonumber \\
 &&{}\nonumber \\
 && {\tilde S}^r = {\bar S}^r = {1\over 2}\, \epsilon^{ruv}\, {\bar
 S}^{uv},\qquad {\tilde S}^{0r} = {{\epsilon^{rsu}\, P^s\, {\bar S}^u}
 \over {Mc + P^o}},
 \label{2.11}
 \eea

\noindent with ${\tilde S}^{\mu\nu}$ function only of the 3-spin
${\bar S}^r$ of the rest spin tensor ${\bar S}_{AB} =
\epsilon^{\mu}_A(u(P))\, \epsilon^{\nu}_B(u(P))\, S_{\mu\nu}$ (both
the spin tensors ${\tilde S}^{\mu\nu}$ and ${\bar S}^{AB}$ satisfy
the Lorentz algebra). Since we assume $P^2 > 0$, the Pauli-Lubanski
invariant is $W^2 = - P^2\, {\vec {\bar S}}^2$.
\bigskip

The main result is that the $P^{\mu}$-dependent gauge fixing
(\ref{2.9}) change the interpretation of the residual gauge
variables, because the remaining 4 first class constraints can be
written in the following form

\bea
 {\tilde {\cal H}}^{\mu}(\tau ) &=& \int d^3\sigma\, {\cal
 H}^{\mu}(\tau ,\vec \sigma ) =\nonumber \\
 &=& P^{\mu} - u^{\mu}(P)\, \int d^3\sigma\, T^{\tau\tau}(\tau ,\vec
 \sigma ) - \epsilon^{\mu}_r(u(P))\, \int d^3\sigma\, T^{\tau
 r}(\tau ,\vec \sigma ) \approx 0,\nonumber \\
 &&{}\nonumber \\
 &&\Downarrow\nonumber \\
 &&{}\nonumber \\
 && Mc = \sqrt{P^2} \approx \int d^3\sigma\, T^{\tau\tau}(\tau ,\vec
 \sigma ),\nonumber \\
 &&{\cal P}^r_{(int)} = \int d^3\sigma\, T^{\tau r}(\tau ,\vec
 \sigma ) \approx 0.
 \label{2.12}
 \eea

The Dirac Hamiltonian is now $H_D = \lambda (\tau )\, \Big(Mc - \int
d^3\sigma\, T^{\tau\tau}(\tau ,\vec \sigma )\Big) + \vec \lambda
(\tau ) \cdot {\vec {\cal P}}_{(int)} + (electromagnetic\,
constraints)$ and the embedding of Wigner hyper-planes is

\beq
 z_W^{\mu}(\tau ,\vec \sigma ) = x^{\mu}(\tau ) +
\epsilon^{\mu}_r(u(P))\, \sigma^r,
 \label{2.a}
 \eeq

\noindent with $x^{\mu}$ function of ${\tilde x}^{\mu}$, $P^{\mu}$
and $S^{\mu\nu}$ (or ${\bar S}^{AB}$) according to Eq.(\ref{2.10}).
The gauge freedom in the choice of $x^{\mu}$ is connected with the
arbitrariness of the spin boosts ${\bar S}^{\tau r}$ (or
$S^{oi}$).\bigskip

Eqs.(\ref{2.12}) imply that the 8 variables ${\tilde x}^{\mu}(\tau
)$ and $P^{\mu}$ are restricted only by a first class constraint
identifying $Mc = \sqrt{P^2}$ with the invariant mass of the
isolated system, evaluated by using its energy-momentum tensor:
therefore these 8 variables are to be reduced to 6 physical
variables describing the external decoupled relativistic center of
mass of the isolated system ($\sqrt{P^2}$ is determined by the
constraint and its conjugate variable, the rest time $u(P) \cdot
\tilde x(\tau )$, is a gauge variable). Therefore these 3 collective
degrees of freedom hidden in the embedding field $z^{\mu}(\tau ,\vec
\sigma )$ become physical variables. However, there are 3 first
class constraints ${\vec {\cal P}}_{(int)} \approx 0$ (the
rest-frame conditions) implying that the Wigner hyper-plane
$\Sigma_{\tau}$ is the rest frame of the isolated system. Therefore
the  remaining 3 gauge degrees of freedom are shifted from the
embeddings to the internal variables inside the Wigner hyper-planes:
now the final 3 gauge variables are the 3-coordinates of the
internal 3-center of mass, which have to be eliminated with a gauge
fixing to the the rest-frame conditions. In this way we avoid a
double counting of the center of mass and the dynamics inside the
Wigner hyper-planes is described only by relative variables.

\subsection{The External Center of Mass and the External Poincare'
Group}

Let us now add the $\tau$-dependent gauge fixing $c\, T_s - \tau
\approx 0$, where $c\, T_s = u(P) \cdot \tilde x = u(P) \cdot x$ is
the Lorentz-scalar rest time, to the first class constraint $Mc -
\int d^3\sigma\, T^{\tau\tau}(\tau ,\vec \sigma ) \approx 0$. After
this imposition (from now on $\tau /c$ is the rest time $T_s$) we
have:

1) The 4-momentum $P^{\mu}$ becomes the 4-momentum of the isolated
"particles plus electro-magnetic field" system: $P^{\mu} = Mc\,
(\sqrt{1 + {\vec h}^2}; \vec h) = Mc\, h^{\mu}$ with $\vec h$
arbitrary a-dimensional 3-velocity and $Mc \equiv \int d^3\sigma\,
T^{\tau\tau}(\tau ,\vec \sigma )$.

2) The 6 physical degrees of freedom describing the external
decoupled relativistic center of mass are the non-evolving Jacobi
data

\beq
 \vec z = Mc\, \Big({\vec {\tilde x}} - {{\vec P}\over {P^o}}\,
 {\tilde x}^o\Big),\qquad \vec h = {{\vec P}\over
 {Mc}},\qquad \{ z^i, h^j \} = \delta^{ij}.
 \label{2.13}
 \eeq

\noindent The 3-vector ${\vec x}_{NW} = \vec{ z}/M\, c$ is the
\textit{external non-covariant canonical 3-center of mass} (the
classical counterpart of the ordinary Newton-Wigner position
operator) and is canonically conjugate to the 3-momentum $M\, c\,
\vec{h}$. We have ${\tilde x}^{\mu}(\tau ) = ({\tilde x}^o(\tau );
{\vec x}_{NW} + {{\vec P}\over {P^o}}\, {\tilde x}^o(\tau
))$.\medskip

From Appendix B of Ref.\cite{17b} we have the following
transformation properties under Poincare' transformations $(a ,
\Lambda )$:

\bea
 h^{\mu} &=& u^{\mu}(P) = (\sqrt{1 + {\vec h}^2}; \vec h)\, \mapsto\, h^{{'}\, \mu} =
\Lambda^{\mu}{}_{\nu}\, h^{\nu},\nonumber \\
 z^i\, &\mapsto& z^{{'}\, i} = \Big(\Lambda^i{}_j - {{\Lambda^i{}_{\mu}\,
h^{\mu}}\over {\Lambda^o{}_{\nu}\, h^{\nu}}}\, \Lambda^o{}_j\Big)\,
z^j + \Big(\Lambda^i{}_{\mu} - {{\Lambda^i{}_{\nu}\, h^{\nu}}\over
{\Lambda^o{}_{\rho}\, h^{\rho}}}\, \Lambda^o{}_{\mu}\Big)\,
(\Lambda^{-1}\, a)^{\mu},\nonumber \\
 \tau\, &\mapsto& \tau^{'} + h_{\mu}\, (\Lambda^{-1}\, a)^{\mu}.
 \label{2.14}
 \eea

\noindent As a consequence, under Lorentz transformations we have
${\vec h}^{'} \cdot {\vec z}^{'} = \vec h \cdot \vec z +
{{\Lambda^o{}_j\, z^j}\over {\Lambda^o{}_{\mu}\, h^{\mu}}}$.\bigskip

3) As shown in Refs.\cite{7b,8b,10b}, for every isolated
relativistic system there are {\it only} three collective variables
(replacing the unique non-relativistic 3-center of mass), which can
be constructed by using {\it only} the generators $P^{\mu}$,
$J^{\mu\nu}$ of the associated realization of the Poincare' group.
They are the external covariant non-canonical Fokker-Pryce 4-center
of inertia $Y^{\mu}(\tau )$, the external non-covariant canonical
4-center of mass (also called center of spin) ${\tilde x}^{\mu}(\tau
)$ and the external non-covariant non-canonical M$\o$ller 4-center
of energy $R^{\mu}(\tau )$ \footnote{In non-relativistic physics,
the center of mass variable take on two essential roles. First, it
is a locally observable 3-vector with the necessary transformation
properties under the Galilei group. Secondly it, together with the
total momentum, form a canonical pair. In relativity the first two
properties are split respectively between the Fokker-Pryce 4-center
of inertia $Y^{\mu}(\tau )$ and ${\tilde x}^{\mu}(\tau )$. Both as
well as $R^{\mu}(\tau )$ move with constant velocity $h^{\mu}$ for
an isolated system in analogy to what happens in non-relativistic
physics.}. All of them have unit 4-velocity $h^{\mu} = u^{\mu}(P)$,
but only $Y^{\mu}(\tau ) = Y^{\mu}(0) + u^{\mu}(P)\, \tau =
Y^{\mu}(0) + \Big(\sqrt{1 + {\vec h}^2}; \vec h\Big)\, \tau$ is a
4-vector, whose world-line can be used as an inertial observer. As
we shall see these collective variables are functions of $\vec z$,
$\vec h$, $Mc$, ${\vec {\bar S}}$ and $\tau$.\medskip

Let us remark that, since the Poincare' generators $P^{\mu}$,
$J^{\mu\nu}$ are {\it global} quantities (they know the whole
instantaneous 3-space), these collective variables, being defined in
terms of them (as shown in Ref.\cite{8b,13b}),  are also global
quantities. They {\it cannot be locally determined}: this is a
fundamental difference from the non-relativistic 3-center of mass.
\bigskip

4) Due to the $\tau$-dependence of the gauge fixing $c\, T_s - \tau
\approx 0$, the Dirac Hamiltonian (\ref{2.7}) becomes $H_D = M\, c +
\vec \lambda (\tau ) \cdot {\vec {\cal P}} + (electromagnetic\,\,
constraints)$.\medskip

5) As shown in Refs.\cite{8b,10b} the final form of the {\it
external Poincare' generators} (\ref{2.5}) of an arbitrary isolated
system in the rest-frame instant form is

\begin{eqnarray}
P^{\mu } &,&\qquad J^{\mu \nu } = {\tilde x}^{\mu }\, P^{\nu } -
{\tilde x}
^{\nu}\, P^{\mu } + {\tilde S}^{\mu\nu },  \nonumber \\
&&{}  \nonumber \\
P^o &=& \sqrt{M^2\, c^2 + {\vec P}^2} = Mc\, \sqrt{1 + {\vec h}^2},
\qquad \vec P = Mc\, \vec h,  \nonumber \\
&&{}  \nonumber \\
 J^{ij} &=& {\tilde x}^i\, P^j - {\tilde x}^j\, P^i + \epsilon ^{iju}\,
{\bar S}^u = z^i\, h^j - z^j\, h^i + \epsilon^{iju}\, {\bar S}^u,  \nonumber \\
 K^i &=& J^{oi} = {\tilde x}^o\, P^i - {\tilde x}^i\, \sqrt{M^2\, c^2
+ {\vec P}^2} - { { \epsilon ^{isu}\, P^s\, {\bar S}^u}\over {M\, c
+ \sqrt{ M^2\, c^2 + {\vec P}^2} }} =\nonumber \\
 &=& - \sqrt{1 + {\vec h}^2}\, z^i + {{({\vec {\bar S}} \times \vec h)^i}\over
 {1 + \sqrt{1 + {\vec h}^2}}}.
  \label{2.15}
\end{eqnarray}

\noindent Note that both ${\tilde L}^{\mu\nu} = {\tilde x}^{\mu}\,
P^{\nu} - {\tilde x} ^{\nu}\, P^{\mu}$ and ${\tilde S}^{\mu\nu} =
J^{\mu\nu} - {\tilde L}^{\mu\nu}$ are conserved.\medskip

It is this external realization which implements the Wigner
rotations  on the Wigner hyper-planes through the last term in the
Lorentz boosts.
\medskip

Let us remark that this realization is universal in the sense that
it depends on the nature of the isolated system only through an
$U(2)$ algebra \cite{18b}, whose generators, which will be defined
in Eqs.(\ref{2.22}), are the invariant mass $M$ (which in turn
depends on the relative variables and on the type of interaction)
and the internal spin ${ \vec {\bar S}}$, which is
interaction-independent being in an instant form of dynamics.
\medskip

Therefore the isolated system of particles plus fields is described
by a canonical external non-covariant 3-center of mass $\vec z$ (
canonically conjugate to $\vec h$), i.e. a decoupled pseudo-particle
of mass $M$ and spin ${\vec {\bar S}}$ \footnote{For N free
positive-energy scalar particles we have $M\, c = \sum_{i=1}^N\,
\sqrt{m_i^2\, c^2 + {\vec \kappa}_i^2}$, ${\vec {\bar S}} =
\sum_{i=1}^N\, {\vec \eta}_i \times {\vec \kappa}_i$ with the
rest-frame conditions $\sum_{i=1}^N\, {\vec \kappa}_i \approx 0$,
${\bar S}^{\tau r} = - \sum_{i=1}^N\, \eta^r_i\, \sqrt{m_i^2\, c^2 +
{\vec \kappa}_i^2} \approx 0$.} (point particle clock), and an
\textit{internal space} spanned by \textit{internal relative
3-variables} living on the Wigner hyper-planes (they identify the
particular isolated system and are Wigner spin-1 3-vectors) and with
the \textit{ internal 3-center of mass eliminated to avoid double
counting}. $M$ and $ {\vec {\bar S}}$ are (Lorentz scalar and Wigner
spin-1 3-vector) functions of the internal relative variables. As a
consequence we have a non-covariant pseudo-particle carrying a
\textit{pole-dipole structure} described by the internal relative
degrees of freedom. The natural inertial observer for this
description is the external Fokker-Pryce center of inertia (the only
covariant collective variable), which is also a function of $\tau$,
$\vec z$ and $\vec h$.
\bigskip

6) To eliminate the residual gauge freedom in the choice of
$x^{\mu}(\tau )$, i.e. of the inertial observer origin of the
3-coordinates on the Wigner hyper-planes, we add the gauge fixings
${\cal K}^r_{(int)} = {\bar S}^{\tau r} = - {\bar S}^{r\tau} \approx
0$. Its preservation in $\tau$ using the previous $H_D$ implies
$\vec \lambda (\tau ) = 0$, since ${\cal K}^r_{(int)}$ has
non-vanishing Poisson bracket with ${\cal P}^s_{(int)}$. Thus the
final Dirac Hamiltonian becomes $H_D = M\, c + (electromagnetic\,\,
constraints)$, i.e. the invariant mass of the isolated system. In
this way $x^{\mu}(\tau )$ is identified with the Fokker-Pryce
4-center of inertia $Y^{\mu}(\tau )$ and the embedding (\ref{2.a})
of the Wigner hyper-planes becomes

\beq
  z_W^{\mu}(\tau ,\vec \sigma ) = Y^{\mu}(\tau ) +
 \epsilon^{\mu}_r(\vec h)\, \sigma^r.
 \label{2.16}
 \eeq

A check of the consistency of this identification can be easily done
by putting Eq.(\ref{2.16}) into Eqs. (\ref{2.2}) and (\ref{2.5}):
Eqs.(\ref{2.15}) will be recovered if the rest-frame conditions
${\vec {\cal P}}_{(int)} \approx 0$ and ${\vec {\cal K}}_{(int)}
\approx 0$, of Eqs.(\ref{2.22}) hold.

\bigskip

The world-lines of the positive-energy particles are by definition
the derived quantities

\beq
 x^{\mu}_i(\tau ) = Y^{\mu}(\tau ) + \epsilon^{\mu}_r(\vec h)\,
\eta^r_i(\tau ),
 \label{2.17}
  \eeq

\noindent i.e. they are described by the Wigner spin-1 3-vectors
${\vec \eta}_i(\tau )$.\medskip

As shown in Ref.\cite{10b} for particle systems and as it will be
clarified in paper II for the system of particles plus
electro-magnetic field, the rest-frame conditions ${\vec {\cal
P}}_{(int)} \approx 0$ and their gauge fixings ${\vec {\cal
K}}_{(int)} \approx 0$ eliminate the internal 3-center of mass of
the whole isolated system. This implies that the 3-positions ${\vec
\eta}_i(\tau )$ (and also their conjugate momenta ${\vec
\kappa}_i(\tau )$) become functionals {\it only of a set of relative
coordinates and momenta} satisfying Hamilton equations governed by
the invariant mass $Mc$. Once these equations are solved and the
orbits ${\vec \eta}_i(\tau )$ are reconstructed, Eqs.(\ref{2.17})
lead to world-lines functions of $\tau$, of the non-evolving Jacobi
data $\vec z$, $\vec h$, and of the solutions for the relative
variables (Eq.(\ref{2.19}) has to be used for $Y^{\mu}(\tau
)$).\medskip

If ${\vec \kappa}_i(\tau )$ are the conjugate momenta,  with Poisson
brackets $\{ \eta^r_i(\tau ), \kappa_{js}(\tau ) \} = \delta_{ij}\,
\delta^r_s$, then the particle 4-momenta $p^{\mu}_i(\tau )$ are the
derived quantities $p_i^{\mu}(\tau ) = \sqrt{m_i^2\, c^2 + {\vec
\kappa}^2_i(\tau )}\, h^{\mu} - \epsilon^{\mu}_r(\vec h)\,
\kappa_{ir}(\tau )$ satisfying $p^2_i(\tau ) = m_i^2\, c^2$.

\bigskip

7) In Refs.\cite{8b} it is shown that the three relativistic
collective variables, originally defined in terms of the Poincare'
generators in Refs. \cite{13b}, \cite{8b}, can be expressed in terms
of the the variables $\tau$, $\vec z$, $\vec h$, $M$ and ${\vec
{\bar S}}$ (the spin or total baricentric angular momentum of the
isolated system) in the following way:\medskip

a) the pseudo-world-line of the canonical non-covariant 4-center of
mass (or center of spin) is

\bea
 {\tilde x}^{\mu}(\tau ) &=& \Big({\tilde x}^o(\tau ); {\tilde {\vec
x}}(\tau )\Big) = \Big(\sqrt{1 + {\vec h}^2}\, (\tau + {{\vec h
\cdot \vec z}\over {Mc}}); {{\vec z}\over {Mc}} + (\tau + {{\vec h
\cdot \vec z}\over {Mc}})\, \vec h\Big) =\nonumber \\
 &=& z^{\mu}_W(\tau
,{\tilde {\vec \sigma}}) = Y^{\mu}(\tau ) + \Big(0; {{ - {\vec {\bar
S}} \times \vec h}\over {Mc\, (1 + \sqrt{1 + {\vec h}^2})}}\Big),
 \label{2.18}
 \eea

\noindent  so that we get $Y^{\mu}(0) = \Big(\sqrt{1 + {\vec h}^2}\,
{{\vec h \cdot \vec z}\over {Mc}}; {{\vec z}\over {Mc}} + {{\vec h
\cdot \vec z}\over {Mc}}\, \vec h + {{{\vec {\bar S}} \times \vec
h}\over {Mc\, (1 + \sqrt{1 + {\vec h}^2})}}\Big)$ (we have used
\cite{8b} ${\tilde {\vec \sigma}} = {{- {\vec {\bar S}} \times \vec
h}\over {Mc\, (1 + \sqrt{1 + {\vec h}^2})}}$);\bigskip

b) the world-line of the non-canonical covariant Fokker-Pryce
4-center of inertia is

\bea
 Y^{\mu}(\tau ) &=& \Big({\tilde x}^o(\tau ); \vec Y(\tau )\Big) =
 \nonumber \\
 &=& \Big(\sqrt{1 + {\vec h}^2}\, (\tau + {{\vec h \cdot \vec z}\over
{Mc}});  {{\vec z}\over {Mc}} + (\tau + {{\vec h \cdot \vec z}\over
{Mc}})\, \vec h + {{{\vec {\bar S}} \times \vec h}\over {Mc\, (1 +
\sqrt{1 +
{\vec h}^2})}} \Big) = z_W^{\mu}(\tau ,\vec 0);\nonumber \\
 &&{}
 \label{2.19}
 \eea

\bigskip

c) the pseudo-world-line of the non-canonical non-covariant
M$\o$ller 4-center of energy is

\bea
 R^{\mu}(\tau ) &=& \Big({\tilde x}^o(\tau ); \vec R(\tau )\Big) =
 \nonumber \\
 &=&\Big(\sqrt{1 + {\vec h}^2}\, (\tau + {{\vec h \cdot \vec z}\over
{Mc}}); {{\vec z}\over {Mc}} + (\tau + {{\vec h \cdot \vec z}\over
{Mc}})\, \vec h - {{ {\vec {\bar S}} \times \vec h}\over {Mc\,
\sqrt{1 + {\vec h}^2}\, (1 + \sqrt{1 + {\vec h}^2})}} \Big) =\nonumber \\
 &=& z^{\mu}_W(\tau ,{\vec \sigma}_R) = Y^{\mu}(\tau ) + \Big(0; {{- \,
{\vec {\bar S}} \times \vec h}\over {Mc\, \sqrt{1 + {\vec
h}^2}}}\Big),
 \label{2.20}
 \eea

\noindent (we have used \cite{8b} ${\vec \sigma}_R = {{-\, {\vec
{\bar S}} \times \vec h}\over {Mc\, \sqrt{1 + {\vec h}^2}}}$).

\subsection{The Internal Poincare' Generators}

Eqs.(\ref{2.2}) show that all the dependence on the isolated system
is hidden in the energy-momentum tensor $T^{AB}(\tau ,\vec \sigma )$
(see also Ref.\cite{10b}), like it happens in general relativity.

For the system "N charged particles plus electro-magnetic field" in
the inertial rest frame it has the following form \cite{14b} (the
Grassmann-valued electric charges eliminate diverging self-energies)
 \footnote{The \textit{internal Lorentz generators} are
determined by the rest-frame spin tensor ${\bar S}^{AB}$. See
Ref.\cite{14b} for the final form of the internal generators: $M$ is
given in Eq. (6.19) [Eq. (6.35) with use of the rest-frame
condition], ${\vec {\cal P}}_{(int)}$ in Eq. (5.49), ${\vec {\cal
J}}_{(int)}$ in Eq. (6.39), ${\vec {\cal K}}_{(int)}$ in Eq.
(6.46).} in the radiation gauge (see Subsection E)

\begin{eqnarray*}
T^{\tau\tau}(\tau ,\vec \sigma )&=& \sum_{i=1}^N\, \delta^3(\vec
\sigma - { \vec \eta}_i(\tau ))\, \sqrt{m^2_i\, c^2 + [{\vec
\kappa}_i(\tau ) - {\frac{{ Q_i}}{c}}\, {\ \vec A}_{\perp}(\tau
,{\vec \eta}_i(\tau ))]^2} + \nonumber
\\
&+&{\frac{1}{2\, c}}\, [\Big( {\vec \pi}_{\perp} + \sum_{i=1}^N\,
Q_i\, { \frac{{\ \vec \partial}}{{\triangle}}}\delta^3(\vec \sigma
-{\vec \eta}
_i(\tau))\Big)^2 + {\vec B}^2](\tau ,\vec \sigma ) =  \nonumber \\
&=& T^{\tau\tau}_{matter}(\tau ,\vec \sigma ) +
T^{\tau\tau}_{em}(\tau ,\vec
\sigma ),  \nonumber \\
 &&{}\nonumber \\
&&T^{\tau\tau}_{em}(\tau ,\vec \sigma ) = {\frac{1}{{2\, c}}}\,
\Big({\vec \pi}^2_{\perp} + {\vec B}^2\Big)(\tau, \vec \sigma ),
 \end{eqnarray*}

\begin{eqnarray*}
 T^{r\tau}(\tau ,\vec \sigma )&=&\sum_{i=1}^N\, \delta^3(\vec \sigma
-{\vec \eta} _i(\tau ))\, [\kappa_i^r(\tau ) - {\frac{{Q_i}}{c}}\,
A_{\perp}^r(\tau
,{\vec \eta}_i(\tau ))] +  \nonumber \\
&+& {\frac{1}{c}}\, [\Big( {\vec \pi}_{\perp} + \sum_{i=1}^N\, Q_i\,
{\frac{{ \vec \partial}}{ {\triangle}}}\, \delta^3(\vec \sigma
-{\vec \eta}_i(\tau ))
\Big)\times \vec B](\tau ,\vec \sigma ) =  \nonumber \\
&=& T^{r\tau}_{matter}(\tau ,\vec \sigma ) + T^{r\tau}_{em}(\tau
,\vec
\sigma ),  \nonumber \\
 &&{}\nonumber \\
&&T^{r\tau}_{em}(\tau ,\vec \sigma ) = {\frac{1}{c}}\, \Big({\vec
\pi} _{\perp} \times \vec B\Big)(\tau ,\vec \sigma ),
\end{eqnarray*}

\bea
 T^{rs}(\tau ,\vec \sigma )&=& \sum_{i=1}^N\, \delta^3(\vec
\sigma - {\vec \eta} _i(\tau ))\, {\frac{{\ [\kappa_i^r(\tau ) -
{\frac{{Q_i}}{c}}\, A_{\perp}^r(\tau ,{\vec \eta}_i(\tau ))]\, [
\kappa_i^s(\tau ) - {\frac{{Q_i} }{c}}\, A_{\perp}^s(\tau ,{\vec
\eta}_i(\tau ))]}}{{\sqrt{m_i^2\, c^2 + [{\ \vec \kappa}_i(\tau ) -
{\frac{{Q_i}}{c}}\, {\vec A}_{\perp}(\tau ,{\vec \eta
}_i(\tau ))]^2} }}} -  \nonumber \\
&-& {\frac{1}{c}}\, \Big[ {\frac{1}{2}}\, \delta^{rs}\, [\Big( {\vec
\pi} _{\perp} + \sum_{i=1}^N\, Q_i\, {\frac{{\vec
\partial}}{{\triangle}}}\, \delta^3(\vec \sigma - {\vec \eta}_i(\tau
))\Big)^2 + {\vec B}^2] -
\nonumber \\
&-&[\Big( {\vec \pi}_{\perp} + \sum_{i=1}^N\, Q_i\, {\frac{{\vec
\partial}}{
{\triangle}}}\, \delta^3(\vec \sigma - {\vec \eta}_i(\tau
))\Big)^r\, \Big( { \vec \pi}_{\perp} + \sum_{i=1}^N\, Q_i\,
{\frac{{\vec \partial}}{{\triangle}}
}\, \delta^3(\vec \sigma - {\vec \eta}_i(\tau ))\Big)^s +  \nonumber \\
&+&B^r\, B^s]\Big] (\tau ,\vec \sigma ) =  \nonumber \\
&=& T^{rs}_{matter}(\tau ,\vec \sigma ) + T^{rs}_{em}(\tau ,\vec
\sigma ),\nonumber \\
&&{}\nonumber \\
 &&T^{rs}_{em}(\tau ,\vec \sigma ) = -
{\frac{1}{c}}\, \Big[{\frac{1}{2}}\, \delta^{rs}\, \Big({\vec
\pi}^2_{\perp} + {\vec B}^2\Big) - \Big( \pi^r_{\perp}\,
\pi^s_{\perp} + B^r\, B^s\Big)\Big](\tau ,\vec \sigma ).
 \label{2.21}
\end{eqnarray}

\bigskip

By using $T^{AB}(\tau ,\vec \sigma )$ we can define an internal
realization of the Poincare' group acting inside the Wigner
hyper-planes. The resulting \textit{internal Poincare' generators},
acting in the internal space, are

\begin{eqnarray*}
\mathcal{E}_{(int)} &=& \mathcal{P}^{\tau }_{(int)}\, c = M\, c^2 =
c\, \int d^3\sigma\, T^{\tau\tau}(\tau ,\vec \sigma ) =  \nonumber \\
&=& c\, \sum_{i=1}^{N}\, \sqrt{ m_{i}^{2}\, c^2 + \Big({\vec{
\kappa}} _i(\tau ) - {\frac{{Q_i}}{c}}\, {\vec{A }}_{\perp }(\tau
,\vec{\eta} _i(\tau ))\Big)^2} +  \nonumber \\
&+&\sum_{i\neq j}\, \frac{Q_{i}\, Q_{j}}{4\pi\, \mid
\vec{\eta}_{i}(\tau ) - \vec{\eta} _{j}(\tau )\mid } +
{\frac{1}{2}}\, \int d^{3}\sigma \, [{\vec{
\pi }}_{\perp }^{2} + {\vec{B}}^{2}](\tau ,\vec{\sigma}) =
 \end{eqnarray*}

 \begin{eqnarray*}
 &=&c\, \sum_{i=1}^N\, \Big(\sqrt{m^2_i\, c^2 + {\vec
\kappa}^2_i(\tau )} - { \ \frac{{Q_i}}{c}}\, {\frac{{{\vec
\kappa}_i(\tau ) \cdot {\vec A} _{\perp}(\tau , {\vec \eta}_i(\tau
))}}{\sqrt{m^2_i\, c^2 + {\vec \kappa}
^2_i(\tau )}}} \Big) +  \nonumber \\
&+&\sum_{i\neq j}\, \frac{Q_{i}\, Q_{j}}{4\pi\, \mid
\vec{\eta}_{i}(\tau ) - \vec{\eta} _{j}(\tau )\mid } +
{\frac{1}{2}}\, \int d^{3}\sigma \, [{\vec{
\pi }}_{\perp }^{2} + {\vec{B}}^{2}](\tau ,\vec{\sigma}),
 \end{eqnarray*}

 \begin{eqnarray*}
 &\rightarrow_{c \rightarrow \infty}& (\sum_i\, m_i)\, c^2 + \sum_i\,
{\frac{{ {\vec \kappa}_i^2(\tau )}}{{2m_i}}} + \sum_{i\neq j}\,
\frac{Q_{i}\, Q_{j}}{ 4\pi\, \mid \vec{\eta}_{i}(\tau ) - \vec{\eta}
_{j}(\tau )\mid } - \nonumber \\
&-& {\frac{1}{c}}\, \sum_i\, Q_i\, {\frac{{{\vec \kappa}_i(\tau
)}}{{m_i}}} \cdot {\vec A}_{\perp}(\tau ,{\vec \eta}_i(\tau )) -
{\frac{1}{{c^2}}}\,
\sum_i\, {\frac{{{\vec \kappa}_i^4(\tau )}}{{8 m_i^3}}} +  \nonumber \\
&+& {\frac{1}{{c^3}}}\, \sum_i\, Q_i\, {\frac{{{\vec
\kappa}_i^2(\tau )}}{{2 m_i^2}}}\, {\frac{{{\vec \kappa}_i(\tau
)}}{{m_i}}} \cdot {\vec A}
_{\perp}(\tau ,{\vec \eta}_i(\tau )) +  \nonumber \\
&+& O(c^{-4}) + {\frac{1}{2}}\, \int d^{3}\sigma \, [{\vec{\pi
}}_{\perp}^{2} + {\vec{B}}^{2}](\tau ,\vec{\sigma}),
 \end{eqnarray*}

\begin{eqnarray*}
 \mathcal{\vec{P}}_{(int)} &=& \int d^3\sigma\, T^{r\tau}(\tau ,\vec
\sigma ) = \sum_{i=1}^N\, {\vec{\kappa}}_i(\tau ) + {\frac{1}{c}}\,
\int d^{3}\sigma\, \lbrack {\vec{\pi}}_{\perp } \times
{\vec{B}}](\tau ,\vec{\sigma}) \approx 0,  \nonumber \\
&&{}  \nonumber \\
\mathcal{J}_{(int)}^r &=& {\bar S}^r = {\frac{1}{2}}\,
\epsilon^{ruv}\, \int d^3\sigma\, \sigma^u\, T^{v\tau}(\tau
,\vec \sigma ) =  \nonumber \\
&=&\sum_{i=1}^{N}\,\Big(\vec{\eta}_{i}(\tau )\times
{\vec{\kappa}}_{i}(\tau ) \Big)^{r} + {\frac{1}{c}}\, \int
d^{3}\sigma (\vec{\sigma}\times \,\Big([{ \vec{\pi}}_{\perp }{\
\times }{\vec{B}}]\Big)^{r}(\tau ,\vec{\sigma}),
 \end{eqnarray*}

\begin{eqnarray*}
 \mathcal{K}_{(int)}^{r} &=&{\bar{S}}^{\tau r} = -
{\bar{S}}^{r\tau } = - \int d^3\sigma\, \sigma^r\, T^{\tau\tau}(\tau
,\vec \sigma ) =\nonumber \\
&=& - \sum_{i=1}^{N}\, \eta^r_{i}(\tau )\, \sqrt{m_{i}^{2}\, c^2 +
\Big({{\ \vec{\kappa} }}_{i}(\tau ) - {\frac{{Q_{i}}}{c}}\,
{\vec{A}}_{\perp }(\tau ,{
\ \ \vec{\eta}}_{i}(\tau))\Big)^2} +  \nonumber \\
&+& {\frac{1}{c}}\, \sum_{i=1}^{N}\, \Big[\sum_{j\not=i}^{1..N}\,
Q_{i}\, Q_{j}\, [{\ \frac{1}{{\ \triangle
_{{\vec{\eta}}_{j}}}}}{\frac{{\partial }}{{
\partial \eta _{j}^{r}}} }c({\vec{\eta}}_{i}(\tau ) - {\vec{\eta}}_{j}(\tau
)) - \eta _{j}^{r}(\tau )\, c({\ \vec{\eta}}_{i}(\tau ) -
{\vec{\eta}}_{j}(\tau ))] +  \nonumber \\
&+& Q_{i}\, \int d^{3}\sigma\, {\pi}_{\perp }^{r}(\tau
,\vec{\sigma})\, c( \vec{\sigma} - {\ \vec{\eta}}_{i}(\tau ))\Big] -
{\frac{1}{2c}}\, \int d^{3}\sigma\, \sigma ^{r}\,
({{\vec{\pi}}}_{\perp }^{2} + {{\vec{B} }}
^{2})(\tau ,\vec{\sigma}) =
 \end{eqnarray*}

 \bea
 &=& - \sum_{i=1}^{N}\, \eta^r_{i}(\tau )\, \Big( \sqrt{m^2_i\, c^2 +
{\vec \kappa}^2_i(\tau )} - {\frac{{Q_i}}{c}}\, {\frac{{{\vec
\kappa}_i(\tau ) \cdot {\vec A}_{\perp}(\tau , {\vec \eta}_i(\tau
))}}{\sqrt{m^2_i\, c^2 + {\
\vec \kappa}^2_i(\tau )}}} \Big) +  \nonumber \\
 &+& {\frac{1}{c}}\, \sum_{i=1}^{N}\, \sum_{j\not=i}^{1..N}\,
Q_{i}\, Q_{j}\, \Big[\int d^3\sigma\, {1\over {4\pi\, |\vec \sigma -
{\vec \eta}_j(\tau )|}}\, {{\partial}\over {\partial\, \sigma^r}}\,
{1\over {4\pi\, |\vec \sigma - {\vec \eta}_i(\tau )|}} +\nonumber \\
 &+&{{\eta^r_j(\tau )}\over {4\pi\, |{\vec \eta}_i(\tau ) - {\vec
\eta}_j(\tau )|}} \Big] -  \nonumber \\
 &-& {1\over c}\, \sum_{i=1}^N\, Q_{i}\, \int d^{3}\sigma\, {{{\pi}_{\perp }^{r}(\tau
,\vec{\sigma})} \over {4\pi\, |\vec{\sigma} - {\
\vec{\eta}}_{i}(\tau )|}} - {\frac{1}{2c}}\, \int d^{3}\sigma\,
\sigma ^{r}\, ({{\vec{\pi}}}_{\perp }^{2} + {{\vec{B} }}
^{2})(\tau ,\vec{\sigma}) \approx 0,  \nonumber \\
&&{}\nonumber \\
 &&{}\nonumber \\
 {\cal K}^r_{(int)} &\rightarrow_{c \rightarrow \infty}& c\, {\cal
 K}^r_{Galilei}  + O(c^{-1}) \approx 0,\nonumber \\
 &&{}\nonumber \\
 {\vec {\cal K}}_{Galilei} &=& - \sum_{i=1}^N\, m_1\, {\vec
 \eta}_i(\tau ) = - \Big(\sum_{i=1}^N\, m_i\Big)\, {\vec \eta}_{12} \approx 0,
  \label{2.22}
\end{eqnarray}

\noindent with $c({\vec{\eta}}_{i}-{\vec{\eta}}_{j}):=-1/(4\pi
|\vec{\eta} _{j}-\vec{ \eta}_{i}|)$.

\bigskip

Note that, as required by the Poincare' algebra in an instant form
of dynamics \cite{10b}, there are interaction terms both in the
internal energy {\it and} in the internal Lorentz boosts.

\bigskip

This realization of the Poincare' group is \textit{unfaithful} due
to the rest-frame conditions ${\vec {\mathcal{P}}}_{(int)} \approx
0$. Moreover also the internal boosts have been put equal to zero as
gauge fixings to the rest-frame conditions ${\vec
{\mathcal{P}}}_{(int)} \approx 0$. Since we have ${\vec
{\mathcal{K}}} _{(int)} = - M\, c\, {\vec R}_+ \approx 0$
\footnote{${\vec R}_+$ is the internal M$\o$ller 3-center of energy
\cite{10b}. Due to the rest-frame condition ${\vec {\cal P}}_{(int)}
\approx 0$, we have ${\vec q}_+ \approx {\vec R}_+ \approx {\vec
y}_+$, where ${\vec q}_+$ is the internal canonical 3-center of mass
and ${\vec y}_+$ is the internal Fokker-Pryce 3-center of inertia
\cite{10b}.}, the internal 3-center of mass (${\vec q}_+ \approx
{\vec R}_+ \approx {\vec y}_+ \approx 0)$ is eliminated \cite{10b},
avoiding a double counting.

\bigskip

The only surviving internal generators are $M$ and ${\vec
{\mathcal{J}}} _{(int)}$, which appear in the external Poincare'
generators.
\medskip

In Eqs.(\ref{2.22}) there is also the non-relativistic limit,
identifying the Galilei generators plus $1/c$ electro-magnetic
corrections in the rest frame and with the non-relativistic center
of mass ${\vec x}_{com}$ put in the origin of the coordinates (only
relative variables survive). From  Eqs.(\ref{2.15}) one can get the
Galilei generators associated with the description of the
non-relativistic center of mass of the particles.

\subsection{The Electro-Magnetic Field in the Radiation Gauge}

In the previous Subsection we used the radiation gauge for the
electro-magnetic field, because it is the one naturally selected by
the Shanmugadhasan canonical transformation identifying Darboux
bases of phase space adapted to Dirac first class constraints
\cite{7b,8b,10b}.\medskip

As already said (see also Ref.\cite{14b}), the electromagnetic
potential in the adapted radar 4-coordinates is $A_A(\tau ,\vec
\sigma ) = {\ \frac{{\partial\, z^{\mu}(\tau ,\vec \sigma
)}}{{\partial\, \sigma^A}}}\, A_{\mu}(z(\tau ,\vec \sigma ))$, so
that $ A_{\mu}(z(\tau ,\vec \sigma )) = {\ \frac{{\partial
\sigma^A(x)}}{{\partial x^{\mu}}}}\, A_A(\tau ,\vec \sigma )$ . In
the rest-frame instant form for the electromagnetic potential we
have (see Eq.(3.39) of Ref.\cite{14b})

\begin{eqnarray}
A^{\mu}\Big(Y^{\alpha}(\tau ) + \epsilon^{\alpha}_r(\vec h)\,
\sigma^r\Big) &=& h^{\mu}\, A^{\tau}(\tau ,\vec \sigma ) -
\epsilon^{\mu}_r(\vec h)\,
A^r(\tau ,\vec \sigma ) =  \nonumber \\
&=& h^{\mu}\, A^{\tau}(\tau ,\vec \sigma ) + \epsilon^{\mu}_r(\vec
h)\, {\ \frac{{\partial^r}}{{\triangle}}}\, \vec
\partial \cdot \vec A(\tau ,\vec \sigma ) - \epsilon^{\mu}_r(\vec h)\,
A^r_{\perp}(\tau ,\vec \sigma ),\nonumber \\
&&{}  \nonumber \\
&&A^{\tau}(\tau ,\vec \sigma ) = h^{\mu}\, A_{\mu}\Big(h^{\alpha}\,
\tau + \epsilon^{\alpha}_r(\vec h)\, \sigma^r\Big),
 \label{2.23}
\end{eqnarray}

\noindent where $\triangle = - {\vec \partial}^2$ and $\vec \partial
\cdot {\vec A}_{\perp}(\tau ,\vec \sigma ) \equiv 0$.
\medskip

In the Dirac Hamiltonian (\ref{2.4}) the term $\int d^3\sigma\,
\Big[\lambda_{\tau}\, \pi^{\tau} - A_{\tau}\, \Gamma\Big](\tau ,\vec
\sigma )$ is the generator of the electro-magnetic gauge
transformations. The two first class constraints (\ref{2.3}) imply
that $A_{\tau}(\tau ,\vec \sigma )$ and one component of $A_r(\tau
,\vec \sigma )$ are gauge variables. The second natural gauge
variable at the Hamiltonian level aside from $A_{\tau}$ is $\eta (\tau ,\vec \sigma )= -
{1\over {\triangle}}\, \vec \partial \cdot \vec A(\tau ,\vec \sigma
)$, since it is canonically conjugate to the Gauss law: $\{ \eta
(\tau ,\vec \sigma ), \Gamma (\tau ,{\vec \sigma}_1) \} =
\delta^3(\vec \sigma - {\vec \sigma}_1))$.\medskip

Therefore in the rest-frame instant form the natural gauge fixing is
$\eta (\tau ,\vec \sigma ) \approx 0$, namely the Coulomb gauge
$\vec \partial \cdot \vec A(\tau ,\vec \sigma ) \approx 0$. Its
preservation in $\tau$ implies the gauge fixing determining the
scalar potential: $A_{\tau}(\tau ,\vec \sigma ) \approx
\sum_{i=1}^N\, {{Q_i}\over {4\pi\, |\vec \sigma - {\vec \eta}_i(\tau
)|}}$. The $\tau$-preservation of this secondary gauge fixing
implies $\lambda_{\tau}(\tau ,\vec \sigma ) = 0$.\medskip

After this elimination of the electro-magnetic gauge degrees of
freedom we remain only with transverse electromagnetic degrees of
freedom (they are a canonical basis of Dirac observables of the
electro-magnetic field). This is  a \textit{rest-frame radiation
gauge} where the final form of the Dirac Hamiltonian is $H_D = Mc$
and where we have

\begin{eqnarray}
&&A^{\tau}(\tau ,\vec \sigma ) = \sum_{i=1}^N\, {{Q_i}\over {4\pi\,
|\vec \sigma - {\vec \eta}_i(\tau )|}},\nonumber \\
&& \Rightarrow\,\, h^{\mu}\, A_{\mu}\Big(h^{\alpha}\, \tau +
\epsilon^{\alpha}_r(\vec h)\, \sigma^r\Big)
= \sum_{i=1}^N\, {{Q_i}\over {4\pi\, |\vec \sigma - {\vec \eta}_i(\tau )|}},\nonumber \\
&&{}  \nonumber \\
&&\vec \partial \cdot \vec A(\tau ,\vec \sigma ) = 0,\,\,
\Rightarrow\,\, - \epsilon^{\mu}_r(\vec h)\,
{\frac{{\partial}}{{\partial\, \sigma^r}}}\,
A_{\mu}\Big(h^{\alpha}\,
\tau + \epsilon^{\alpha}_r(\vec h)\, \sigma^r\Big) =\nonumber \\
&&\qquad = - \sum_r\, \epsilon^{\mu}_r(\vec h)\,
\epsilon^{\nu}_r(\vec h)\, {\ \frac{{\partial\,
A_{\mu}(x)}}{{\partial\, x^{\nu}}}} = \Big(\eta^{\mu\nu} - h^{\mu}\,
h^{\nu}\Big)\, {\frac{{\partial\, A_{\mu}(x)}}{{\partial\,
x^{\nu}}}} =  \nonumber \\
&&\qquad = \partial_{\mu}\, A^{\mu}(x){|}_{x = z(\tau ,\vec \sigma
)} = 0.
 \label{2.24}
\end{eqnarray}

Therefore this radiation gauge is a particular case of Lorentz gauge
where $A^{\mu}\Big(Y^{\alpha}(\tau ) + \epsilon^{\alpha}_r(\vec h)\,
\sigma^r\Big) = h^{\mu}\, \sum_{i=1}^N\, {{Q_i}\over {4\pi\, |\vec
\sigma - {\vec \eta}_i(\tau )|}} - \epsilon^{\mu}_r(\vec h)\,
A^r_{\perp}(\tau ,\vec \sigma )$.

\bigskip

If ${\vec \pi}_{\perp}(\tau ,\vec \sigma ) =  {\vec E}_{\perp}(\tau
,\vec \sigma )\,$ ($ \cir\, - {{\partial\, {\vec A}_{\perp}(\tau
,\vec \sigma )}\over {\partial\, \tau}}$ due to the first half of
Hamilton equations) is the momentum conjugate to ${\vec
A}_{\perp}(\tau ,\vec \sigma )$, we have the Poisson brackets

\beq
 \{ A^r_{\perp}(\tau ,\vec \sigma ), \pi^s_{\perp}(\tau ,{\vec
 \sigma}_1) \} = - c\, P^{rs}_{\perp}(\vec \sigma )\, \delta^3(\vec
 \sigma - {\vec \sigma}_1),
 \label{2.25}
 \eeq

\noindent where $P^{rs}_{\perp}(\vec \sigma ) = \delta^{rs} +
{{\partial^r\, \partial^s}\over {\triangle}}$, $\triangle = - {\vec
\partial}^2$, $\partial^r = - \partial_r = - {{\partial}\over {\partial \sigma^r}}$.

\subsection{The M$\o$ller Radius and the Comparison with other
Approaches to Relativistic Mechanics}

As said in Refs.\cite{7b,8b}, if in an arbitrary fixed Lorentz frame
we draw  the pseudo-world-lines corresponding to the position of
${\tilde x}^{\mu}(\tau)$ and $R^{\mu}(\tau)$ in all possible
inertial frames \footnote{In each Lorentz frame one has different
pseudo-world-lines describing $R^{\mu}(\tau)$ and ${\tilde
x}^{\mu}(\tau)$: the canonical 4-center of mass ${\tilde
x}^{\mu}(\tau)$ {\it lies in between} $Y^{\mu}(\tau)$ and
$R^{\mu}(\tau)$ in every (non rest)-frame.}, it turns out that they
fill a world-tube \cite{11b}, the M$\o$ller world-tube, around the
world-line $Y^{\mu}(\tau)$ of the covariant non-canonical
Fokker-Pryce 4-center of inertia $Y^{\mu}(\tau)$. The {\it invariant
radius} of the tube is $\rho =\sqrt{- W^2}/p^2 = |\vec
S|/\sqrt{P^2}$ where ($W^2 = - P^2\, {\vec {\bar S}}^2$ is the
Pauli-Lubanski invariant when $P^2
> 0$). This classical intrinsic radius delimitates the
non-covariance effects (the pseudo-world-lines) of the canonical
4-center of mass ${\tilde x}^{\mu}(\tau)$ \footnote{In the
rest-frame the world-tube is a cylinder: in each instantaneous
3-space there is a disk of possible positions of the canonical
3-center of mass orthogonal to the spin. In the non-relativistic
limit the radius $\rho$ of the disk tends to zero and we recover the
non-relativistic center of mass.}.\bigskip

The existence of the M$\o$ller world-tube for rotating systems is a
consequence of the Lorentz signature of Minkowski space-time. It
identifies a region which cannot be explored without a breaking of
manifest Lorentz covariance: this implies a limitation on the
localization of the canonical center of mass due to its
frame-dependence. Moreover it leads to the identification of a {\it
fundamental length}, the M$\o$ller radius, associated with every
configuration of an isolated system and built with its {\it global}
Poincare' Casimirs. This unit of length is really remarkable for the
following two reasons:\bigskip

A) At the quantum level $\rho$ becomes the Compton wavelength of the
isolated system times its spin eigenvalue $\sqrt{s(s+1)}$ , $\rho
\mapsto \hat \rho = \sqrt{s(s+1)} \hbar /M = \sqrt{s(s+1)}
\lambda_M$ with $M = \sqrt{P^2}$ the invariant mass and
$\lambda_M=\hbar /M$ its associated Compton wavelength. Therefore
the region of frame-dependent localization of the canonical center
of mass is also the region where classical relativistic physics is
no longer valid, because any attempt to make a localization more
precise than the Compton wavelength at the quantum level leads to pair
production. The interior of the classical M$\o$ller world-tube must
be described by using quantum mechanics!\medskip

In string theory extended Heisenberg relations $\triangle\, x =
{{\hbar}\over {\triangle\, p}} + {{\triangle\, p}\over {\alpha^{'}}}
= {{\hbar}\over {\triangle\, p}} + {{l_s^2\, \triangle\, p}\over
{\hbar}}$ ($l_s = \sqrt{\hbar\, \alpha^{'}}$ is the fundamental
string length) have been proposed \cite{19b} to get the lower bound
$\triangle\, x > l_s$ (due to the $y + 1/y$ form) forbidding the
exploration of distances below $l_s$. By replacing $l_s$ with the
M$\o$ller radius of an isolated system and $x$ with its
Newton-Wigner 3-center of mass ${\vec x}_{NW} = \vec z/Mc$ in the
modified Heisenberg relations, $\triangle\, x^r_{NW} = {{\hbar}\over
{\triangle\, p^r}} + {{\rho^2\, \triangle\, p^r}\over {\hbar}}$, we
could obtain the impossibility ($\triangle\, x^r > \rho$) to explore
the interior of the M$\o$ller world-tube of the isolated system
\cite{7b}. This would be compatible with a non-self-adjoint
Newton-Wigner position operator after quantization.\medskip

Moreover, the M$\o$ller radius of a field configuration (think of
the radiation field discussed in the next Subsection H) could be a
candidate for a physical (configuration-dependent) ultraviolet
cutoff in QFT \cite{7b}.

\bigskip

B) As shown in Refs. \cite{11b}, where the M$\o$ller world-tube was
introduced in connection with the M$\o$ller center of energy
$R^{\mu}(\tau )$, an extended rotating
relativistic isolated system with the material radius smaller than its
intrinsic radius $\rho$ one has: i) its peripheral rotation velocity
can exceed the velocity of light; ii) its classical energy density
cannot be positive definite everywhere in every frame
\footnote{Classically, energy density is always positive and the
stress-energy tensor for all classical fields satisfies the weak
energy condition $T_{\mu\nu}\, u^{\mu}\, u^{\nu} \geq 0$, where
$u^{\mu}$ is any time-like or null vector. In a sense the M$\o$ller
world-tube is a classical version of the Epstein, Glaser, Jaffe theorem
\cite{20b} in QFT: if a field $Q(x)$ satisfies $< \Psi | Q(x) | \Psi
>\, \geq 0$ for all states and if $< \Omega | Q(x) | \Omega >\, = 0$
for the vacuum state, then $Q(x) = 0$. Therefore in QFT the weak
energy condition does not hold for the renormalized stress-energy
tensor. Since it has by definition a null vacuum expectation value,
there are states $| Y >$ such that $< Y | T_{\mu\nu}\, u^{\mu}\,
u^{\nu} | Y > \, <\, 0$. This holds both for the scalar field and
for the squeezed state of the electromagnetic field (see also
Ref.\cite{6b}).}. Therefore the M$\o$ller radius $\rho$ is also a
remnant of the energy conditions of general relativity in flat
Minkowski space-time \cite{7b}.

\bigskip

Let us add some comments clarifying the connection of the rest-frame
instant form of dynamics with the other existing forms of
relativistic mechanics.\medskip

In Refs.\cite{10b,21b} there are reviews of the various approaches
to classical relativistic mechanics induced by the No-Interaction
theorem \cite{22b}, which shows the existence of a conflict among
the following three requirements: i) Hamiltonian dynamics; ii)
relativistic invariance; iii) manifestly covariant world-lines;
except in the case of free particles \footnote{This theorem appears
also in non-relativistic mechanics \cite{23b}, if one reformulates
it as a many-time theory, namely with an independent time variable
for each particle plus suitable first-class constraints. The gauge
fixings of these constraints, identifying  of the particle times
with the Newton time, imply the recovering of the standard one-time
Newton mechanics. The problem in special relativity is the absence
of an absolute time and of an absolute instantaneous 3-space.}. The
way out from this theorem turns out to be the non-covariance of the
relativistic canonical (Newton-Wigner) 3-center of mass.\medskip

The existing approaches to relativistic mechanics besides the
present one may be classified in three main groups (see
Refs.\cite{10b,21b} for more bibliography):\medskip

A) The models in which each particle is described in phase space by
8 canonical variables $x^{\mu}_i(\tau )$, $p^{\mu}_i(\tau )$ (with
$x^{\mu}_i(\tau )$ assumed to describe the world-line of particle
$i$) and by a modified mass-shell first-class constraint $\chi_i =
p^2_i - m_i^2\, c^2 + V_i \approx 0$. These constraints eliminate
the variables $p^o_i$ and imply that the time variables $x^o_i$ are
gauge variables (absence of physical relative times) to be fixed
with gauge fixings (the clock synchronization problem). The
non-linear first-class property $\{ \chi_i, \chi_j \} \approx 0$
determines the arguments of the allowed potentials $V_i$. Only for
the two-body case solutions  are known. The simplest solution
\cite{24b}, \cite{25b}, is $V_1 = V_2 = V(r^2_{\perp})$ with
$r^{\mu}_{\perp} = (\eta^{\mu\nu} - {{P^{\mu}\, P^{\nu}}\over
{P^2}})\, (x_{1\nu} - x_{2\nu})$, $P^{\mu} = p_1^{\mu} + p_2^{\mu}$:
therefore the interaction is instantaneous in the rest frame (see
Ref.\cite{17b} for its quantization). However, since we have $\{
x^{\mu}_i, \chi_j \} \not= 0$ for all $i,j$, the canonical variables
$x^{\mu}_i(\tau )$ (and therefore also the world-lines) are not
Dirac observables. As a consequence, each gauge fixing on the times
$x^o_i$'s generates different world-lines, whose totality spans a
so-called world-sheet (see Komar in Ref.\cite{24b}). Moreover the
gauge fixings must satisfy the so-called {\it world-line conditions}
(WLC) \cite{26b}, according to which the manifest covariance of the
world-lines after a good gauge fixing  is saved by correcting the
Lorentz transformations with suitable gauge transformations
generated by the first-class constraints $\chi_i \approx 0$.\medskip

Let us remark that this class of models can be connected to the
rest-frame instant form by leaving the {\it derived} world-lines as
in Eq.(\ref{2.17}) but modifying the {\it derived} momenta from the
form given after Eqs.(\ref{2.17}) to the new form ${\tilde
p}^{\mu}_i(\tau ) = \Big(\sqrt{m_i^2\, c^2 + \Big[{\vec
\kappa}_i(\tau ) - Q_i\, {\vec A}_{\perp}(\tau ,{\vec \eta}_i(\tau
))\Big]^2} - Q_i\, V(\tau, {\vec \eta}_i(\tau))\Big)\, h^{\mu} -
\epsilon^{\mu}_r(\vec h)\, \kappa_{ir}(\tau)$, where $V(\tau ,\vec
\sigma) = \sum_j\, {{Q_j}\over {4\pi\, |\vec \sigma - {\vec
\eta}_i(\tau )|}}$ is the radiation gauge term along $h^{\mu}$ of
the gauge potential $A^{\mu}(Y^{\alpha}(\tau) +
\epsilon^{\mu}_r(\vec h)\, \sigma^r) = V(\tau ,\vec \sigma )\,
h^{\mu} - \epsilon^{\mu}_r(\vec h)\, A^r_{\perp}(\tau ,\vec \sigma
)$ given after Eq.(\ref{2.24}), because in this case we have
$\Big({\tilde P}_i(\tau) - Q_i\, A(x_i(\tau))\Big)^2 = m_i^2\, c^2$.

\medskip

B) If we add $N-1$ gauge fixings to the N first-class constraints in
A), implying that the N particles are simultaneously interacting in
a particular inertial or non-inertial frame, then the world-lines
are well defined and a Lagrangian description becomes possible
\cite{27b}. Since the natural, intrinsically defined, frame is the
rest frame, these models are the ancestors of the rest-frame instant
form of dynamics.\medskip

Both the approaches A) and B) have to face the problem of {\it
separability or cluster decomposition property} \cite{28b} (see also
Todorov in Ref.\cite{24b}) essential for scattering theory at
the quantum level: with non-confining potentials falling off at big
inter-particle separations we must recover the world-lines of free
particles. The models B) are natural for confining potentials.

C) A third, non-Hamiltonian approach started with the Currie-Hill
WLC \cite{29b} and led to {\it predictive mechanics} \cite{30b},
where it is emphasized that the world-line 4-coordinates
$q^{\mu}_i(\tau_i)$ are each one labeled by the proper time of
particle $i$ also in presence of interactions, so that a many-time
formulation of relativistic mechanics with a non-linear realization
of the Poincare' algebra can be formulated. Droz Vincent \cite{25b}
found the Hamiltonian reformulation of predictive mechanics and
showed that the covariant predictive 4-coordinates $q^{\mu}_i(\tau
)$ are not canonical, $\{ q^{\mu}_i, q^{\nu}_j \} \not= 0$ for any
pair $i,j$, (except in the free case), so that they cannot coincide
with the canonical 4-coordinates $x^{\mu}_i(\tau )$ of the approach
A).\medskip

The rest-frame instant form has the independent canonical variables
$\vec z$, $\vec h$  (with $\vec z$ non-covariant, thus avoiding the
No-Interaction theorem), ${\vec \rho}_a(\tau )$, ${\vec \pi}_a(\tau
)$, $a = 1,..,N-1$ (the relative canonical variables), and rebuilds
the particle world-lines as derived quantities, $x^{\mu}_i(\tau ) =
z^{\mu}(\tau , {\vec \eta}_i(\tau ))$ with the ${\vec \eta}_i(\tau
)$ function of the relative variables due to the rest-frame
conditions. It turns out that what we denoted with $x^{\mu}_i(\tau
)$ are not the canonical 4-coordinates of approach A), but must be
interpreted as a realization of the covariant non-canonical
predictive 4-coordinates $q^{\mu}_i(\tau )$. Moreover, our derived
world-lines satisfy the cluster decomposition property when the
potentials in the internal mass $M$ are non-confining
\footnote{Since they depend upon the relative variables, on the
solution of the Hamilton equations with Hamiltonian the internal
mass $M$ they know the type of potential. For non-confining
potentials, modulo tails the on-shell relative variables tend to
their value for free particles.}, even if our independent canonical
variables give a spatially non-separable description of the isolated
system (induced by the notion of relativistic center of mass, by the
structure of the Poincare' group and by the clock synchronization
problem).

\medskip

Let us also remark that, as shown in Ref.\cite{8b}, the 4-vector
$Y^{\mu}(\tau )$ is non canonical since $\{ Y^{\mu}, Y^{\nu} \}
\not= 0$ (it is a function of $P^{\mu}$ and $\vec S$): as a
consequence there is a non-commutative structure associated to it
already at the classical level. The same happens for the 4-vectors
$x^{\mu}_i(\tau )$. Are these non-commutative structures playing any
role at the quantum level?

\bigskip

The advantage of the rest-frame instant form of dynamics is to allow
the explicit treatment of N-body problems also in presence of
interactions, while in all the previously quoted approaches it was
possible to study in detail only the $N = 2$ case. Moreover only
with this approach  has it been possible to make contact with the
treatment of relativistic bound states in such a way to recover the
Darwin and Salpeter potentials starting from the classical theory
\cite{14b,15b}.

\subsection{The Non-Relativistic Limit of the Rest-Frame Instant
Form}

Let us consider the non-relativistic limit of two positive-energy
scalar free particles, disregarding the electro-magnetic field with
its two {\it electric} and {\it magnetic} limits \cite{4b}.\bigskip

The particles are described by the Newtonian canonical variables
${\vec x}_{(n)\, i}$, ${\vec p}_{(n)\, i}$, $i=1,2$, or by the
canonically equivalent center-of-mass and relative variables ${\vec
x}_{(n)}$, ${\vec p}_{(n)}$, ${\vec r}_{(n)}$, ${\vec q}_{(n)}$ (see
Ref.\cite{31b} for the case of N particles)

\bea
 {\vec x}_{(n)} &=& {1\over m}\, \sum_{i=1}^2\, m_i\, {\vec
 x}_{(n)\, i},\qquad {\vec p}_{(n)} = \sum_{i=1}^2\, {\vec p}_{(n)\,
 i},\qquad m = m_1 + m_2,\nonumber \\
 {\vec r}_{(n)} &=& {\vec x}_{(n)\, 1} - {\vec x}_{(n)\, 2},\qquad
 {\vec q}_{(n)} = {1\over m}\, \Big(m_2\, {\vec p}_{(n)\, 1} - m_1\,
 {\vec p}_{(n)2}\Big),\nonumber \\
 &&{}\nonumber \\
 {\vec x}_{(n)\, 1} &=& {\vec x}_{(n)} + {{m_2}\over m}\, {\vec
 r}_{(n)},\qquad {\vec x}_{(n)\, 2} = {\vec x}_{(n)} - {{m_1}\over
 m}\, {\vec r}_{(n)},\nonumber \\
 {\vec p}_{(n)\, 1} &=& {{m_1}\over m}\, {\vec p}_{(n)} + {\vec
 q}_{(n)},\qquad {\vec p}_{(n)\, 2} = {{m_2}\over m}\, {\vec
 p}_{(n)} - {\vec q}_{(n)}.
 \label{2.26}
 \eea

The generators of the centrally extended Galilei algebra are (we
have changed the sign of the Galilei boosts with respect to
Refs.\cite{32b})

 \bea
 E_{Galilei} &=& \sum_{i=1}^2\, {{{\vec p}_{(n)\, i}^2}\over {2m_i}} =
 {{{\vec p}^2_{(n)}}\over {2m}} + {{{\vec q}^2_{(n)}}\over {2\mu}},\qquad {1\over
 {\mu}} = {1\over {m_1}} + {1\over {m_2}},\nonumber \\
 {\vec P}_{Galilei} &=& {\vec p}_{(n)} = \sum_{i=1}^2\, {\vec p}_{(n)\,
 i},\nonumber \\
 {\vec J}_{Galilei} &=& \sum_{i=1}^2\, {\vec x}_{(n)\, i} \times
 {\vec p}_{(n)\, i} = {\vec x}_{(n)} \times {\vec p}_{(n)} + {\vec
 S}_{(n)},\qquad {\vec S}_{(n)} = {\vec r}_{(n)} \times {\vec
 q}_{(n)},\nonumber \\
 {\vec K}_{Galilei} &=& t\, {\vec p}_{(n)} - m\, {\vec
 x}_{(n)},\nonumber \\
 &&{}\nonumber \\
 \{ E_{Galilei}, {\vec K}_{Galilei} \} &=& {\vec P}_{Galilei},\qquad
 \{ P^i_{Galilei}, K^j_{Galilei} \} = m\, \delta^{ij},\qquad \{
 K^i_{Galilei}, K^j_{Galilei} \} = 0,\nonumber \\
 \{ A^i, J^j_{Galilei} \} &=& \epsilon^{ijk}\, A^k, \qquad \vec A =
 {\vec P}_{Galilei}, {\vec J}_{Galilei}, {\vec K}_{Galilei}.
 \label{2.27}
 \eea

The main property of the Galilei algebra is that the presence of
interactions changes the energy, $E_{Galilei}\, \rightarrow\,
E^{'}_{Galilei} = E_{Galilei} + V({\vec r}_{(n)})$ but not the
Galilei boosts \footnote{This is the reason why there is no
"No-Interaction Theorem" in Newtonian mechanics, so that Newtonian
kinematics is trivial. However, as already said, this theorem
reappears when we make a many-time reformulation of Newtonian
mechanics \cite{23b}.}.
\bigskip

Also at the non-relativistic level the 2-body system can be
presented as a decoupled particle, the external center of mass
${\vec x}_{(n)}(t)$ with momentum ${\vec p}_{(n)}$, of mass $m$ in
the absolute Euclidean 3-space carrying an internal space of
relative variables (${\vec r}_{(n)}(t)$, ${\vec q}_{(n)}(t)$) with
Hamiltonian $H_{rel} = {{{\vec q}_{(n)}^2}\over {2\, \mu}}$ and rest
spin ${\vec S}_{(n)}$. The external center of mass  is associated
with an external realization of the Galilei group with generators
$E_{Galilei} = {{{\vec p}_{(n)}^2}\over {2m}} + H_{rel}$, ${\vec
P}_{Galilei} = {\vec p}_{(n)}$, ${\vec J}_{Galilei} = {\vec x}_{(n)}
\times {\vec p}_{(n)} + {\vec S}_{(n)}$, ${\vec K}_{Galilei} = t\,
{\vec p}_{(n)} - m\, {\vec x}_{(n)}(t)$. The internal space can be
identified with the rest frame (${\vec p}_{(n)} \approx 0$) if we
choose the origin of 3-coordinates in the external center of mass
(${\vec x}_{(n)}(t) \approx 0$): in it the particles variables are
${\vec \eta}_{(n)i}(t) = {\vec x}_{(n)i}(t){|}_{{\vec x}_{(n)} =
{\vec p}_{(n)} = 0}$, ${\vec \kappa}_{(n)i}(t) = {\vec
p}_{(n)i}(t){|}_{{\vec x}_{(n)} = {\vec p}_{(n)} = 0}$ (they are the
non-relativistic counterpart of the variables ${\vec \eta}_i(\tau
)$, ${\vec \kappa}_i(\tau )$ on the instantaneous Wigner 3-spaces).
With this identification we get a unfaithful internal realization of
the Galilei group with generators ${\cal E}_{Galilei} = H_{rel}$,
${\vec {\cal P}}_{Galilei} = {\vec p}_{(n)} \approx 0$ (the
rest-frame conditions), ${\vec {\cal J}}_{Galilei} = {\vec
S}_{(n)}$, ${\vec {\cal K}}_{Galilei} = t\, {\vec p}_{(n)} - m\,
{\vec x}_{(n)}(t) \approx 0$ (the gauge fixings to the rest-frame
conditions implying ${\vec x}_{(n)}(t) \approx 0$). Inside the
internal space we have ${\vec x}_{(n)1} \approx {\vec \eta}_{(n)1} =
{{m_2}\over m}\, {\vec r}_{(n)}$, ${\vec x}_{(n)2} \approx {\vec
\eta}_{(n)2} = - {{m_1}\over m}\, {\vec r}_{(n)}$, ${\vec p}_{(n)1}
\approx {\vec \kappa}_{(n)1} = {\vec q}_{(n)}$, ${\vec p}_{(n)2}
\approx {\vec \kappa}_{(n)2} = - {\vec q}_{(n)}$ and we can
introduce the following auxiliary variables (having an obvious
relativistic counterpart) ${\vec \rho}_{(n)12} = {\vec \eta}_{(n)1}
- {\vec \eta}_{(n)2} = {\vec r}_{(n)}$, ${\vec \pi}_{(n)12} =
{{m_2}\over m}\, {\vec \kappa}_{(n)1} - {{m_1}\over m}\, {\vec
\kappa}_{(n)2} = {\vec q}_{(n)}$, ${\vec \eta}_{(n)12} = {{m_1}\over
m}\, {\vec \eta}_{(n)1} + {{m_2}\over m}\, {\vec \eta}_{(n)2}
\approx 0$, ${\vec \kappa}_{(n)12} = {\vec \kappa}_{(n)1} + {\vec
\kappa}_{(n)2} \approx 0$.

\bigskip

In the relativistic rest-frame instant form the two-particle system
is described by\medskip

1) the external center-of-mass frozen Jacobi data $\vec z$, $\vec
h$,, carrying the internal mass $M\, c = \sum_{i=1}^2\,
\sqrt{m_i^2\, c^2 + {\vec \kappa}_i^2}$ and the spin $\vec S =
\sum_{i=1}^2\, {\vec \eta}_i \times {\vec \kappa}_i$;\medskip

2) the two pairs of Wigner 3-vectors ${\vec \eta}_i$, ${\vec
\kappa}_i$, $i=1,2$, or by the canonically equivalent variables

\bea
 {\vec \eta}_{12} &=& {1\over m}\, \sum_{i=1}^2\, m_i\, {\vec \eta}_i,
 \qquad {\vec \kappa}_{12} = \sum_{i=1}^2\, {\vec \kappa}_i,
 \nonumber \\
  {\vec \rho}_{12} &=& {\vec \eta}_1 - {\vec \eta}_2,\qquad {\vec \pi}_{12} =
{1\over m}\, \Big(m_2\, {\vec \kappa}_1 - m_1\, {\vec
\kappa}_2\Big),
 \label{2.28}
 \eea

\noindent restricted by the rest-frame conditions ${\vec
\kappa}_{12} \approx 0$ (so that ${\vec \pi}_{12} \approx {\vec
\kappa}_1 \approx - {\vec \kappa}_2$) and $- \sum_{i=1}^2\, {\vec
\eta}_i\, \sqrt{m_i^2\, c^2 + {\vec \kappa}_i^2} \approx 0$
(elimination of the internal 3-center of mass).\medskip

In terms of these variables we can rebuild the world-lines
$x^{\mu}_i$ of Eq.(\ref{2.17}) and the 4-momenta $p^{\mu}_i$.

\bigskip

Since in the non-relativistic limit we have $\vec P = {\vec
p}_{(n)}$, $\vec h = {{\vec P}\over {M\, c}}\, \rightarrow_{c
\rightarrow \infty}\, 0$, implying $u^{\mu}(P)\, \rightarrow_{c
\rightarrow \infty}\, \Big( 1; \vec 0\Big)$ and
$\epsilon^{\mu}_r(u(P))\, \rightarrow_{c \rightarrow \infty}\, \Big(
0; \delta^i_r\Big)$, it turns out that $\tau/c$, ${\tilde x}^o/c$,
$Y^o/c$, $R^o/c$  and $x^o_i/c$ all become the absolute Newton time
$t$.
\medskip

Moreover from Subsection C we have the following results:\medskip

A) In the reference inertial system we get ${\tilde {\vec x}}(\tau
),\, \vec Y(\tau ),\, \vec R(\tau )\, \rightarrow_{c \rightarrow
\infty}\, {\vec x}_{(n)}(t)$, ${\vec x}_{NW} = {{\vec z}\over
{Mc}}\, \rightarrow_{c \rightarrow \infty}\, {\vec x}_{(n)}(0)$
because Eq.(\ref{2.13}) implies $\vec z\, \rightarrow_{c \rightarrow
\infty}\, \infty$ and $\vec h \cdot \vec z \, \rightarrow_{c
\rightarrow \infty}\, {\vec p}_{(n)} \cdot \Big({\vec x}_{(n)}(t) -
{{{\vec p}_{(n)}}\over m}\, t\Big) = {\vec p}_{(n)} \cdot {\vec
x}_{(n)}(0)$ (it is a Jacobi data of the non-relativistic theory).
\medskip

B) In the inertial rest frame, ${\vec p}_{(n)} \approx 0$, we get
${\vec \eta}_i(\tau )\, \rightarrow_{c \rightarrow \infty}\, {\vec
\eta}_{(n)i}(t)$, ${\vec \kappa}_i(\tau )\, \rightarrow_{c
\rightarrow \infty}\, {\vec \kappa}_{(n)i}(t)$, ${\vec x}_i(\tau )\,
\rightarrow_{c \rightarrow \infty}\, {\vec x}_{(n)}(t ) + {\vec
\eta}_{(n)i}(t )$, ${\vec p}_i(\tau )\, \rightarrow_{c \rightarrow
\infty}\, {\vec \kappa}_{(n)i}(t )$, $p_i^o\, \rightarrow_{c
\rightarrow \infty}\, m_i\, c + {{{\vec \kappa}_{(n)i}^2(t )}\over
{2m_i}}$.\bigskip

The matter part of the internal Poincare' generators (\ref{2.22})
has the limit

\bea
 M\, c &\rightarrow_{c \rightarrow \infty}& m\, c + \sum_{i=1}^2\,
 {{{\vec \kappa}_{(n)i}^2}\over {2\, m_i}} \approx m\, c + {{{\vec \pi}_{(n)12}^2}
 \over {2\, \mu}} = m\, c + H_{rel},\nonumber \\
 {\vec {\cal P}}_{(int)} &\rightarrow_{c \rightarrow \infty}\,&
 {\vec \kappa}_{(n)12} \approx 0,\nonumber \\
 \vec S &\rightarrow_{c \rightarrow \infty}& \sum_{i=1}^2\, {\vec
 \eta}_{(n)i} \times {\vec \kappa}_{(n)i} \approx {\vec \rho}_{(n)12} \times
 {\vec \pi}_{(n)12} = {\vec S}_{(n)},\nonumber \\
 {\vec {\cal K}}_{(int)} &\rightarrow_{c \rightarrow \infty}\,&
 - \sum_{i=1}^2\, m_i\, {\vec \eta}_{(n)i}
 = - m\, {\vec \eta}_{(n)12} \approx 0,
 \label{2.29}
 \eea

\noindent while the limit of the external Poincare' generators
(\ref{2.15}) is

\bea
 \vec P &=& {\vec p}_{(n)} = {\vec P}_{Galilei},\nonumber \\
 P^o &\rightarrow_{c \rightarrow \infty}& m\, c + {{{\vec p}^2_{(n)}}\over {2m}}
 + \sum_{i=1}^2\, {{{\vec \kappa}_{(n)i}^2}\over {2m_i}} \approx
  m\, c + {{{\vec p}^2_{(n)}}\over {2m}}
 + {{{\vec \pi}_{(n)12}^2}\over {2m_i}} = m\, c + E_{Galilei},\nonumber \\
 \vec J &\rightarrow_{c \rightarrow \infty}& {\vec x}_{(n)} \times
 {\vec p}_{(n)} + {\vec S}_{(n)} = {\vec J}_{Galilei},\nonumber \\
 {\vec K}/c &\rightarrow_{c \rightarrow \infty}& t\, {\vec p}_{(n)}
 - m\, {\vec x}_{(n)} = {\vec K}_{Galilei}.
 \label{2.30}
 \eea

\bigskip

Therefore the non-relativistic limit of the rest-frame instant form
leads to the following presentation of the Newton 2-body problem:

1) we have a decoupled external center of mass described by the
canonical variables ${\vec x}_{(n)}$, ${\vec p}_{(n)}$ and carrying
an internal space of relative variables coinciding with the
non-relativistic rest frame centered on the center of mass, ${\vec
p}_{(n)} \approx 0$ and ${\vec x}_{(n)}(t) \approx 0$ with the
Hamiltonian $H_{rel}$ and the rest spin ${\vec S}_{(n)}$;

2) in the internal space we have two pairs of variables ${\vec
\eta}_{(n)i}$, ${\vec \kappa}_{(n)i}$, or the canonically equivalent
${\vec \eta}_{(n)12} \approx 0$, ${\vec \kappa}_{(n)12} \approx 0$,
${\vec \rho}_{(n)12}$, ${\vec \pi}_{(n)12}$, and, as a consequence
from Eqs. (\ref{2.17}) and (\ref{2.26}) we have the following
identifications

\bea
  {\vec \rho}_{12}(\tau ) &=& {\vec \eta}_1(\tau ) - {\vec
  \eta}_2(\tau )\, \rightarrow_{c \rightarrow \infty}\, {\vec
  \rho}_{(n)12}(t) = {\vec \eta}_{(n)1}(t) - {\vec \eta}_{(n)2}(t) =
  {\vec r}_{(n)}(t),\nonumber \\
 {\vec \pi}_{12}(\tau ) &=&  {{m_2}\over m}\, {\vec \kappa}_1(\tau ) -
  {{m_1}\over m}\, {\vec \kappa}_2(\tau)\, \rightarrow_{c \rightarrow \infty}\,
  {\vec \pi}_{(n)12}(t) = {{m_2}\over m}\, {\vec \kappa}_{(n)1}(\tau ) -
  {{m_1}\over m}\, {\vec \kappa}_{(n)2}(\tau) =  {\vec q}_{(n)}(t),\nonumber \\
  &&{}\nonumber \\
  &&\Downarrow\nonumber \\
  &&{}\nonumber \\
  {\vec x}_1(\tau ) &\rightarrow_{c \rightarrow \infty}& {\vec x}_{(n)}(t) + {\vec
  \eta}_{(n)1}(t) =  {\vec x}_{(n)1}(t),\nonumber \\
   {\vec x}_2(\tau ) &\rightarrow_{c \rightarrow \infty}& {\vec x}_{(n)}(t) + {\vec
  \eta}_{(n)2}(t ) {\vec x}_{(n)2}(t).
 \label{2.31}
 \eea
\medskip

Let us remark that, while at the relativistic level the rest-frame
world-lines (\ref{2.17}) depend upon the 4-momentum $P^{\mu}$ of the
external 4-center of mass (because it identifies the instantaneous
Wigner 3-space in every inertial frame, being orthogonal to it), the
non-relativistic  trajectories ${\vec x}_{(n)\, i}(t)$ do not depend
upon ${\vec p}_{(n)}$, but only on ${\vec x}_{(n)}$ (the
non-relativistic definition of center of mass and relative variables
does not mix coordinates and momenta).

\subsection{The Radiation Field in the Radiation Gauge}

Till now we have emphasized the description of particles. Only in
Subsection E  have we given the description of the electro-magnetic
field in the radiation gauge. If we eliminate the particles we get
the rest-frame description of a transverse radiation field in the
radiation gauge, solution of $\Box\, {\vec A}_{\perp rad}(\tau \vec
\sigma ) {\buildrel \circ \over {{=}}} 0$.\medskip

In this Subsection we give the rest-frame parametrization (Fourier
coefficients and their Poisson brackets) of the radiation field in
the radiation gauge by using  the results of Ref.\cite{33b}. The
needed transverse polarization vectors are given in Appendix A. This
material will be needed in Section III together with the
Lienard-Wiechert solution in the radiation gauge of the rest-frame
Hamilton equations, reviewed in the  Subsection J.
\medskip

Instead in Subsection I we will study the connection between the
electro-magnetic and the radiation field in presence of charges by
adapting to the radiation gauge the Coulomb gauge treatment of
atomic physics \cite{1b}.

\medskip

On the Wigner hyperplane we have \footnote{$\sigma^A =
(\sigma^{\tau} = \tau ; \sigma^r)$; $k^A= (k^{\tau} = |\vec k| =
\omega (\vec k); k^r)$, $k^2 = 0$ , with $\vec k$ Wigner spin-1
3-vector and $k^{\tau}$ Lorentz scalar; $ d\tilde k =
{\frac{{d^3k}}{{2\, \omega (\vec k)\, (2 \pi )^3}}}$, $\Omega (\vec
k) = 2\, \omega (\vec k)\, (2\pi)^3$, $[d\tilde k] = [l^{-2}]$.}

\begin{eqnarray*}
{\vec A}_{\perp rad}(\tau ,\vec \sigma )\, &{\buildrel \circ \over
{{=}}} \,& \int d\tilde k\, \sum_{\lambda =1,2}\, {\vec
\epsilon}_{\lambda}(\vec k)\, \Big[ a_{\lambda}(\vec k)\, e^{- i\,
[\omega (\vec k)\, \tau - \vec k \cdot \vec \sigma]} +
a^{*}_{\lambda}(\vec k)\, e^{i\, [\omega (\vec k)\,
\tau - \vec k \cdot \vec \sigma]}\Big],  \nonumber \\
&&{}  \nonumber \\
{\vec \pi}_{\perp rad}(\tau ,\vec \sigma )&=&  {\vec E} _{\perp
rad}(\tau ,\vec \sigma )\, {\buildrel \circ \over {{=}}} -
{\frac{{\partial}}{{\partial \tau}}}\, {\vec A}_{\perp rad}(\tau
,\vec \sigma ) =\nonumber \\
 &=& i \int d\tilde k \, \omega (\vec k)\, \sum_{\lambda =1,2}\,
{\vec \epsilon}_{\lambda}(\vec k)\, \Big[ a_{\lambda}(\vec k)\, e^{-
i\, [\omega (\vec k)\, \tau - \vec k \cdot \vec \sigma]} -
a^{*}_{\lambda}(\vec k)\, e^{i\, [\omega (\vec k)\, \tau - \vec k
\cdot \vec \sigma]}\Big],  \nonumber \\
 &&{}  \nonumber \\
{\vec B}_{rad}(\tau ,\vec \sigma ) &=& \vec \partial \times {\vec
A}_{\perp rad}(\tau ,\vec \sigma ) =  \nonumber \\
&=& i\, \int d\tilde k\, \sum_{\lambda}\, \vec k \times {\vec
\epsilon} _{\lambda}(\vec k)\, \Big[a_{\lambda}(\vec k)\, e^{- i\,
[\omega (\vec k)\, \tau - \vec k \cdot \vec \sigma ]} -
a^{*}_{\lambda}(\vec k)\, e^{i\, [\omega (\vec k)\, \tau - \vec k
\cdot \vec \sigma ]}\Big],
 \end{eqnarray*}

\begin{eqnarray*}
 {\vec {\tilde A}}_{\perp rad}(\tau ,\vec k) &=& \int d^3\sigma\, {\vec A}
_{\perp rad}(\tau ,\vec \sigma )\, e^{- i\, \vec k \cdot \vec
\sigma} =\nonumber \\
 &=& {\frac{1}{{2\, \omega (\vec k)}}}\, \sum_{\lambda =1,2}\,
\Big[{\vec \epsilon}_{\lambda}(\vec k)\, a_{\lambda}(\vec k)\, e^{-
i\, \omega (\vec k)\, \tau} + {\vec \epsilon}_{\lambda}(- \vec k)\,
a^*_{\lambda}(- \vec k)\, e^{i\, \omega (\vec k)\, \tau}\Big],  \nonumber \\
{\vec {\tilde \pi}}_{\perp rad}(\tau ,\vec k) &=& \int d^3\sigma\,
{\vec \pi} _{\perp rad}(\tau ,\vec \sigma )\, e^{- i\, \vec k \cdot
\vec \sigma} =\nonumber \\
 &=& {\frac{i}{2}}\, \sum_{\lambda =1,2}\, \Big[{\vec \epsilon}
_{\lambda}(\vec k)\, a_{\lambda}(\vec k)\, e^{- i\, \omega (\vec
k)\, \tau} - {\vec \epsilon}_{\lambda}(- \vec k)\, a^*_{\lambda}(-
\vec k)\, e^{i\, \omega (\vec k)\, \tau}\Big],  \nonumber \\
{\vec {\tilde B}}_{rad}(\tau ,\vec k) &=& \int d^3\sigma\, {\vec B}
_{rad}(\tau ,\vec \sigma )\, e^{- i\, \vec k \cdot \vec \sigma} =
\nonumber \\
&=& {\frac{i}{{2\, \omega (\vec k)}}}\, \vec k \times \sum_{\lambda
=1,2}\, \Big[{\vec \epsilon}_{\lambda}(\vec k)\, a_{\lambda}(\vec
k)\, e^{- i\, \omega (\vec k)\, \tau} + {\vec \epsilon}_{\lambda}(-
\vec k)\, a^*_{\lambda}(- \vec k)\, e^{i\, \omega (\vec k)\,
\tau}\Big],
 \end{eqnarray*}

\bea
 a_{\lambda}(\vec k)&=& \int d^3\sigma\, {\vec
\epsilon}_{\lambda}(\vec k)\cdot \Big[ \omega (\vec k)\, {\vec
A}_{\perp rad}(\tau ,\vec \sigma ) - i\, {\vec \pi} _{\perp
rad}(\tau ,\vec \sigma ) \Big]\, e^{- i\, \vec k
\cdot \vec \sigma},  \nonumber \\
&&{}  \nonumber \\
&&{}  \nonumber \\
&&\{ A^r_{\perp rad}(\tau ,\vec \sigma ), \pi^s_{\perp rad}(\tau
,{\vec\sigma}_1)\} = - c\, P^{rs}_{\perp}(\vec \sigma )\,
\delta^3(\vec \sigma - {
\vec \sigma}_1),  \nonumber \\
&&{}  \nonumber \\
\{ a_{\lambda}(\vec k), a^{*}_{\lambda^{^{\prime}}}({\vec
k}^{^{\prime}}) \} &=& - i\, 2\, \omega (\vec k)\, c\, (2\pi )^3\,
\delta_{\lambda \lambda^{^{\prime}}}\, \delta^3(\vec k - {\vec
k}^{^{\prime}}) { \buildrel {def}\over {{=}}} - i\, \Omega (\vec
k)\, c\, \delta_{\lambda
\lambda^{^{\prime}}}\, \delta^3(\vec k - {\vec k}^{^{\prime}}),  \nonumber \\
\{ a_{\lambda}(\hat k), a_{\lambda^{^{\prime}}}({\hat
k}^{^{\prime}}) \} &=& \{ a^{*}_{\lambda}(\hat k),
a^{*}_{\lambda^{^{\prime}}}({\hat k}
^{^{\prime}})\} = 0,  \nonumber \\
&&{}  \nonumber \\
\delta^{ij} &=& \sum_{\lambda =1,2}\, \epsilon^i_{\lambda}(\vec k)\,
\epsilon^j_{\lambda}(\vec k) + {\frac{{k^i\, k^j}}{{|\vec
k|^2}}},\quad\quad
\vec k \cdot {\vec \epsilon}_{\lambda}(\vec k)=0,  \nonumber \\
&&{}  \nonumber \\
&& {\vec \epsilon}_{\lambda}(\vec k) \cdot {\vec \epsilon}
_{\lambda^{^{\prime}}}(\vec k) = \delta_{\lambda
\lambda^{^{\prime}}},\qquad {\frac{{\vec k}}{{|\vec k|}}} \cdot
\Big[{\vec \epsilon}_1(\vec k) \times {\ \vec \epsilon}_2(\vec
k)\Big] = 1.
  \label{2.32}
\end{eqnarray}

\noindent See Appendix A for a choice of the polarization 3-vectors
${\vec \epsilon} _{\lambda}(\vec k) $. In the circular basis
(\ref{a6}) with ${\vec \epsilon} _{\pm}(\vec k)$ we have
\footnote{In Ref.\cite{33b} it is shown the following result
$|a_{\sigma}(\vec k)|\, {\rightarrow}_{|\vec k| \rightarrow 0}\,
|\vec k|^{-1+\gamma}$ with $\gamma > 0$.}

\begin{eqnarray*}
{\vec A}_{\perp rad}(\tau ,\vec \sigma)&=& \int d\tilde k\,
\sum_{\sigma =\pm} \Big[ {\vec \epsilon}_{\sigma}(\vec k)
a_{\sigma}(\vec k) e^{-i {\hat k }^A\, \sigma_A}+ {\vec
\epsilon}^{*}_{\sigma}(\vec k)
a^{*}_{\sigma}(\vec k) e^{i {\hat k}^A\, \sigma_A}\Big],  \nonumber \\
{\vec \pi}_{\perp rad}(\tau ,\vec \sigma ) &=&i \int d\tilde k\,
\omega (\vec k) \sum_{\sigma =\pm} \Big[ {\vec
\epsilon}_{\sigma}(\vec k) a_{\sigma}(\vec k) e^{-i {\hat k}^A\,
\sigma_A}- {\vec \epsilon} ^{*}_{\sigma}(\vec k) a^{*}_{\sigma}(\vec
k) e^{i {\hat k}^A\, \sigma_A}
\Big],  \nonumber \\
{\vec B}_{rad}(\tau ,\vec \sigma ) &=& i\, \int d\tilde k\,
\sum_{\sigma =\pm} \vec k \times \Big[ {\vec \epsilon}_{\sigma}(\vec
k) a_{\sigma}(\vec k) e^{-i {\hat k}^A\, \sigma_A}- {\vec
\epsilon}^{*}_{\sigma}(\vec k) a^{*}_{\sigma}(\vec k) e^{i {\hat
k}^A\, \sigma_A}\Big],
\end{eqnarray*}

\begin{eqnarray}
a_{\pm}(\vec k)&=& {\frac{1}{\sqrt{2}}}\, \Big[ a_1(\vec k) \mp i\,
a_2(\vec
k)\Big],  \nonumber \\
&&{}  \nonumber \\
a_1(\vec k)&=& {\frac{1}{\sqrt{2}}}\, \Big[ a_{+}(\vec k) +
a_{-}(\vec k) \Big],\quad a_2(\vec k) = {\frac{i}{\sqrt{2}}}\, \Big[
a_{+}(\vec k)
- a_{-}(\vec k)\Big],  \nonumber \\
&&{}  \nonumber \\
&&\{ a_{\sigma}(\vec k), a^{*}_{\sigma^{^{\prime}}}({\vec
k}^{^{\prime}}) \} = - i\, \Omega (\vec k)\, c\, \delta_{\sigma
\sigma^{^{\prime}}}\, \delta^3(\vec k - {\vec k}^{^{\prime}}),
\nonumber \\
&& \{ a^{*}_{\sigma}(\vec k), a^{*}_{\sigma^{^{\prime}}}({\vec
k}^{^{\prime}}) \} = \{ a_{\sigma}(\vec k),
a_{\sigma^{^{\prime}}}({\vec k}^{^{\prime}}) \} = 0,  \nonumber \\
&&{}  \label{2.33}
\end{eqnarray}

\bigskip

By eliminating the particles in Eq.(\ref{2.22}) we get the following
expression for the internal Poincare' generators of the radiation
field in rest-frame instant form [ $ {\cal P}^A_{rad} = ({\cal
P}^{\tau}_{rad} = {\cal E}_{rad}/c = M_{rad}\, c; {\vec {\cal
P}}_{rad})$, $ {\cal J}^u_{rad} = {\frac{1}{2}}\, \epsilon^{urs}\,
{\cal J}^{rs}_{rad}$]

\begin{eqnarray*}
M_{rad}\, c^2 &=& {\cal E}_{rad} = c\, {\cal P}^{\tau}_{rad}=
{\frac{1}{2}}\, \int d^3\sigma\, \Big[ {\vec \pi} _{\perp rad}^2 +
{\vec B}_{rad}^2\Big] (\tau ,\vec \sigma ) = \sum_{\lambda =1,2}\,
\int d\tilde k\, \omega (\vec k)\,
a^{*}_{\lambda}(\vec k)\, a_{\lambda}(\vec k),  \nonumber \\
&&{}  \nonumber \\
{\vec {\cal P}}_{rad}&=& {\frac{1}{c}}\,\int d^3\sigma\, \Big[ {\vec
\pi}_{\perp rad} \times {\vec B }_{rad}\Big] (\tau ,\vec \sigma ) =
{\frac{1}{c}}\, \sum_{\lambda =1,2}\, \int d\tilde k\, \vec k\,
a^{*}_{\lambda}(\vec k)\, a_{\lambda}(\vec k) \approx 0,
\end{eqnarray*}

\begin{eqnarray*}
 {\vec {\cal J}}_{rad} &=& {\vec {\bar S}}_{rad} = {\frac{1}{c}}\, \int d^3\sigma\, \vec \sigma
\times \Big({ \vec \pi}_{\perp rad} \times {\vec B}_{rad}\Big)(\tau
,\vec \sigma ) =\nonumber \\
&=& {\frac{i}{c}}\, \sum_{\lambda}\, \int d\tilde k\,
a^*_{\lambda}(\vec k)\, \vec k \times {\frac{{\partial}}{{\partial\,
\vec k}}}\,a_{\lambda}(\vec k) +  \nonumber \\
&+& {\frac{i}{2\, c}}\, \sum_{\lambda \lambda^{^{\prime}}}\, \int
d\tilde k\, \Big[ a_{\lambda}(\vec k)\,
a^{*}_{\lambda^{^{\prime}}}(\vec k) - a^{*}_{\lambda}(\vec k)\,
a_{\lambda^{^{\prime}}}(\vec k)\Big]\, {\vec
\epsilon}_{\lambda}(\vec k) \cdot \Big( \vec k\, \times
{\frac{{\partial}}{{
\partial\, \vec k}}}\Big)\, {\vec \epsilon}_{\lambda^{^{\prime}}}(\vec k) -
\nonumber \\
&-& {\frac{i}{c}}\, \sum_{\lambda \lambda^{^{\prime}}}\, \int
d\tilde k\, { \vec \epsilon} _{\lambda}(\vec k) \times {\vec
\epsilon}_{\lambda^{^{ \prime}}}(\vec k)\, a^*_{\lambda}(\vec k)\,
a_{\lambda^{^{\prime}}}(\vec k),
\end{eqnarray*}

\begin{eqnarray}
{\cal K}^r_{rad}&=& {\cal J}^{\tau r}_{rad} = - {\frac{1}{2\, c}}\,
\int d^3\sigma \, \sigma^r\, \Big[ {\vec \pi}_{\perp rad}^2 + {\vec
B}_{rad}^2\Big]
(\tau ,\vec \sigma ) =  \nonumber \\
&=& {\frac{i}{c}}\, \int d\tilde k\, a^{*}_{\lambda}(\vec k)\,
\omega (\vec k)\, {\frac{{\partial}}{{\partial k^r}}}\,
a_{\lambda}(\vec k) +  \nonumber \\
&+& {\frac{i}{2\, c}}\, \sum_{\lambda ,\lambda^{^{\prime}}=1,2}\,
\int d\tilde k\, \Big[ a_{\lambda}(\vec k)\,
a^{*}_{\lambda^{^{\prime}}}(\vec k) - a^{*}_{\lambda}(\vec k)\,
a_{\lambda^{^{\prime}}}(\vec k)\Big]\, {\vec
\epsilon}_{\lambda}(\vec k) \cdot \omega (\vec k)\, {\frac{{\partial
{\vec \epsilon}_{\lambda^{^{\prime}}}(\vec k)}}{{\partial k^r}}}
 \approx 0,  \nonumber \\
 && {}  \nonumber \\
 &&{}  \nonumber \\
 h_{rad} &=& {\frac{{{\vec {\cal J}}_{rad} \cdot \vec k}}{{|\vec k|}}} = {\frac{i}{c}
}\, \int d\tilde k\, \Big[ a^{*}_2(\vec k)\, a_1(\vec k) -
a^{*}_1(\vec k)\,
a_2(\vec k)\Big]\,\, =  \nonumber \\
&=&{\frac{1}{c}}\, \int d\tilde k\, \Big[ a^{*}_{+}(\vec k)
a_{+}(\vec k)- a^{*}_{-}(\vec k)a_{-}(\vec k)\Big],
  \label{2.34}
\end{eqnarray}

\noindent where in the last line we defined the \textit{helicity}.

\bigskip

One needs \cite{33b} $a_{\lambda}(\hat k),\,\, \vec \partial
a_{\lambda}(\hat k) \in L_2(R^3, d^3k)$ for the existence of the
previous ten integrals (and of the occupation number $N_{\lambda} =
\int d\tilde k\, a^*_{\lambda}(\vec k)\, a_{\lambda}(\vec k)$) as
finite quantities. Moreover one can show \cite{33b} the existence of
the following behavior: i) $|a_{\lambda}(\hat k)|\, {\
\rightarrow}_{|\vec k| \rightarrow \infty} \, |\vec
k|^{-{\frac{3}{2}}-\rho}$ with $\rho > 0$; ii) $|a_{\lambda}(\hat
k)|\, {\rightarrow}_{|\vec k| \rightarrow 0}\, |\vec
k|^{-{\frac{3}{2}}+\epsilon}$ with $\epsilon > 0$.

\subsection{The Connection between the Electro-Magnetic and
Radiation Fields in the Radiation Gauge}

In this Subsection we discuss the treatment of the electromagnetic
field in presence of dynamical charges used by atomic physicists in
the Coulomb gauge (we follow chapter I of Ref.\cite{1b}). \bigskip

Let us remark that in our rest-frame radiation gauge (a special case
of Lorentz gauge) we have Wigner covariance, so that quantities like
$\vec k \cdot \vec \sigma$ and $d^3k$ and ${\vec k}^2$ are Lorentz
scalars, since the 3-vectors are Wigner spin-1 3-vectors.

\bigskip

In the radiation gauge we have $A^{\tau}(\tau ,\vec \sigma ) =
\sum_{i=1}^N\, {{Q_i}\over {4\pi\, |\vec \sigma - {\vec \eta}_i(\tau
)|}}$ and $\vec \partial \cdot \vec A(\tau ,\vec \sigma ) = 0$
(Coulomb gauge), so that we have the following fields: ${\vec
A}_{\perp}(\tau ,\vec \sigma )$, $\vec B(\tau ,\vec \sigma ) = \vec
\partial \times {\vec A}_{\perp}(\tau ,\vec \sigma )$, ${\vec
\pi}_{\perp}(\tau ,\vec \sigma ) =  {\vec E} _{\perp}(\tau ,\vec
\sigma )\, {\buildrel \circ \over {=}}\,- {\frac{{
\partial\, {\vec A}_{\perp}(\tau ,\vec \sigma )}}{{\partial\, \tau}}}$. We
used ${\buildrel \circ \over {=}}$ in the last equation to make
explicit that this result is equivalent to the kinematical first
half of Hamilton equations, whose complete set in the radiation
gauge is obtained from Eqs. ( \ref{4.11}), (\ref{4.15}) of Section
IV and from the transverse matter current ${\vec j}_{\perp}(\tau
,\vec \sigma )$ of Eq.(\ref{b7})

\begin{eqnarray}
\partial_{\tau}\, {\vec A}_{\perp}(\tau ,\vec \sigma ) &{\buildrel
\circ \over {=}}& - {\vec \pi}_{\perp}(\tau ,\vec \sigma ),\quad
(kinematical),\nonumber \\
\partial_{\tau}\, {\vec \pi}_{\perp}(\tau ,\vec \sigma ) &{\buildrel
\circ \over {=}}& - {\vec \partial}^2\, {\vec A}_{\perp}(\tau ,\vec
\sigma ) - {\vec j}_{\perp}(\tau ,\vec \sigma ),\quad (dynamical),
\nonumber \\
 &&{}\nonumber \\
&&\Rightarrow\quad \Box\, {\vec A}_{\perp}(\tau ,\vec \sigma ) {
\buildrel \circ \over {=}} {\vec j}_{\perp}(\tau ,\vec \sigma ).
\label{2.35}
\end{eqnarray}
\medskip

From the point of view of Maxwell equation the use of the potentials
in the radiation gauge automatically satisfies the two equations

\beq \vec \partial \cdot \vec B(\tau ,\vec \sigma )\, {\buildrel
\circ \over {=}}\, 0,  \qquad \vec \partial \times {\vec
\pi}_{\perp}(\tau ,\vec \sigma )\, { \buildrel \circ \over {=}}\, -
\partial_{\tau}\, \vec B(\tau ,\vec \sigma ).
 \label{2.36}
\eeq

In the radiation gauge the following equation is trivial and
sourceless  (the charge density, connected to the longitudinal
electric field, has been reabsorbed due to the presence of the
Coulomb potential among the charges)

\begin{equation}
\vec \partial \cdot {\vec \pi}_{\perp}(\tau ,\vec \sigma )\, {
\buildrel \circ \over {=}}\, 0.
 \label{2.37}
\end{equation}

The real dynamical Maxwell equation in the radiation gauge is

\beq
 \vec \partial \times \vec B(\tau ,\vec \sigma )\, {\buildrel
\circ \over {= }}\, \partial_{\tau}\, {\vec \pi}_{\perp}(\tau ,\vec
\sigma ) + {\vec j}_{\perp}(\tau ,\vec \sigma ),
  \qquad \Rightarrow\quad \Box\, {\vec A}_{\perp}(\tau ,\vec \sigma
) { \buildrel \circ \over {=}} {\vec j}_{\perp}(\tau ,\vec \sigma ).
 \label{2.38}
\eeq

In the last line we used $\vec \partial \times \vec B = \vec
\partial \times (\vec \partial \times {\vec A}_{\perp}) = - {\vec \partial}^2\,
{\vec A} _{\perp}$ and the kinematical Hamilton equation ${\vec
\pi}_{\perp}\, { \buildrel \circ \over {=}}\, - \partial_{\tau}\,
{\vec A}_{\perp}$
\bigskip

Let us now define the following Fourier transforms of the fields

\begin{eqnarray*}
{\vec A}_{\perp}(\tau ,\vec \sigma ) &=& {\frac{1}{{(2\pi)^3}}}\,
\int d^3k\, {\vec {\tilde A}}_{\perp}(\tau ,\vec k)\, e^{i\, \vec k
\cdot \vec \sigma} =  \nonumber \\
&=& {\frac{1}{{(2\pi )^3}}}\, \int d^3k\, \sum_{\lambda = 1,2}\,
{\vec \epsilon}_{\lambda}(\vec k)\, \Big[b_{em\, \lambda}(\tau ,\vec
k)\, e^{i\, \vec k \cdot \vec \sigma}\, + b^*_{em\, \lambda}(\tau
,\vec k)\, e^{- i\, \vec k \cdot \vec \sigma}\Big],  \nonumber \\
&&{}  \nonumber \\
{\vec \pi}_{\perp}(\tau ,\vec \sigma ) &=& {\vec E} _{\perp}(\tau
,\vec \sigma )\, = {\frac{1}{{(2\pi)^3}}}\, \int d^3k\, {\vec {
\tilde \pi}}_{\perp}(\tau ,\vec k)\, e^{i\, \vec k \cdot \vec
\sigma}\, { \buildrel \circ \over {=}}\, - {\frac{{\partial\, {\vec
A}_{\perp}(\tau ,\vec \sigma )}}{{\partial\, \tau}}} =  \nonumber \\
&=& - {\frac{1}{{(2\pi )^3}}}\, \int d^3k\, \sum_{\lambda = 1,2}\,
{\vec \epsilon}_{\lambda}(\vec k)\, \Big[{\frac{{\partial\, b_{em\,
\lambda}(\tau ,\vec k)}}{{\partial\, \tau}}}\, e^{i\, \vec k \cdot
\vec \sigma} + {\frac{{
\partial\, b^*_{em\, \lambda}(\tau ,\vec k)}}{{\partial\, \tau}}}\, e^{- i\,
\vec k \cdot \vec \sigma}\Big],  \nonumber \\
&&{}  \nonumber \\
\vec B(\tau ,\vec \sigma ) &=& {\frac{1}{{(2\pi)^3}}}\, \int d^3k\,
{\vec { \tilde B}}(\tau ,\vec k)\, e^{i\, \vec k \cdot \vec \sigma}
=\nonumber \\
 &=& {\frac{i}{{(2\pi )^3}}}\, \int d^3k\, \sum_{\lambda
=1,2}\, \vec k \times {\vec \epsilon}_{\lambda}(\vec k)\,
\Big[b_{em\, \lambda}(\tau ,\vec k)\, e^{i\, \vec k \cdot \vec
\sigma} - b^*_{em\, \lambda}(\tau ,\vec k)\,
e^{- i\, \vec k \cdot \vec \sigma}\Big],  \nonumber \\
&&{}  \nonumber \\
&&{}  \nonumber \\
&&\vec \partial \cdot {\vec A}_{\perp}(\tau ,\vec \sigma )\,\,
\Rightarrow\,\, \vec k \cdot {\vec \epsilon}_{\lambda}(\vec k) =
\vec k \cdot {\vec \epsilon}_{\lambda}(- \vec k) = 0,
\end{eqnarray*}

\begin{eqnarray*}
{\vec {\tilde A}}_{\perp}(\tau ,\vec k) &=& \int d^3\sigma\, {\vec
A} _{\perp}(\tau ,\vec \sigma )\, e^{- i\, \vec k \cdot \vec \sigma}
= {\vec {
\tilde A}}^*(\tau , - \vec k) =  \nonumber \\
&=& \sum_{\lambda = 1,2}\, \Big[{\vec \epsilon}_{\lambda}(\vec k)\,
b_{em\, \lambda}(\tau ,\vec k) + {\vec \epsilon}_{\lambda}(- \vec
k)\, b^*_{em\,
\lambda}(\tau ,-s \vec k)\Big],  \nonumber \\
&&\vec k \cdot {\vec {\tilde A}}_{\perp}(\tau ,\vec k) = 0,
 \end{eqnarray*}

\begin{eqnarray*}
 {\vec {\tilde \pi}}_{\perp}(\tau ,\vec k) &=& \int d^3\sigma\, {\vec
\pi} _{\perp}(\tau ,\vec \sigma )\, e^{- i\, \vec k \cdot \vec
\sigma} = {\vec {\tilde \pi}}^*_{\perp}(\tau , - \vec k) =  \nonumber \\
&=& - \sum_{\lambda = 1,2}\, \Big[{\vec \epsilon}_{\lambda}(\vec
k)\, {\frac{ {\partial\, b_{em\, \lambda}(\tau ,\vec
k)}}{{\partial\, \tau}}} + {\vec \epsilon}_{\lambda}(- \vec k)\,
{\frac{{\partial\, b^*_{em\, \lambda}(\tau
,- \vec k)}}{{\partial\, \tau}}}\Big] =  \nonumber \\
&{\buildrel {def}\over {=}}& {\frac{i}{2}}\, \Big[\vec \alpha(\tau
,\vec k)
- {\vec \alpha}^*(\tau ,- \vec k)\Big],  \nonumber \\
&& \vec k \cdot {\vec {\tilde \pi}}_{\perp}(\tau ,\vec k) =
{\frac{i}{2}}\, \vec k \cdot \Big[\vec \alpha (\tau ,\vec k) - {\vec
\alpha}^*(\tau ,- \vec k)\Big],
 \end{eqnarray*}

\bea
 {\vec {\tilde B}}(\tau ,\vec k) &=& \int d^3\sigma\, \vec B(\tau
,\vec \sigma )\, e^{- i\, \vec k \cdot \vec \sigma} = {\vec {\tilde
B}}^*(\tau ,- \vec k) =  \nonumber \\
&=&i\, \sum_{\lambda = 1,2}\, \vec k \times \Big[{\vec \epsilon}
_{\lambda}(\vec k)\, b_{em\, \lambda}(\tau ,\vec k) + {\vec
\epsilon} _{\lambda}(- \vec k)\, b^*_{em\, \lambda}(\tau , - \vec
k)\Big] = \nonumber \\
&{\buildrel {def}\over {=}}& {\frac{i}{2}}\, {\frac{{\vec k}}{{|\vec
k|}}} \, \times \Big[\vec \alpha(\tau ,\vec k) + {\vec
\alpha}^*(\tau ,- \vec k)\Big],  \nonumber \\
 &&{}\nonumber \\
\vec k \cdot {\vec {\tilde B}}(\tau ,\vec k) \, = 0,&&  \quad \vec k
\times \Big[\vec \alpha(\tau ,\vec k) + {\vec \alpha}^*(\tau ,- \vec
k)\Big]\, {\buildrel \circ \over {=}}\, - 2\, i\, |\vec k|\, {\vec {
\tilde B}}(\tau ,\vec k).
  \label{2.39}
\end{eqnarray}

We see that the Fourier coefficients $b_{em\, \lambda} (\tau ,\vec
k)$ do not enjoy of the nice properties of the Fourier coefficients
$a_{\lambda}(\vec k)$ of the free radiation field (\ref{2.32}). Now
the electric field ${\vec \pi}_{\perp} =  {\vec E}_{\perp}$ depends
upon ${{\partial\, b_{em}\, \lambda (\tau ,\vec k) }\over
{\partial\, \tau}}$.

\bigskip

As a consequence, following Ref.\cite{1b}, in Eqs.(\ref{2.39}) we
introduced the following function $\vec \alpha (\tau ,\vec k)$

\begin{eqnarray*}
\vec \alpha (\tau ,\vec k) &=& - i\, \Big[{\vec {\tilde
\pi}}_{\perp} - { \frac{{\vec k}}{{|\vec k|}}}\, \times {\vec
{\tilde B}}\Big](\tau ,\vec k) =
\nonumber \\
&=& \sum_{\lambda = 1,2}\, \Big[{\vec \epsilon}_{\lambda}(\vec k)\,
\Big( |\vec k|\, b_{em\, \lambda}(\tau ,\vec k) + i\,
{\frac{{\partial\, b_{em\,
\lambda}(\tau ,\vec k)}}{{\partial\, \tau}}}\Big) +  \nonumber \\
&+& {\vec \epsilon}_{\lambda}(- \vec k)\, \Big(|\vec k|\, b^*_{em\,
\lambda}(\tau ,- \vec k) + i\, {\frac{{\partial\, b^*_{em\,
\lambda}(\tau ,
- \vec k)}}{{\partial\, \tau}}}\Big) \Big],  \nonumber \\
&&{}  \nonumber \\
{\vec \alpha}^*(\tau ,- \vec k) &=& i\, \Big[{\vec {\tilde
\pi}}_{\perp} + { \frac{{\vec k}}{{|\vec k|}}}\, \times {\vec
{\tilde B}}^* \Big](\tau ,\vec k) = i\, \Big[{\vec {\tilde
\pi}}_{\perp}^* - {\frac{{(- \vec k)}}{{|\vec k|}
}} \times {\vec {\tilde B}}^*\Big](\tau ,- \vec k),  \nonumber \\
&&{}  \nonumber \\
&&\vec k \cdot \vec \alpha (\tau ,\vec k) = 0,
 \end{eqnarray*}

\bea
&&\Downarrow  \nonumber \\
&&{}  \nonumber \\
{\vec {\tilde \pi}}(\tau ,\vec k) &=& {\frac{i}{2}}\, \Big[\vec
\alpha(\tau
,\vec k) - {\vec \alpha}^*(\tau ,- \vec k)\Big],  \nonumber \\
{\vec {\tilde B}}(\tau ,\vec k) &=& {\frac{i}{2}}\, {\frac{{\vec
k}}{{|\vec k|}}}\, \times \Big[\vec \alpha(\tau ,\vec k) + {\vec
\alpha}^*(\tau ,- \vec k)\Big].
  \label{2.40}
\end{eqnarray}
\bigskip

The Fourier transform of Eqs.(\ref{2.36}), (\ref{2.37}) and
(\ref{2.38}) is

\begin{eqnarray}
i\, \vec k \cdot {\vec {\tilde B}}(\tau ,\vec k) &{\buildrel \circ
\over {=}}& 0,  \qquad i\, \vec k \times {\vec {\tilde
\pi}}_{\perp}(\tau ,\vec k)\, { \buildrel \circ \over {=}} -
\partial_{\tau}\, {\vec {\tilde
B}}(\tau ,\vec k),\nonumber \\
&&{}  \nonumber \\
i\, \vec k \cdot {\vec {\tilde \pi}}_{\perp}(\tau ,\vec k)
&{\buildrel \circ \over {=}}& 0,  \qquad  i\, \vec k \times {\vec
{\tilde B}}(\tau ,\vec k)\, {\buildrel \circ \over {= }}\,
\partial_{\tau}\, {\vec {\tilde \pi}}_{\perp}(\tau
,\vec k) + {\vec { \tilde j}}_{\perp}(\tau ,\vec k),  \nonumber \\
&&{}  \nonumber \\
\Rightarrow&& - i\, \vec k \times {\vec {\tilde B}}^*(\tau ,- \vec
k)\, - \partial_{\tau}\, {\vec {\tilde \pi}}^*_{\perp}(\tau ,- \vec
k)\, { \buildrel \circ \over {=}}\, {\vec {\tilde j}}_{\perp}(\tau
,\vec k) = {\vec { \tilde j}}^*_{\perp}(\tau ,- \vec k),
  \label{2.41}
\end{eqnarray}

\noindent where ${\vec {\tilde j}}_{\perp}(\tau ,\vec k)$ is the
Fourier transform of ${\vec j}_{\perp}(\tau ,\vec \sigma )$.

\medskip

A consequence of Eqs.(\ref{2.41}) is ($\omega (\vec k) = |\vec k|$)

\beq
\partial_{\tau}\, \Big[{\vec {\tilde \pi}}_{\perp} \pm {\frac{{\vec k}}{{
|\vec k|}}} \times {\vec {\tilde B}}\Big](\tau ,\vec k)\, {\buildrel
\circ \over {=}}\, \pm \omega (\vec k)\, \Big[{\vec {\tilde
\pi}}_{\perp} \pm { \frac{{\vec k}}{{|\vec k|}}} \times {\vec
{\tilde B}}\Big](\tau ,\vec k) -  {\vec {\tilde j}}_{\perp}(\tau
,\vec k).
 \label{2.42}
\eeq

\noindent At the level of Maxwell equations in the radiation gauge
one uses the dynamical equation and one of the two automatic ones.

\bigskip

The form of the function

\beq
 \vec \alpha (\tau ,\vec k)\, {\buildrel {def} \over {=}}\,
\sum_{\lambda = 1,2}\, {\vec \epsilon}_{\lambda}(\vec k)\, a_{em\,
\lambda}(\tau ,\vec k),
 \label{2.43}
 \eeq

\noindent is suggested by Eqs.(\ref{2.42}). Eqs.(\ref{2.41} ) imply
the following equations of motion for $\vec \alpha (\tau ,\vec k)$

\begin{eqnarray}
\partial_{\tau}\, \vec \alpha (\tau ,\vec k)\, &{\buildrel \circ \over {=}}
& - i\, \omega (\vec k)\, \vec \alpha (\tau ,\vec k) + i\, {\vec
{\tilde j}}_{\perp}(\tau ,\vec k),  \nonumber \\
\partial_{\tau}\, {\vec \alpha}^* (\tau , - \vec k)\, &{\buildrel
\circ \over {=}}& i\, \omega (\vec k)\, {\vec \alpha}^* (\tau , -
\vec k) - i\, {\vec {\tilde j}}_{\perp}(\tau , - \vec k),  \nonumber \\
&&{}  \nonumber \\
\Rightarrow&& \partial_{\tau}\, a_{em\, \lambda}(\tau ,\vec k)\, {
\buildrel \circ \over {=}}\, - i\, \omega (\vec k)\, a_{em\,
\lambda}(\tau ,\vec k) + i\, {\vec \epsilon}_{\lambda}(\vec k) \cdot
{\vec {\tilde j}} _{\perp}(\tau ,\vec k).
 \label{2.44}
\end{eqnarray}
\bigskip

Instead Eqs.(\ref{2.44}) imply the following equations of motion for
$ b_{em\, \lambda}(\tau ,\vec k)$ [we put ${\vec {\tilde
j}}_{\perp}(\tau ,\vec k) = {\frac{1}{2}}\, [{\vec {\tilde
j}}_{\perp}(\tau ,\vec k) + {\vec { \tilde j}}^*_{\perp}(\tau ,-
\vec k)]$ due to Eqs.(\ref{2.41})]

\begin{equation}
{\frac{{\partial^2\, b_{em\, \lambda}(\tau ,\vec k)}}{{\partial\,
\tau^2}}} \, {\buildrel \circ \over {=}}\, - \omega^2(\vec k)\,
b_{em\, \lambda}(\tau ,\vec k) + {\frac{1}{2}}\, {\vec {\tilde
j}}(\tau ,\vec k).
  \label{2.45}
\end{equation}

\bigskip

Since the relation $\vec \partial \times \vec B(\tau ,\vec \sigma )
= \vec \partial \times \Big(\vec \partial \times {\vec
A}_{\perp}(\tau ,\vec \sigma )\Big) = - {\vec \partial}^2\, {\vec
A}_{\perp}(\tau ,\vec \sigma )$ implies $i\, \vec k \times {\vec
{\tilde B}}(\tau ,\vec k) = {\vec k}^2\, {\vec { \tilde
A}}_{\perp}(\tau ,\vec k)$, we can write the following
representation of the function $\vec \alpha (\tau ,\vec k)$

\begin{equation}
\vec \alpha (\tau ,\vec k)\, =\, \Big[|\vec k|\, {\vec {\tilde A}}_{
\perp} - i\, {\vec {\tilde \pi}}_{\perp}\Big](\tau ,\vec k).
 \label{2.46}
\end{equation}

\bigskip

As a consequence, we arrive at the following representation of the
fields formally identical to Eqs.(\ref{2.32}) for the radiation
field by sending $a_{\lambda}(\vec k)$ into $b_{em}\, \lambda (\tau
,\vec k)$

\begin{eqnarray}
{\vec {\tilde A}}_{\perp}(\tau ,\vec k) &=& {\frac{i}{{{\vec
k}^2}}}\, \vec k \times {\vec {\tilde B}}(\tau ,\vec k) =
 {\frac{1}{{2\, |\vec k|}}}\, \Big[\vec \alpha (\tau ,\vec k) +
{\vec \alpha}^*(\tau ,- \vec k)\Big],  \nonumber \\
{\vec A}_{\perp}(\tau ,\vec \sigma )&=& {\frac{1}{{(2\pi)^3}}}\,
\int {\frac{ {d^3k}}{{2\, \omega (\vec k)}}}\, \sum_{\lambda =
1,2}\, {\vec \epsilon} _{\lambda}(\vec k)\, \Big[a_{em\,
\lambda}(\tau ,\vec k)\, e^{i\, \vec k \cdot \vec \sigma} +
a^*_{em\, \lambda}(\tau ,\vec k)\, e^{- i\, \vec k
\cdot \vec \sigma}\Big],  \nonumber \\
&&{}  \nonumber \\
&&{}  \nonumber \\
{\vec \pi}_{\perp}(\tau ,\vec \sigma ) &=& {\frac{i}{{2\,
(2\pi)^3}}}\, \int d^3k\, \sum_{\lambda = 1,2}\, {\vec
\epsilon}_{\lambda}(\vec k)\, \Big[ a_{em\, \lambda}(\tau ,\vec k)\,
e^{i\, \vec k \cdot \vec \sigma} - a^*_{em\, \lambda}(\tau ,\vec
k)\, e^{- i\, \vec k \cdot \vec \sigma}\Big] =
\nonumber \\
&{{\buildrel \circ \over {=}}}_{dyn}& - {\frac{{\partial\, {\vec A}
_{\perp}(\tau ,\vec \sigma )}}{{\partial\, \tau}}},  \nonumber \\
&&{}  \nonumber \\
\vec B(\tau ,\vec \sigma ) &=& {\frac{i}{{2\, (2\pi)^3}}}\, \int
{\frac{{d^3k }}{{\omega (\vec k)}}}\, \vec k \times \sum_{\lambda =
1,2}\, {\vec \epsilon} _{\lambda}(\vec k)\, \Big[a_{em\,
\lambda}(\tau ,\vec k)\, e^{i\, \vec k \cdot \vec \sigma} -
a^*_{em\, \lambda}(\tau ,\vec k)\, e^{- i\, \vec k
\cdot \vec \sigma}\Big],  \nonumber \\
&&{}  \nonumber \\
&&{}  \nonumber \\
a_{em\, \lambda}(\tau ,\vec k) &=& \int d^3\sigma\, {\vec \epsilon}
_{\lambda}(\vec k)\cdot \Big[\omega (\vec k)\, {\vec A}_{\perp}(\tau
,\vec \sigma ) - i\, {\vec \pi}_{\perp}(\tau ,\vec \sigma )\Big]\,
e^{- i\, \vec k \cdot \vec \sigma}.
  \label{2.47}
\end{eqnarray}

The peculiarity of this new representation (in terms of $a_{em\,
\lambda}(\tau ,\vec k)$ and not of $b_{em\, \lambda}(\tau ,\vec k)$)
is that it is only by using the dynamical equations (\ref{2.44})
(i.e. the dynamical Hamilton equations) and the property ${\vec
{\tilde j}} _{\perp}(\tau ,\vec k) = {\vec {\tilde
j}}^*_{\perp}(\tau ,- \vec k)$ of Eqs.(\ref{2.41}) that we can show
the validity of the kinematical Hamilton equations ${\vec
\pi}_{\perp}(\tau ,\vec \sigma )\, {\buildrel \circ \over { =}}\, -
\partial_{\tau}\, {\vec A}_{\perp}(\tau ,\vec \sigma )$, which were
definitory of the old representation (\ref{2.39}). With this
representation the kinematical Hamilton equations hold only in the
space of the solutions of the dynamical Hamilton equations.\medskip

However, due to the formal identity of Eqs.(\ref{2.47}) and
(\ref{2.32}), the Poisson brackets $\{ A^r_{\perp}(\tau ,{\vec
\sigma}_1), \pi^s_{\perp}(\tau ,{\vec \sigma}_2)\} = - c\,
P^{rs}_{\perp}({\vec \sigma} _1)\, \delta^3({\vec \sigma}_1 - {\vec
\sigma}_2)$ imply the Poisson brackets $\{ a_{em\, \lambda}(\tau
,\vec k), a^*_{em\, \lambda^{^{\prime}}}(\tau , {\vec
k}^{^{\prime}}) \} = - i\, \Omega (\vec k)\, c\, \delta_{\lambda
\lambda^{^{\prime}}}\, \delta^3(\vec k - {\vec k} ^{^{\prime}})$
like in Eqs.(\ref{2.32}) for the radiation field.

\bigskip

By using the last of Eqs.(\ref{2.47}), Eqs. (\ref{2.22}) and
(\ref{b14}) and (\ref{b15}) of Appendix B  we have from ${\vec
A}_{\perp}(\tau ,\vec \sigma )\, { \buildrel \circ \over {=}}\,
{\vec A}_{\perp S}(\tau ,\vec \sigma ) + {\vec A} _{\perp rad}(\tau
,\vec \sigma )$

\begin{eqnarray}
 &&a_{em\, \lambda} (\tau ,\vec k) {\buildrel \circ \over
{=}}_{dyn}\,\, a_{\lambda}(\vec k)\, e^{- i\, \omega (\vec k)\, \tau}
+  \nonumber \\
 &&{}\nonumber \\
&&+ \omega (\vec k)\, {\vec \epsilon}_{\lambda}(\vec k) \cdot
\sum_{i=1}^2\, Q_i\, {\frac{{e^{- i\, \vec k \cdot {\vec
\eta}_i(\tau )}} }{{{\vec k}^4}}} \, {\frac{{\vec k \times
\Big({\vec \kappa}_i(\tau ) \times \vec k\Big)\, \sqrt{m_i^2\, c^2 +
{\vec \kappa}_i^2(\tau )}}}{{m_i^2\, c^2 + {\vec \kappa} _i^2(\tau )
- \Big({\vec \kappa}_i(\tau ) \cdot {\frac{{\vec k}}{{|\vec k|}}}
\Big)^2}}}  \, \Big[1 + {\frac{{{\vec \kappa}_i(\tau ) \cdot \vec
k}}{\sqrt{m^2_i\, c^2
+ {\vec \kappa}_i^2(\tau )}}}\Big].  \nonumber \\
&&{}  \label{2.48}
\end{eqnarray}
\medskip

This already shows that on shell an arbitrary electro-magnetic field
in the radiation gauge can be describe as a transverse radiation
field plus particle terms coming from the Lienard-Wiechert fields.

\bigskip

Moreover we have the following expression for the terms of the
internal Poincare' generators (\ref{2.22}) involving only the
electromagnetic field

\begin{eqnarray*}
c\, p^{\tau}_{em}(\tau ) &=& e_{em}(\tau ) = M_{em}(\tau )\, c^2 =
{\frac{1}{ 2}}\, \int d^3\sigma\, \Big({\vec \pi}^2_{\perp} + {\vec
B}^2\Big)(\tau ,\vec \sigma ) = {\frac{1}{2}}\, \int d^3k\,
\Big({\vec {\tilde \pi}}
_{\perp}^2 + {\vec {\tilde B}}^2\Big)(\tau ,\vec k) =  \nonumber \\
&=& \sum_{\lambda =1,2}\, \int d\tilde k\, \omega (\vec k)\,
a^*_{em\, \lambda}(\tau ,\vec k)\, a_{em\, \lambda}(\tau ,\vec k),  \nonumber \\
{\vec p}_{em}(\tau ) &=& {\frac{1}{c}}\, \int d^3\sigma\, \Big({\vec
\pi} _{\perp} \times \vec B\Big)(\tau ,\vec \sigma ) =
{\frac{1}{c}}\, \int d^3k\, \Big({\vec {\tilde \pi}}_{\perp} \times
{\vec {\tilde B}}\Big)(\tau ,\vec k) =  \nonumber \\
&=&{\frac{1}{c}}\, \sum_{\lambda =1,2}\, \int d\tilde k\, \vec k\,
a^*_{em\, \lambda}(\tau ,\vec k)\, a_{em\, \lambda}(\tau ,\vec k),
  \nonumber \\
 {\vec j}_{em}(\tau ) &=& {\frac{1}{c}}\, \int d^3\sigma\, \vec
\sigma \times \Big({\vec \pi}_{\perp} \times \vec B\Big)(\tau ,\vec
\sigma ) = \nonumber \\
&=&{\frac{i}{c}}\, \sum_{\lambda}\, \int d\tilde k\, a^*_{em\,
\lambda}(\vec k)\, \vec k \times {\frac{{\partial}}{{\partial\, \vec
k}}}\, a_{em\, \lambda}(\vec k) +  \nonumber \\
&+& {\frac{i}{2\, c}}\, \sum_{\lambda \lambda^{^{\prime}}}\, \int
d\tilde k\, \Big[ a_{em\, \lambda}(\vec k)\, a^{*}_{em\,
\lambda^{^{\prime}}}(\vec k) - a^{*}_{em\, \lambda}(\vec k)\,
a_{em\, \lambda^{^{\prime}}}(\vec k)\Big] \, {\vec
\epsilon}_{\lambda}(\vec k) \cdot \Big( \vec k\, \times {\frac{{
\partial}}{{\partial\, \vec k}}}\Big)\, {\vec \epsilon}_{\lambda^{^{
\prime}}}(\vec k) -  \nonumber \\
&-& {\frac{i}{c}}\, \sum_{\lambda \lambda^{^{\prime}}}\, \int
d\tilde k\, { \vec \epsilon} _{\lambda}(\vec k) \times {\vec
\epsilon}_{\lambda^{^{ \prime}}}(\vec k)\, a^*_{em\, \lambda}(\vec
k)\, a_{em\, \lambda^{^{\prime}}}(\vec k),
 \end{eqnarray*}

\bea
 k^r_{em}(\tau )&=& - {\frac{1}{2\, c}}\, \int d^3\sigma \, \sigma^r\, \Big[ {
\vec \pi}_{\perp}^2 + {\vec B}^2\Big] (\tau ,\vec \sigma ) =  \nonumber \\
&=& {\frac{i}{c}}\, \int d\tilde k\, a^{*}_{em\, \lambda}(\vec k)\,
\omega (\vec k)\, {\frac{{\partial}}{{\partial k^r}}}\, a_{em\,
\lambda}(\vec k) +\nonumber \\
&+& {\frac{i}{2\, c}}\, \sum_{\lambda ,\lambda^{^{\prime}}=1,2}\,
\int d\tilde k\, \Big[ a_{em\, \lambda}(\vec k)\, a^{*}_{em\,
\lambda^{^{\prime}}}(\vec k) - a^{*}_{em\, \lambda}(\vec k)\,
a_{em\, \lambda^{^{\prime}}}(\vec k)\Big]\, {\vec
\epsilon}_{\lambda}(\vec k) \cdot \omega (\vec k)\, {\frac{{\partial
{\vec \epsilon}_{\lambda^{^{\prime}}}(
\vec k)}}{{\partial k^r}}},  \nonumber \\
&& {}  \nonumber \\
&&{}  \nonumber \\
h_{em}(\tau ) &=& {\frac{{{\vec j}_{em}(\tau ) \cdot \vec k}}{{|\vec
k|}}} = {\frac{i}{c}}\, \int d\tilde k\, \Big[ a^{*}_{em\, 2}(\vec
k)\, a_{em\, 1}(\vec k) - a^{*}_{em\,1}(\vec k)\, a_{em\, 2}(\vec
k)\Big]\,\, =  \nonumber \\
&=&{\frac{1}{c}}\, \int d\tilde k\, \Big[ a^{*}_{em\, +}(\vec k)
a_{em\, +}(\vec k)- a^{*}_{em\, -}(\vec k)a_{em\, -}(\vec k)\Big],
 \label{2.49}
\end{eqnarray}

\bigskip

Therefore at the end the internal Poincare- generators (\ref{2.22})
$M$, ${ \vec {\bar S}} = {\vec {\mathcal{J}}}_{(int)}$, ${\vec
{\mathcal{P}}} _{(int)} \approx 0$, ${\vec {\mathcal{K}}}_{(int)}
\approx 0$, will depend on the particles and on $a_{em\,
\lambda}(\tau ,\vec k)$.

\subsection{The Lienard-Wiechert Electro-Magnetic Potential and
Field associated with the Particles}

In Ref.\cite{14b} we obtained the following expression for the
Lienard-Wiechert transverse electromagnetic potential, electric
field and magnetic field inhomogeneous solution of the equations
$\Box\, A^r_{\perp}(\tau ,\vec \sigma )\, { \buildrel \circ \over
{=}}\, j^r_{\perp}(\tau ,\vec \sigma ) = \sum_i\, Q_i\,
P^{rs}_{\perp}(\vec \sigma )\, {\dot \eta}^s_i(\tau ) \delta^3(\vec
\sigma - {\vec \eta}_i(\tau ))$ (see Eq.(\ref{b7}) for the
Hamiltonian expansion of ${ \vec j}_{\perp}(\tau ,\vec \sigma )$)

\medskip

\begin{eqnarray}
&&{\vec{A}}_{\perp S}(\tau ,\vec{\sigma})\, {\buildrel \circ \over
{{=}}}\, \sum_{i=1}^{N}Q_{i}{\vec{A}} _{\perp
Si}(\vec{\sigma}-{\vec{\eta}}_{i}(\tau
),{\vec{\kappa}}_{i}(\tau )),  \nonumber \\
&&{}  \nonumber \\
&&\vec{A}_{\perp Si}(\vec{\sigma} -
\vec{\eta}_{i},{\vec{\kappa}}_{i}) = {\ \frac{1}{4\pi |\vec{\sigma}
- \vec{\eta}_{i}|}} {\frac{1}{{\sqrt{m^2_i\, c^2 + {\vec
\kappa}^2_i} + \sqrt{ m_{i}^{2}\, c^2 + (\vec{\kappa}_{i} \cdot {\
\frac{{\vec{\sigma} - \vec{\eta}_{i}}}{{| \vec{\sigma} -
\vec{\eta}_{i}|}}}
)^{2}}}}}\times  \nonumber \\
&&\Big[{\vec \kappa}_i + {\frac{{[\vec{\kappa}_{i} \cdot
(\vec{\sigma} - \vec{\eta}_{i})]\, (\vec{ \sigma} -
\vec{\eta}_{i})}}{{|\vec{\sigma} - \vec{ \eta}_{i}|^{2}}}}\,
{\frac{\sqrt{ m_{i}^{2}\, c^2 + {{\vec{\kappa}_{i}}^{2}}}
}{\sqrt{m_{i}^{2}\, c^2 + (\vec{\kappa}_{i} \cdot {\
\frac{{\vec{\sigma} - \vec{\eta}_{i}}}{{|\vec{\sigma} -
\vec{\eta}_{i}|}}})^{2}}}} \Big] =
\nonumber \\
&=& \Big({\frac{{\alpha_{i1}}}{c}} + {\frac{{\alpha_{i3}}}{{c^3}}} +
\sum_{k=2}^{\infty}\, {\frac{{\alpha_{i\,
2k+1}}}{{c^{2k+1}}}}\Big)(\tau ,\vec \sigma ).
 \label{2.50}
\end{eqnarray}
\bigskip

\begin{eqnarray*}
 \vec{E}_{\perp S}(\tau ,\vec{\sigma}) &=&{\vec{\pi}}_{\perp
S}(\tau ,\vec{ \sigma}) = - {\frac{\partial \vec{A}_{\perp S}(\tau
,\vec{
\sigma})}{\partial \tau }} =  \nonumber \\
&=&\sum_{i=1}^{N}\, Q_{i}\, {\vec{\pi}}_{\perp Si}(\vec{\sigma} -
{\vec{\eta} }_{i}(\tau ),{\vec{\kappa}}_{i}(\tau )) =
 \end{eqnarray*}

\begin{eqnarray*}
 &=&\sum_{i=1}^{N}\, Q_{i}\, {\frac{{{\vec{\kappa}}_{i}(\tau ) \cdot
{\vec{\partial}} _{\sigma }}}{\sqrt{m_{i}^{2}\, c^2 +
{\vec{\kappa}}_{i}^{2}(\tau ) }}}\, {\vec{A}} _{\perp
Si}(\vec{\sigma}-{\vec{\eta}}_{i}(\tau
),{\vec{\kappa }}_{i}(\tau )) =  \nonumber \\
&=& - \sum_{i=1}^{N}\, Q_{i} \times  \nonumber \\
&&{\frac{1}{{4\pi |\vec{\sigma} - {\vec{\eta}}_{i}(\tau )|^{2}}}}\,
\Big[{\ \vec{ \kappa}}_{i}(\tau )\, [{\vec{\kappa}}_{i}(\tau ) \cdot
{\frac{{\vec{ \sigma} - {\ \vec{\eta}}_{i}(\tau )}}{{|\vec{\sigma} -
{\vec{\eta}}_{i}(\tau )|}}}]\, {\frac{ \sqrt{m_{i}^{2}\, c^2 +
{\vec{\kappa}}_{i}^{2}(\tau )}}{ [m_{i}^{2}\, c^2 + ({\vec{\kappa}}
_{i}(\tau ) \cdot {\frac{{\vec{\sigma} - { \ \vec{\eta}}_{i}(\tau
)}}{{|\vec{\sigma} - {\vec{\eta}}_{i}(\tau )|}}}
)^{2}]^{3/2}}}+  \nonumber \\
&+& {\frac{{\vec{\sigma} - {\vec{\eta}}_{i}(\tau )}}{{|\vec{\sigma}
- {\vec{ \eta}} _{i}(\tau )|}}}\, \Big({\frac{{\
{\vec{\kappa}}_{i}^{2}(\tau ) + ({\ \vec{\kappa}} _{i}(\tau ) \cdot
{\frac{{\vec{\sigma} - {\vec{\eta}}_{i}(\tau )}}{{|\vec{\sigma } -
{\vec{\eta}}_{i}(\tau )|}}})^{2}}}{{{\vec{\kappa}} _{i}^{2}(\tau ) -
({\vec{ \kappa}}_{i}(\tau ) \cdot {\frac{{\vec{\sigma} - { \
\vec{\eta}}_{i}(\tau )}}{{| \vec{\sigma} - {\vec{\eta}}_{i}(\tau
)|}}})^{2} }} }\, ({\frac{\sqrt{m_{i}^{2}\, c^2 +
{\vec{\kappa}}_{i}^{2}(\tau )}}{\sqrt{ m_{i}^{2}\, c^2 +
({\vec{\kappa}}_{i}(\tau ) \cdot {\frac{{\vec{\sigma} - {\
\vec{\eta}}_{i}(\tau )}}{{|\vec{\sigma} - {\vec{ \eta}}_{i}(\tau
)|}}})^{2}}}} - 1) +
 \end{eqnarray*}

\bea
 &+& {\frac{{({\vec{\kappa}}_{i}(\tau ) \cdot {\frac{{\vec{\sigma} -
{\vec{ \eta}} _{i}(\tau )}}{{|\vec{\sigma} - {\vec{\eta}}_{i}(\tau
)|}}})^{2}\, \sqrt{m_{i}^{2}\, c^2 + {\vec{\kappa}}_{i}^{2}(\tau
)}}}{{[m_{i}^{2}\, c^2 + ({\vec{\kappa}}_{i}(\tau ) \cdot
{\frac{{\vec{\sigma} - {\vec{\eta}} _{i}(\tau )}}{{|\vec{\sigma} -
{\vec{\eta}} _{i}(\tau )|}}})^{2}\,]^{3/2}}}}
\Big)\Big] =  \nonumber \\
&=& \Big({\frac{{\beta_{i2}}}{{c^2}}} + \sum_{k=2}^{\infty}\,
{\frac{{ \beta_{i\, 2k}}}{{c^{2k}}}}\Big)(\tau ,\vec \sigma ).
  \label{2.51}
\end{eqnarray}

\bigskip

\begin{eqnarray}
\vec{B}_{S}(\tau ,\vec{\sigma}) &=& \vec \partial \times {\vec
A}_{\perp S}(\tau ,\vec \sigma ) =\sum_{i=1}^N\, Q_{i}\,
{\vec{B}}_{Si}( \vec{\sigma
} - {\vec{\eta}}_{i}(\tau ), {\vec{\kappa}}_{i}(\tau ))=  \nonumber \\
&=&\sum_{i=1}^{N}\, Q_{i}\, {\frac{1}{{4\pi |\vec{\sigma} -
{\vec{\eta}} _{i}(\tau )|^{2}}}}\, {\frac{{m_{i}^{2}\, c^2\,
{\vec{\kappa}}_{i}(\tau ) \times {\frac{{\vec{ \sigma} -
{\vec{\eta}}_{i}(\tau )}}{{|\vec{\sigma} - {\ \vec{\eta}}_{i}(\tau
)|}}}} }{{[m_{i}^{2}\, c^2 + ({\vec{\kappa}}_{i}(\tau ) \cdot
{\frac{{\vec{\sigma} - {\vec{\eta }}_{i}(\tau )}}{{|\vec{\sigma} -
{\ \vec{\eta}}_{i}(\tau )|}}})^{2}\,]^{3/2}}}} =  \nonumber \\
&=& \Big({\frac{{\gamma_{i1}}}{c}} + {\frac{{\gamma_{i3}}}{{c^3}}} +
\sum_{k=2}^{\infty}\, {\frac{{\gamma_{i\,
2k+1}}}{{c^{2k+1}}}}\Big)(\tau ,\vec \sigma ).
 \label{2.52}
\end{eqnarray}
\medskip

See Appendix B, Subsection 1 for the coefficients of the expansions
and other properties, and Subsection 2 for the Fourier transform of
the Lienard-Wiechert quantities.

\bigskip

In Ref.\cite{14b} the final internal Poincare' generators for $N =
2$ in absence of the radiation field were \footnote{$V_{DARWIN} =
V_D({\tilde {\vec \eta}}_1 - {\tilde {\vec \eta}}_2, {\tilde {\vec
\kappa}} _1, {\tilde {\vec \kappa}}_2)$ is given in Eq.(6.19) of
Ref.\cite{14b} and in Eq.(\ref{4.5}). ${\tilde {\vec \eta}}_i$,
${\tilde { \vec \kappa}}_i$ are dressed particle canonical variables
given in Eqs.(5.51) of Ref.\cite{14b}, where ${\vec
{\mathcal{J}}}_{(int)}$ is given in Eq.(6.40) and ${\vec
{\mathcal{K}}}_{(int)}$ in Eq.(6.46). ${\tilde { \mathcal{K}}}_{ij}$
are a-dimensional quantities given in Eq.(5.35) of Ref. \cite{14b},
given in Eq.(\ref{3.9}).}

\begin{eqnarray*}
\mathcal{E}_{(int)} &=& M\, c^2 = c\, \sum_{i=1}^2\, \sqrt{m^2_i\,
c^2 + { \tilde {\vec \kappa}}^2_i} + {\frac{{Q_1\, Q_2}}{{4\pi\,
|{\tilde { \vec \eta }}_1 - {\tilde {\vec \eta}}_2|}}} +
V_{DARWIN}({\tilde {\vec \eta}}_1(\tau ) - {\tilde {\vec
\eta}}_2(\tau ); {\tilde {\vec \kappa}}_i(\tau )),  \nonumber \\
{\vec {\mathcal{P}}}_{(int)} &=& {\tilde {\vec \kappa}}_1 + {\tilde
{\vec \kappa}}_2 \approx 0,  \nonumber \\
{\vec {\mathcal{J}}}_{(int)} &=& \sum_{i=1}^2\, {\tilde {\vec
\eta}}_i \times {\tilde {\vec \kappa}}_i,
 \end{eqnarray*}

\bea
 {\vec {\mathcal{K}}}_{(int)} &=& -\sum_{i=1}^2\,
\widetilde{\vec{\eta}_{i}} \, \Big[ \sqrt{m_{i}^{2}\, c^2
+\widetilde{\vec{\kappa}_{i}}^{2}}+ \nonumber \\
&+&\frac{\widetilde{\vec{\kappa}_{i}}\cdot \sum_{j\neq
i}Q_{i}Q_{j}[\vec{\partial}_{\tilde{\eta}_{i}}\frac{1}{2}
\widetilde{\mathcal{K}}_{ij}( \widetilde{\vec{\kappa}_{i}},
\widetilde{\vec{\kappa}_{j}},\widetilde{\vec{ \eta}_{i}}-
\widetilde{\vec{\eta}_{j}})-2\vec{A}{{_{\perp Sj}(}}\widetilde{
\vec{\kappa}_{j}},\widetilde{\vec{\eta}_{i}} -
\widetilde{\vec{\eta}_{j}})]}{ 2\, c\, \sqrt{ m_{i}^{2}\, c^2 +
\widetilde{\vec{\kappa}_{i}}^{2}}}\Big]
-\nonumber \\
&-& \frac{1}{2\, c}\, \sum_{i=1}^2\, \sum_{j\neq i}Q_i\, Q_j\,
\sqrt{ m_{i}^{2}\, c^2 + \widetilde{\vec{\kappa}_{i}}^{2}}
\vec{\partial}_{\tilde{ \kappa}_{i}}\widetilde{
\mathcal{K}}_{ij}(\widetilde{\vec{\kappa}_{i}},
\widetilde{\vec{\kappa}_{j}}, \widetilde{\vec{\eta}_{i}} -
\widetilde{\vec{\eta}_{j}})+  \nonumber \\
&+&\sum_{i=1}^2\, \sum_{j\not=i}\frac{Q_{i}\, Q_{j}}{8\pi\, c }\,
\frac{ \widetilde{\vec{\eta}_{i}} -
\widetilde{\vec{\eta}_{j}}}{|\widetilde{\vec{\eta }_{i}}-
\widetilde{\vec{\eta}_{j}}|} - \sum_{i=1}^2\, \sum_{j\not=i}\,
\frac{ Q_{i}\, Q_{j}}{ 4\pi\, c }\int d^{3}\sigma
\frac{{\vec{\pi}}_{\perp Sj}(\vec{ \sigma}-
\widetilde{\vec{\eta}_{j}} ,\widetilde{\vec{\kappa}_{j}})}{|\vec{
\sigma}- \widetilde{\vec{\eta}_{i}}|}-  \nonumber \\
&-&{\frac{1}{2\, c}}\sum_{i=1}^2\, \sum_{j\neq i}Q_{i}Q_{j}\int
d^{3}\sigma \vec{ \sigma}[{\vec{\pi}_{\perp
Si}}(\vec{\sigma}-\widetilde{\vec{\eta}_{i}} ,
\widetilde{\vec{\kappa}_{i}})\cdot {\vec{\pi}}_{\perp
Sj}(\vec{\sigma}-\widetilde{\vec{\eta}_{j}},
\widetilde{\vec{\kappa}_{j}})+  \nonumber \\
&+&\vec{B}{_{Si}}(\vec{\sigma}-\widetilde{\vec{\eta}_{i}},\widetilde{\vec{
\kappa}_{i}})\cdot {\vec{B}}_{Sj}(\vec{\sigma} -
 \widetilde{\vec{\eta}_{j}}, \widetilde{\vec{\kappa}_{j}})] \approx 0.
 \label{2.53}
 \end{eqnarray}

\vfill\eject

\section{The Canonical Transformation.}

In this Section we will define a canonical transformation allowing
the identification of the radiation electro-magnetic field in the
radiation gauge, as an alternative to the method of Rf.\cite{1b}
(see also Eq.(4)-(5) of the Appendix of Ref.\cite{2b}) valid in the
Coulomb gauge. The non-radiative part of the electro-magnetic field
will become a dressing of charged particles and will imply the
replacement of the Coulomb potential with the Darwin one (it is a
modification beyond the $O(1/c)$ semi-relativistic standard atomic
physics). The only interaction between the dressed particles and the
radiation field is due to the rest-frame conditions and its gauge
fixings (vanishing of the internal boosts), eliminating the internal
3-center of mass. This is possible due to $Q^2_i = 0 $.

\bigskip

Knowing the Lienard-Wiechert (LW) solution of Eqs. (\ref{2.50}),
(\ref {2.51}), (\ref{2.52}), let us look for a canonical
transformation to a new canonical basis ${\vec A}_{\perp rad}(\tau
,\vec \sigma )$, ${\vec \pi} _{\perp rad}(\tau ,\vec \sigma )$,
${\hat {\vec \eta}}_i(\tau )$, ${\hat { \vec \kappa }}_i(\tau )$
such that the new transverse electromagnetic field be the radiation
field of Eqs.(\ref{2.32})

\begin{eqnarray}
{\vec A}_{\perp rad}(\tau ,\vec \sigma ) &=& {\vec A}_{\perp }(\tau
,\vec \sigma ) - {\vec A}_{\perp S}(\tau ,\vec \sigma ),  \nonumber \\
{\vec \pi}_{\perp rad}(\tau ,\vec \sigma ) &=& {\vec \pi}_{\perp
}(\tau ,\vec \sigma ) - {\vec \pi}_{\perp S}(\tau ,\vec \sigma ),  \nonumber \\
{\hat {\vec \eta}}_i(\tau ) &=& {\vec \eta}_i(\tau ) + Q_i\, {\vec
K}_i(\tau ),  \nonumber \\
{\hat {\vec \kappa}}_i(\tau ) &=& {\vec \kappa}_i(\tau ) + Q_i\,
{\vec W}_i(\tau ),  \nonumber \\
&&{}  \nonumber \\
&&\{ A^r_{\perp}(\tau ,\vec \sigma ), \pi^s_{\perp}(\tau ,{\vec
\sigma}_1)\} = - c\, P^{rs}_{\perp}(\vec \sigma )\, \delta^3(\vec
\sigma - {\vec \sigma} _1),\quad \{ \eta^r_i(\tau ),
\kappa_{js}(\tau )\} = \delta_{ij}\, \delta^r_s,  \nonumber \\
&&{}  \nonumber \\
&&\Downarrow  \nonumber \\
&&{}  \nonumber \\
&&\{ A^r_{\perp rad}(\tau ,\vec \sigma ), \pi^s_{\perp rad}(\tau
,{\vec \sigma}_1)\} = - c\, P^{rs}_{\perp}(\vec \sigma )\,
\delta^3(\vec \sigma - { \vec \sigma}_1),\quad \{ {\hat
\eta}^r_i(\tau ), {\hat \kappa}_{js}(\tau )\}
= \delta_{ij}\, \delta^r_s,  \nonumber \\
&&{}  \nonumber \\
or && \{ a_{\lambda}(\vec k), a^*_{\lambda^{^{\prime}}}({\vec
k}_1)\} = - i\, \Omega (\vec k)\, c\, \delta_{\lambda
\lambda^{^{\prime}}}\, \delta^3(\vec k - {\vec k}_1),
 \label{3.1}
\end{eqnarray}

\noindent where Eqs.(\ref{2.32}) have been used.

\bigskip

If this canonical transformation exists, then, consistently,
Eqs.(4.17) of Ref.\cite{14b} imply [see also Eq.(\ref{b7})]

\begin{equation}
\Box\, {\vec A}_{\perp}(\tau ,\vec \sigma )\,\, {\buildrel \circ
\over {{=}} } \,\, {\vec j}_{\perp}(\tau ,\vec \sigma ),\quad
\Rightarrow\quad \Box\, {\ \vec A}_{\perp rad}(\tau ,\vec \sigma
)\,\, {\buildrel \circ \over {{=}}} \,\, 0.
  \label{3.2}
\end{equation}

\bigskip

Let us build the canonical transformation. For the field sector we
have (the Poisson brackets of the LW fields vanish because only the
terms proportional to $Q^2_i = 0$ survive due to Eqs.(\ref{2.50})
and (\ref{2.51}))

\begin{eqnarray}
0 &=& \{ A^r_{\perp rad}(\tau ,\vec \sigma ), A^s_{\perp rad}(\tau
,{\vec \sigma}_1) \} = \sum_{i\not = j}^{1..N}\, Q_i\, Q_j\, \{
A^r_{\perp Si}(\tau ,\vec\sigma ), A^s_{\perp Sj}(\tau ,{\vec \sigma}_1)
\} = 0,  \nonumber \\
 0 &=& \{ \pi^r_{\perp rad}(\tau ,\vec \sigma ), \pi^s_{\perp
rad}(\tau ,{\ \vec \sigma}_1) \} = \sum_{i\not = j}^{1..N}\, Q_i\,
Q_j\, \{ \pi^r_{\perp Si}(\tau ,\vec \sigma ), \pi^s_{\perp Sj}(\tau
,{\vec \sigma}_1) \} = 0,  \nonumber \\
- c\, P^{rs}_{\perp}(\vec \sigma )\, \delta^3(\vec \sigma - {\vec
\sigma}_1) &=& \{ A^r_{\perp rad}(\tau ,\vec \sigma ), \pi^s_{\perp
rad}(\tau ,{\vec \sigma} _1) \} = - c\, P^{rs}_{\perp}(\vec \sigma
)\, \delta^3(\vec \sigma - {\vec \sigma }_1) +  \nonumber \\
&+& \sum_{i\not = j}^{1..N}\, Q_i\, Q_j\, \{ A^r_{\perp Si}(\tau
,\vec \sigma ), \pi^s_{\perp Sj}(\tau ,{\vec \sigma}_1) \} = - c\,
P^{rs}_{\perp}(\vec \sigma )\, \delta^3(\vec \sigma - {\vec
\sigma}_1).
  \label{3.3}
\end{eqnarray}

\bigskip

Let us now determine ${\hat {\vec \eta}}_i(\tau )$ and ${\hat {\vec
\kappa}} _i(\tau )$. To this end let introduce the following
functionals

\begin{equation}
T_i(\tau ) = \int d^3\sigma\, \Big[{\vec \pi}_{\perp} \cdot {\vec
A}_{\perp Si} - {\vec A}_{\perp} \cdot {\vec \pi}_{\perp
Si}\Big](\tau ,\vec \sigma ),
 \label{3.4}
\end{equation}

\noindent which have the property ($\mathcal{K}_{ij}$ had been
defined in Ref.\cite{14b} to find the particle canonical basis of
the Dirac brackets)

\begin{eqnarray}
Q_i\, Q_j\, \{ T_i(\tau ), T_j(\tau ) \} &=& - c\, Q_i\, Q_j\,
\mathcal{K}
_{ij}(\tau ) = c\, Q_i\, Q_j\, \mathcal{K}_{ji}(\tau ),  \nonumber \\
&&{}  \nonumber \\
\mathcal{K}_{ij}(\tau ) &=& \int d^3\sigma\, \Big[{\vec A}_{\perp
Si} \cdot { \ \vec \pi}_{\perp Sj} - {\vec \pi}_{\perp Si} \cdot
{\vec A}_{\perp Sj} \Big] (\tau ,\vec \sigma ),
  \label{3.5}
\end{eqnarray}

\noindent and let us make the following ansatz

\begin{eqnarray}
{\hat \eta}^r_i(\tau ) &=& \eta^r_i(\tau ) + {\frac{{Q_i}}{c}}\,
{\frac{{\partial\, T_i(\tau )}}{{\partial\, \kappa_{ir}}}} -
{\frac{1}{2}}\, {\frac{{ Q_i}}{c}}\, \sum_{k\not= i}^{1..N}\, Q_k\,
{\frac{{\partial\, \mathcal{K}_{ik}(\tau
)}}{{\partial\, \kappa_{ir}}}},  \nonumber \\
{\hat \kappa}_{ir}(\tau ) &=& \kappa_{ir}(\tau ) -
{\frac{{Q_i}}{c}}\, { \frac{{\partial\, T_i(\tau )}}{{\partial\,
\eta^r_i}}} + {\frac{1}{2}}\, { \frac{{Q_i}}{c}}\, \sum_{k\not=
i}^{1..N}\, Q_k\, {\frac{{\partial\, \mathcal{K}
_{ik}(\tau )}}{{\partial\, \eta_i^r} }},  \nonumber \\
&&Q_i\, {\hat {\vec \eta}}_i = Q_i\, {\vec \eta}_i,\qquad Q_i\,
{\hat {\vec \kappa}}_i = Q_i\, {\vec \kappa}_i.
 \label{3.6}
\end{eqnarray}
\bigskip

Let us verify whether the ansatz defines a canonical transformation.
We have immediately

\begin{eqnarray}
0 &=& \{ A^r_{\perp rad}(\tau ,\vec \sigma ), {\hat \eta}^s_i(\tau )
\} = Q_i\, \Big[{\frac{1}{c}}\, \{ A^r_{\perp}(\tau ,\vec \sigma ),
{\frac{{\partial\, T_i(\tau )} }{{\partial\, \kappa_{is}}}}\} +
{\frac{{\partial\, A^r_{\perp Si}(\tau ,\vec \sigma )} }{{\partial\,
\kappa_{is}}}}\Big] =\nonumber \\
&=& Q_i\, \Big[{\frac{{\partial}}{{\partial\, \kappa_{is}}}}\, \int
d^3\sigma_1\, \Big(- P^{ru}_{\perp}(\vec \sigma )\, \delta^3(\vec
\sigma - {\vec \sigma}_1)\Big)\, A^u_{\perp Si}(\tau ,{\vec \sigma}_1) +  \nonumber \\
&+& {\frac{{\partial\,A^r_{\perp Si}(\tau ,\vec \sigma
)}}{{\partial\, \kappa_{is}}}} \Big] = 0,  \nonumber \\
&&{}  \nonumber \\
0 &=& \{ \pi^r_{\perp rad}(\tau ,\vec \sigma ), {\hat \eta}^s_i(\tau
) \} = Q_i\, \Big[{\frac{1}{c}}\, \{ \pi^r_{\perp}(\tau ,\vec \sigma
), {\frac{{\partial\, T_i(\tau )}}{{\partial\, \kappa_{is}}}}\} +
{\frac{{\partial\, \pi^r_{\perp Si}(\tau ,\vec \sigma )}
}{{\partial\, \kappa_{is}}}}\Big] = 0,\nonumber \\
&&{}  \nonumber \\
0 &=& \{ A^r_{\perp rad}(\tau ,\vec \sigma ), {\hat \kappa}_{is} \}
= Q_i\, \Big[- {\frac{1}{c}}\, \{ A^r_{\perp}(\tau ,\vec \sigma ),
{\frac{{\partial\, T_i(\tau )}}{{\ \partial\, \eta_i^s}}}\} -
{\frac{{\partial\, A^r_{\perp Si}(\tau ,\vec \sigma )}}{{\partial\,
\eta_i^s}}}\Big] = 0,\nonumber \\
&&{}  \nonumber \\
0 &=& \{ \pi^r_{\perp rad}(\tau ,\vec \sigma ), {\hat \kappa}_{is}
\} = Q_i\, \Big[- {\frac{1}{c}}\, \{ \pi^r_{\perp}(\tau ,\vec \sigma
), {\frac{{\partial\, T_i(\tau )}}{{\partial\, \eta_i^s}}}\} -
{\frac{{\partial\, \pi^r_{\perp Si}(\tau ,\vec \sigma
)}}{{\partial\, \eta_i^s}}}\Big] = 0.
 \label{3.7}
\end{eqnarray}

\bigskip

Finally we have

\begin{eqnarray}
\delta_{ij}\, \delta^r_s &=& \{ {\hat \eta}^r_i(\tau ), {\hat
\kappa}_{js}(\tau )\} = \delta_{ij}\, \delta^r_s +  \nonumber \\
&+& {\frac{{\partial}}{{\partial\, \kappa_{ir}}}}\, \Big(-
{\frac{{Q_j}}{c}} \, {\frac{{\partial T_j(\tau )}}{{\partial\,
\eta^s_j}}} + {\frac{1}{2}}\, { \frac{{Q_j}}{c}}\, \sum_{k\not=
j}^{1..N}\, Q_k\, {\frac{{\partial\, \mathcal{K}
_{jk}(\tau )}}{{\ \partial\, \eta^s_j}}}\Big) +  \nonumber \\
&+& {\frac{{\partial}}{{\partial\, \eta^s_j}}}\,
\Big({\frac{{Q_i}}{c}}\, { \frac{{\partial T_i(\tau )}}{{\partial\,
\kappa_{ir}}}} - {\frac{1}{2}}\, { \frac{{Q_i}}{c}}\, \sum_{k\not=
i}\, Q_k\, {\frac{{\partial\, \mathcal{K} _{ik}(\tau )}}{{\partial\,
\kappa_{ir}}}} \Big) - {\frac{{Q_i\, Q_j}}{{c^2}}} \, \{
{\frac{{\partial\, T_i(\tau )}}{{\ \partial\, \kappa_{ir}}}},
{\frac{{\partial\, T_j(\tau )}}{{\partial\, \eta^s_j}}} \} =  \nonumber \\
&&{}  \nonumber \\
&=& \delta_{ij}\, \delta^r_s - {\frac{1}{2}}\, \delta_{ij}\,
{\frac{{Q_i}}{c} }\, \sum_{k\not= i}\, Q_k\,
\Big({\frac{{\partial^2\, \mathcal{K}_{ik}(\tau ) }}{{\ \partial\,
\eta^s_i\, \partial\, \kappa_{ir}}}} - {\frac{{\partial^2\,
\mathcal{K}_{ik}(\tau )}}{{\partial\, \kappa_{ir}\, \partial\,
\eta^s_i}}}\Big) +  \nonumber \\
&+& {\frac{1}{2}}\, {\frac{{Q_i\, Q_j}}{c}}\,
\Big({\frac{{\partial^2\, \mathcal{K} _{ji}(\tau )}}{{\partial\,
\eta^s_j\, \partial\, \kappa_{ir}}}} - {\frac{{\ \partial^2\,
\mathcal{K}_{ij}(\tau )}}{{\partial\, \kappa_{ir}\,
\partial\, \eta^s_j}}} \Big) + {\frac{{Q_i\, Q_j}}{c}}\, {\frac{{
\partial^2\, \mathcal{K}_{ij}(\tau )} }{{\partial\, \kappa_{ir}\, \partial\,
\eta^s_j}}} = \delta_{ij}\, \delta^r_s.
  \label{3.8}
\end{eqnarray}

Analogously we have that the conditions $0 = \{ {\hat \eta}^r_i,
{\hat \eta} ^s_j \} = \{ {\hat \kappa}_{ir}, {\hat \kappa}_{js} \}$
are satisfied.

\bigskip

Since $Q_i\, {\hat {\vec A}}_{\perp Si}$ and $Q_i\, {\hat {\vec
\pi}}_{\perp Si}$ are the same functions (\ref{2.50}) and
(\ref{2.51}) of ${\hat {\vec \eta} }_i$, ${\hat {\vec \kappa}}_i$ as
$Q_i\, {\vec A}_{\perp Si}$ and $ Q_i\, {\ \vec \pi}_{\perp SI}$
were of ${\vec \eta}_i$, ${\vec \kappa}_i$ and since we have

\begin{eqnarray}
Q_i\, Q_j\, \mathcal{K}_{ij}(\tau ) &=& Q_i\, Q_j\, {\hat
{\mathcal{K}}} _{ij}(\tau ) = Q_i\, Q_j\, \int d^3\sigma\,
\Big[{\hat {\vec A}}_{\perp Si} \cdot {\hat {\vec \pi}}_{\perp Sj} -
{\hat {\vec \pi}}_{\perp Si} \cdot {\ \hat {\vec A}}_{\perp Sj}\Big](\tau
,\vec \sigma ),  \nonumber \\
&&{}  \nonumber \\
Q_i\, T_i(\tau ) &=& Q_i\, {\hat T}_i(\tau ) + Q_i\, \sum_{k\not=
i}^{1..N}\, Q_k\, {\hat {\mathcal{K}}}_{ik}(\tau ),\quad with  \nonumber \\
{\hat T}_i(\tau ) &=& \int d^3\sigma\, \Big[{\vec \pi}_{\perp rad}
\cdot {\ \hat {\vec A}}_{\perp Si} - {\vec A}_{\perp rad} \cdot
{\hat {\vec \pi}} _{\perp Si}\Big](\tau ,\vec \sigma ),
  \label{3.9}
\end{eqnarray}

\noindent the inverse of the canonical transformation (\ref{3.1}),
(\ref{3.6} ) is

\begin{eqnarray}
{\vec A}_{\perp}(\tau ,\vec \sigma ) &=& {\vec A}_{\perp rad} +
{\hat {\vec A}}_{\perp S}(\tau ,\vec \sigma ),  \nonumber \\
{\vec \pi}_{\perp}(\tau ,\vec \sigma ) &=& {\vec \pi}_{\perp rad} +
{\hat {\vec \pi}}_{\perp S}(\tau ,\vec \sigma ),  \nonumber \\
&&{}  \nonumber \\
\eta^r_i(\tau ) &=& {\hat \eta}^r_i(\tau ) - {\frac{{Q_i}}{c}}\,
{\frac{{\partial\, {\hat T }_i(\tau )}}{{\partial\, {\hat
\kappa}_{ir}}}} - {\frac{1 }{2}}\, {\frac{{Q_i}}{c}}\, \sum_{k\not=
i}^{1..N}\, Q_k\, {\frac{{\partial\, { \hat {\mathcal{K}}}_{ik}(\tau
)}}{{\partial\,{\hat \kappa}_{ir}}}},\nonumber \\
\kappa_{ir}(\tau ) &=& {\hat \kappa}_{ir}(\tau ) +
{\frac{{Q_i}}{c}}\, { \frac{{\partial\, {\ \hat T}_i(\tau
)}}{{\partial\, {\hat \eta}^r_i}}} + { \frac{1}{2}}\,
{\frac{{Q_i}}{c}}\, \sum_{k\not= i}^{1..N}\, Q_k\, {\frac{{
\partial\, {\hat {\mathcal{K}}}_{ik}(\tau )} }{{\partial\, {\hat \eta}_i^r}}}
.
  \label{3.10}
\end{eqnarray}

\bigskip

The generating function of this canonical transformation is

\beq
 S = {1\over c}\,\, \sum_i^{1..N}\, Q_i\, T_i[{\vec A}_{\perp}, {\vec \pi}_{\perp},
 {\vec \eta}_i, {\vec \kappa}_i].
 \label{3.11}
 \eeq

Due to the Grassmann-valued charges for every phase space variable
$A = {\vec A}_{\perp}(\tau ,\vec \sigma )$, ${\vec \pi}_{\perp}(\tau
,\vec \sigma )$, ${\vec \eta}_i(\tau )$, ${\vec \kappa}_i(\tau )$,
we have the following truncation

\beq
 e^{\{ ., S\}}\, A = A + \{ A, S\} + {1\over 2}\, \{\, \{ A, S \} ,
 S \}.
 \label{3.12}
 \eeq
\medskip

Eqs.(\ref{3.1}) and (\ref{3.6}) are reproduced by means of
Eqs.(\ref{3.12})

\bea {\hat{\eta}}_{i}^{r}(\tau ) &=& e^{\{\cdot ,S\}}\, \eta
_{i}^{r}(\tau ), \nonumber \\
{\hat{\kappa}}_{ir}(\tau ) &=& e^{\{\cdot ,S\}}\, \kappa _{ir}(\tau
)_{i}^{r}(\tau ), \nonumber \\
{\vec{A}}_{\perp rad}(\tau ,\vec{\sigma}) &=& e^{\{\cdot ,S\}}\,
{\vec{A}}_{\perp }(\tau ,\vec{\sigma}), \nonumber \\
{\vec{\pi}}_{\perp rad}(\tau ,\vec{\sigma}) &=& e^{\{\cdot ,S\}}\,
{\vec{\pi}}_{\perp }(\tau ,\vec{\sigma}).
 \label{3.13}
  \eea

\bigskip

This result says that, at least at the classical level, the isolated
system of "N particles with Grassmann-valued electric charges and
mutual Coulomb interaction plus the transverse electro-magnetic
field" is canonically equivalent to a free radiation field plus a
system of Coulomb-dressed charged particles mutually interacting
through a Coulomb plus Darwin potential. This is possible because
the Grassmann-valued charges imply the following properties:\medskip

a) A regularization of Coulomb self-energies, namely a ultraviolet
cutoff.

b) The emergence of the Darwin potential, namely of the main
ingredient for the theory of relativistic bound states in scalar
QED.

c) Transverse Lienard-Wiechert potential and electric field
depending only on the positions and momenta of the charged particles
and not on the higher accelerations. Therefore they are the
action-at-a-distance potentials hidden in all the possible
(retarded, advanced, symmetric,..) solutions of Maxwell equations.
The Coulomb plus Darwin potentials give the Cauchy problem
formulation of the interaction hidden in the one-photon exchange
Feynman diagrams. All the loop diagrams and the soft and hard photon
emission diagrams are eliminated. Therefore there is also an
infrared regularization. Also the classical problems with causality
violations (either runaway solutions or pre-accelerations) are
absent. There is no Larmor emission of radiation from a single
Grasmmann-valued charge. However, as shown in Ref.\cite{b}, in the
wave zone of a system of Grassmann-valued charges one can obtain the
Larmor formula for the radiated energy as a result of he
interference terms $Q_i\, Q_j$ with $i \not= j$ (${{dE}\over
{d\tau}} = {2\over 3}\, \sum_{i \not= j}\, {{Q_i\, Q_j}\over {(4\pi
)^2}}\, {\ddot {\vec \eta}}_i(\tau ) \cdot {\ddot {\vec
\eta}}_j(\tau )$).

d) In the rest-frame instant form the isolated system of "N
particles with Grassmann-valued electric charges and with mutual
Coulomb interaction plus the transverse electro-magnetic field" is
described as a non-covariant external decoupled center of mass
carrying the internal mass and the spin of the system. The system
lives inside the instantaneous Euclidean Wigner 3-spaces, where the
canonical equivalence with a radiation field and a system of
Coulomb-dressed Grassmann-valued charges with mutual Coulomb plus
Darwin interaction holds. However, the two non-interacting
subsystems are connected by the rest-frame condition ${\vec {\cal
P}}_{(int)} \approx 0$ and by the elimination, ${\vec {\cal
K}}_{(\int)} \approx 0$, of the internal 3-center of mass. As a
consequence, the radiation field {\it knows} the particle subsystem:
strictly speaking it is {\it not a free field}. Therefore the
existence of the canonical equivalence may be a classical
implementation of the mechanism which allows the Haag-Ruelle
scattering theory (see for instance Ref.\cite{16b}) to avoid the
Haag theorem. As clearly said in Ref.\cite{a}, this happens when it
is possible to define in the Hilbert space of the interacting fields
"surrogates" of the free field states in the asymptotic $t
\rightarrow\, \pm\, \infty$ regimes (these surrogates are not
unitarily equivalent to free fields). Moreover in our case there is
the universal breaking of Lorentz covariance connected to the
decoupled external center of mass, which is present in the
rest-frame instant form even in absence of particles and even if the
No-Interaction theorem does not hold in field theory \cite{22b}.
This is the price for having this instant form emerging from the
general description of isolated systems in non-inertial frames.

\vfill\eject

\section{The Internal Poincare' Generators and the Hamilton Equations after the
Canonical Transformation}

\subsection{The New Expression of the Internal Poincare' Generators}

For the internal 3-momentum of Eq.(\ref{2.22}) in the new canonical
basis Eqs. (\ref{c1})-(\ref{c3}) imply the following expression

\begin{eqnarray}
{\vec {\mathcal{P}}}_{(int)} &=& \sum_{i=1}^N\, {\hat {\vec
\kappa}}_i(\tau ) + { \frac{1}{c}}\, \int d^3\sigma\, \Big({\vec
\pi}_{\perp rad} \times {\vec B}
_{rad} \Big) (\tau ,\vec \sigma ) +  \nonumber \\
&+& \sum_{i=1}^N\, {\frac{{Q_i}}{c}}\, \Big[{\frac{{\partial\, {\hat
T}_i(\tau )}}{ {\partial {\ \hat {\vec \eta}}_i}}} + \int
d^3\sigma\, \Big({\vec \pi} _{\perp rad} \times {\hat {\vec B}}_{Si}
+ {\hat {\vec \pi}}_{\perp Si}
\cdot {\vec B}_{rad}\Big) (\tau ,\vec \sigma )\Big] +  \nonumber \\
&+& \sum_{i \not= j}^{1..N}\, {\frac{{Q_i\, Q_j}}{c}}\, \Big( \int
d^3\sigma\, \Big( {\hat {\vec \pi }}_{\perp Si} \times {\hat {\vec
B}}_{Sj} \Big)(\tau ,\vec \sigma ) + {\frac{ 1}{2}}\,
{\frac{{\partial\, {\hat {\mathcal{K}}}_{ij}}}{{
\partial\, {\hat { \vec \eta}}_i}}}\Big) =  \nonumber \\
&=& \sum_{i=1}^N\, {\hat {\vec \kappa}}_i(\tau ) + {\frac{1}{c}}\,
\int d^3\sigma\, \Big({\vec \pi} _{\perp rad} \times {\vec B}_{rad}
\Big)(\tau ,\vec \sigma ) =  \nonumber \\
&=& {\vec {\mathcal{P}}}_{matter} + {\vec
{\mathcal{P}}}_{rad}\approx 0.
 \label{4.1}
\end{eqnarray}
\medskip

For the internal angular momentum of Eq.(\ref{2.22}) Eq.(\ref{c4})
implies

\begin{eqnarray}
{\vec {\mathcal{J}}}_{(int)} &=& \sum_{i=1}^N\, {\hat {\vec \eta}}_i
\times {\hat { \vec \kappa}}_i + {\frac{1}{c}}\, \int d^3\sigma\,
\vec \sigma \times \Big({ \vec \pi}_{\perp rad} \times {\vec
B}_{rad}\Big)(\tau ,\vec \sigma ) +\nonumber \\
 &+& \sum_{i=1}^N\, {\frac{{Q_i}}{c}}\, \Big[\Big({\hat {\vec \eta}}_i(\tau
) \times {\frac{{\partial }}{{\partial {\hat {\vec \eta}}_i}}} +
{\hat {\vec \kappa}}_i(\tau ) \times {\frac{{\partial}}{{\partial\,
{\hat {\vec \kappa}}_i}}}\Big)\, {\hat T}_i(\tau ) +  \nonumber \\
&+& \int d^3\sigma\, \vec \sigma \times \Big({\vec \pi}_{\perp rad}
\times { \ \hat {\vec B}}_{Si} + {\hat {\vec \pi}}_{\perp Si} \times
{\vec B}_{rad}\Big) (\tau ,\vec \sigma )\Big] +  \nonumber \\
&+& \sum_{i \not= j}^{1..N}\, {\frac{{Q_i\, Q_j}}{c}}\,
\Big[{\frac{1}{2}}\, \Big({ \hat {\vec \eta }}_i \times
{\frac{{\partial }}{{\partial\, {\hat {\vec \eta} }_i}}} + {\hat {
\vec \kappa}}_i \times {\frac{{\partial }}{{\partial\, { \hat {\vec
\kappa}}_i }}} \Big) {\hat {\mathcal{K}}}_{ij}(\tau ) + \nonumber \\
&+& \int d^3\sigma\, \vec \sigma \times \Big({\hat {\vec
\pi}}_{\perp Si} \times {\hat {\vec B}}_{Sj} \Big)(\tau
,\vec \sigma )\Big] =  \nonumber \\
&=& \sum_{i=1}^N\, {\hat {\vec \eta}}_i \times {\hat {\vec
\kappa}}_i + {\frac{1}{c }}\, \int d^3\sigma\, \vec \sigma \times
\Big({\vec \pi}_{\perp rad} \times {
\vec B} _{rad}\Big)(\tau ,\vec \sigma ) =  \nonumber \\
&=& {\vec {\mathcal{J}}}_{matter} + {\vec {\mathcal{J}}}_{rad}.
 \label{4.2}
\end{eqnarray}

\bigskip

By adding the second class constraints ${\vec A}_{\perp rad}(\tau
,\vec \sigma ) \approx 0$ and ${\vec \pi}_{\perp rad}(\tau ,\vec
\sigma ) \approx 0 $ we recover the expression of Eqs.(\ref{2.53})
for the internal 3-momentum and angular momentum

\bigskip

Since we have

\begin{eqnarray}
c\, \sqrt{m^2_i\, c^2 + {\vec \kappa}^2_i} &=& c\, \sqrt{m^2_i\, c^2
+ { \hat {\vec \kappa}}^2_i + {\frac{{2\, Q_i}}{c}}\, {\hat {\vec
\kappa}}_i \cdot \Big({\frac{{\partial\, {\hat T}_i} }{{\partial\,
{\hat {\vec \eta}}_i} }} + {\frac{1}{2}}\, \sum_{k\not= i}^{1..N}\,
Q_k\, {\frac{{\partial\, {\hat { \mathcal{K}}}_{ik}}}{{\partial\,
{\hat{\vec \eta}}_i}}}\Big)} = \nonumber \\
&=& c\, \sqrt{m^2_i\, c^2 + {\hat {\vec \kappa}}^2_i} + Q_i\,
{\frac{{{\hat { \vec \kappa}}_i \cdot \Big({\frac{{\partial\, {\hat
T}_i}}{{\partial\, { \hat { \vec \eta}}_i}}} + {\frac{1}{2}}\,
\sum_{k\not= i}^{1..N}\, Q_k\, {\frac{{\ \partial\, {\hat
{\mathcal{K}}}_{ik}}}{{\partial\, {\hat {\vec \eta}}_i}}}
\Big)}}{\sqrt{m^2_i\, c^2 + {\hat {\vec \kappa}}^2_i}}},  \nonumber \\
&&{}  \nonumber \\
&&Q_i\, \sqrt{m^2_i\, c^2 + {\vec \kappa}^2_i} = Q_i\, \sqrt{m^2_i\,
c^2 + { \hat {\vec \kappa}}^2_i},
  \label{4.3}
\end{eqnarray}

\noindent the internal energy of (\ref{2.22}) takes the following
form

\begin{eqnarray*}
\mathcal{E}_{(int)}
&=&M\,c^{2}=c\,\sum_{i=1}^N\,\sqrt{m_{i}^{2}\,c^{2}+{\ \hat{
\vec{\kappa}}}_{i}^{2}(\tau )}+  \nonumber \\
&+&\sum_{i=1}^N\, Q_i\, {\frac{{{\hat{\vec{\kappa}}}_{i}(\tau )
}}{\sqrt{ m_{i}^{2}\,c^{2}+{\hat{\vec{\kappa}}} _{i}^{2}(\tau )}}}
\cdot \Big[ \Big({ \frac{{\ \partial \,{\hat{T}}_{i}}}{{\partial
\,{\hat{\vec{\eta}}}_{i}}}}+{ \frac{1}{2}
}\,\sum_{j\not=i}\,Q_{j}\,{\frac{{\partial \,{\hat{\mathcal{K}}}
_{ij}}}{{\ \partial \,{\hat{\vec{\eta}}}_{i}}}}\Big)-  \nonumber \\
&-&{\vec{A}}_{\perp rad}(\tau ,{\hat{\vec{\eta}}}_{i}(\tau
))-\sum_{j\not=i}\,Q_{j}\,{\hat{\vec{A}}}_{\perp Sj}(\tau
,{\hat{\vec{\eta}}} _{i}(\tau ))\Big] + \sum_{i\not=j}^{1..N}\,
{\frac{{Q_{i}\,Q_{j}}}{{4\pi \,|{\hat{\vec{
\eta}}}_{i}(\tau )-{\hat{\vec{\eta}}}_{j}(\tau )|}}}+  \nonumber \\
&+&{\frac{1}{2}}\,\int d^{3}\sigma \,\Big({\vec{\pi}}_{\perp
rad}^{2}+{\vec{B }}_{rad}^{2}\Big)(\tau
,\vec{\sigma})+\sum_{i}\,Q_{i}\,\int d^{3}\sigma \,
\Big({\vec{\pi}}_{\perp rad}\cdot {\hat{\vec{\pi }}}_{\perp
Si}+{\vec{B}}_{rad}\cdot {\hat{\vec{B}}}_{Si}\Big)(\tau
,\vec{ \sigma})+  \nonumber \\
&+&{\frac{1}{2}}\,\sum_{i\not=j}^{1..N}\,Q_{i}\,Q_{j}\,\int
d^{3}\sigma \,\Big( { \hat{\vec{\pi}}}_{\perp Si}\cdot
{\hat{\vec{\pi}}}_{\perp Sj}+{\hat{\vec{B}}}
_{Si}\cdot {\hat{\vec{B}}}_{Sj}\Big)(\tau ,\vec{\sigma})=  \nonumber \\
&&{}  \nonumber \\
&&{}  \nonumber \\
 &=&c\,\sum_{i=1}^N\,\sqrt{m_{i}^{2}\,c^{2}+{\hat{\vec{\kappa}}}_{i}^{2}(\tau
)} +\sum_{i\not=j}^{1..N}\,{\frac{{Q_{i}\,Q_{j}}}{{4\pi
\,|{\hat{\vec{\eta}}}_{i}(\tau
)-{\hat{\vec{\eta}}}_{j}(\tau )|}}}+V_{DARWIN}+  \nonumber \\
&+&\sum_{i=1}^N\,Q_{i}\,\Big(\int d^{3}\sigma \,\Big[{\vec{\pi}}
_{\perp rad}\cdot \Big({\hat{\vec{\pi}}}_{\perp Si}+({\frac{c\, {{\
\hat{\vec{\kappa}}}_{i}}}{\sqrt{m_{i}^{2}\,c^{2}+{\hat{\vec{\kappa}}}_{i}^{2}
}}\cdot }\,{ \frac{{\partial }}{{\partial
\,{\hat{\vec{\eta}}}_{i}}}})\,{\hat{ \vec{A}}}
_{\perp Si}\Big)-  \nonumber \\
&-&{\vec{A}}_{\perp rad}\cdot
\Big({\vec{\partial}}^{2}\,{\hat{\vec{A}}} _{\perp Si}+({\frac{c\,
{{\hat{\vec{\kappa}}}_{i}}}{\sqrt{m_{i}^{2}\,c^{2}+{ \hat{
\vec{\kappa}}}_{i}^{2}}}\cdot }\,\ \ {\frac{{\partial }}{{\partial
\,{ \hat{ \vec{\eta}}}_{i}}}})\,{\hat{\vec{\pi}}}_{\perp
Si}\Big)\Big](\tau ,\vec{\sigma})-  \nonumber \\
&-&{\frac{{{\hat{\vec{\kappa}}}_{i}\cdot {\vec{A}}_{\perp rad}(\tau
,{\hat{ \vec{\eta}}}_{i}(\tau ))}}{\sqrt{m_{i}^{2}\,c^{2} +
{\hat{\vec{\kappa}}} _{i}^{2}}}}\Big)+{\frac{1}{2}}\,\int
d^{3}\sigma \,\Big({\vec{\pi}}_{\perp rad}^{2} +
{\vec{B}}_{rad}^{2}\Big)(\tau ,\vec{\sigma}) =
 \end{eqnarray*}

\bea
 &=& c\,\sum_{i=1}^N\,\sqrt{m_{i}^{2}\,c^{2}+{\hat{\vec{
\kappa}}}_{i}^{2}(\tau )} + \sum_{i\not=j}^{1..N}\,
{\frac{{Q_{i}\,Q_{j}}}{{4\pi \,|{\ \hat{\vec{\eta}}}
_{i}(\tau )-{\hat{\vec{\eta}}}_{j}(\tau )|}}}+V_{DARWIN}+  \nonumber \\
&+& {\frac{1}{2}}\,\int d^{3}\sigma \,\Big({\vec{ \pi}}_{\perp
rad}^{2}+{\vec{B}}_{rad}^{2}\Big)(\tau ,\vec{\sigma}) =  \nonumber \\
&=& \mathcal{P}^{\tau}_{matter} + \mathcal{P}^{\tau}_{rad},
  \label{4.4}
\end{eqnarray}

\noindent with the following expression for the Darwin potential

\begin{eqnarray}
V_{DARWIN}({\hat{\vec{\eta}}}_{1}(\tau )-{\hat{\vec{\eta}}}_{2}(\tau
);{\hat{ \vec{\kappa}}}_{i}(\tau ))
&=&\sum_{i\not=j}^{1..N}\,Q_{i}\,Q_{j}\,\Big({\frac{{{\
\hat{\vec{\kappa}}}_{i}\cdot {\hat{\vec{A}}}_{\perp Sj}(\tau
,{\hat{\vec{ \eta }}}_{i}(\tau
))}}{\sqrt{m_{i}^{2}\,c^{2}+{\hat{\vec{\kappa}}}_{i}^{2}}}}
+  \nonumber \\
&+&\int d^{3}\sigma \,\Big[{\frac{1}{2}}\,\Big({\hat{\vec{\pi}
}}_{\perp Si}\cdot {\hat{\vec{\pi}}}_{\perp
Sj}+{\hat{\vec{B}}}_{Si}\cdot {\ \hat{\vec{
B}}}_{Sj}\Big)+  \nonumber \\
&+&({\frac{{{\hat{\vec{\kappa}}}_{i}}}{\sqrt{m_{i}^{2}\,c^{2}+{\hat{\vec{
\kappa}}}_{i}^{2}}}}\cdot {\frac{{\partial }}{{\partial
\,{\hat{\vec{\eta}}} _{i}}}})\, \Big({\hat{\vec{A}}}_{\perp Si}\cdot
{\hat{\vec{\pi}}} _{\perp Sj}-{\hat{\vec{\pi}}}_{\perp Si}\cdot
{\hat{\vec{A}}}_{\perp Sj}\Big) \Big]
(\tau ,\vec{\sigma})\Big).  \nonumber \\
&&{}  \label{4.5}
\end{eqnarray}

\bigskip

By adding the second class constraints ${\vec A}_{\perp rad}(\tau
,\vec \sigma ) \approx 0$ and ${\vec \pi}_{\perp rad}(\tau ,\vec
\sigma ) \approx 0 $ (implying ${\hat T}_i(\tau ) \approx 0$) we
recover the internal energy of Eqs.(\ref{2.53}). \bigskip

Since the canonical transformation is not explicitly
$\tau$-dependent, $M\, c^2$ is still the hamiltonian and both the
energy ${\cal P}^{\tau}_{matter}$ of the dressed particles and the
energy ${\cal P}^{\tau}_{rad}$ of the radiation field are constants
of motion.\bigskip

See Subsection 3 of Appendix B for the $1/c$ expansion of
Eq.(\ref{4.4} ).\bigskip

\bigskip

For the internal boosts, we substitute expressions for the new
canonical
variables given in Eqs.(\ref{3.10}) into the boost expression given in Eq.(%
\ref{2.23}) to obtain

\begin{eqnarray}
{\vec{\mathcal{K}}}_{(int)} &=&-\sum_{i=1}^{N}\,{\hat{\vec{\eta}}}_{i}(\tau
)\,\Big(\sqrt{m_{i}^{2}\,c^{2}+{\hat{\vec{\kappa}}}_{i}^{2}(\tau )}+
\nonumber \\
&+&{\frac{{Q_{i}}}{{c\,\sqrt{m_{i}^{2}\,c^{2}+{\hat{\vec{\kappa}}}_{i}(\tau )%
}}}}\,{\hat{\vec{\kappa}}}_{i}(\tau )\cdot \Big[{\frac{{\partial \,{\hat{T}}%
_{i}(\tau )}}{{\partial \,{\hat{\vec{\eta}}}_{i}}}}+{\frac{1}{2}}%
\,\sum_{j\not=i}^{1..N}\,Q_{j}\,{\frac{{\partial \,{\hat{\mathcal{K}}}%
_{ij}(\tau )}}{{\partial \,{\hat{\vec{\eta}}}_{i}}}}-  \nonumber \\
&-&{\vec{A}}_{\perp rad}(\tau ,{\hat{\vec{\eta}}}_{i}(\tau
))-\sum_{j\not=i}\,Q_{j}\,{\hat{\vec{A}}}_{\perp Sj}(\tau ,{\hat{\vec{\eta}}}%
_{i}(\tau ))\Big]\Big)+  \nonumber \\
&+&\sum_{i=1}^{N}\,{\frac{{Q_{i}}}{c}}\,\sqrt{m_{i}^{2}\,c^{2}+{\hat{\vec{%
\kappa}}}_{i}^{2}(\tau )}\,{\frac{{\partial \,{\hat{T}}_{i}(\tau )}}{{%
\partial \,{\hat{\vec{\kappa}}}_{i}}}}-{\frac{1}{2}}\,\sum_{i\not=j}\,{\frac{%
{Q_{i}\,Q_{j}}}{c}}\,\sqrt{m_{i}^{2}\,c^{2}+{\hat{\vec{\kappa}}}%
_{i}^{2}(\tau )}\,{\frac{{\ \partial \,{\hat{\mathcal{K}}}_{ij}(\tau )}}{{%
\partial \,{\hat{\vec{\kappa}}}_{i}}}}+  \nonumber \\
&+&{\frac{1}{c}}\,\sum_{i=1}^{N}\,Q_{i}\,\Big[\sum_{j\not=i}^{1..N}\,Q_{j}\,[%
{\frac{1}{{\triangle _{{\ \hat{\vec{\eta}}}_{j}}}}}{\frac{{\partial }}{{%
\partial {\hat{\eta}}_{j}^{r}}}}c({\hat{\vec{\eta}}}_{i}(\tau )-{\hat{\vec{%
\eta}}}_{j}(\tau ))-{\hat{\eta}}_{j}^{r}(\tau )\,c({\hat{\vec{\eta}}}%
_{i}(\tau )-{\hat{\vec{\eta}}}_{j}(\tau ))]+  \nonumber \\
&+&\int d^{3}\sigma \,c(\vec{\sigma}-{\hat{\vec{\eta}}}_{i}(\tau ))\,\Big({%
\vec{\pi}}_{\perp rad}+\sum_{j\not=i}\,\,Q_{j}\,{\ \hat{\vec{\pi}}}_{\perp
Sj}\Big)(\tau ,\vec{\sigma})\Big]-  \nonumber \\
&&-{\frac{1}{2c}}\int d^{3}\sigma \,\vec{\sigma}\,\,\Big({\ \vec{\pi}}%
_{\perp rad}^{2}+{\vec{B}}_{rad}^{2}\Big)(\tau ,\vec{\sigma})-  \nonumber \\
&&-\sum_{i=1}^{N}\,\frac{Q_{i}}{c}\,\int d^{3}\sigma \,\vec{\sigma}\,\,\Big({%
\vec{\pi}}_{\perp rad}\cdot {\hat{\vec{\pi}}}_{\perp Si}+{\vec{B}}%
_{rad}\cdot {\hat{\vec{B}}}_{Si}\Big)(\tau ,\vec{\sigma})-  \nonumber \\
&&-{\frac{1}{2c}}\,\sum_{i\not=j}^{1..N}\,Q_{i}\,Q_{j}\,\int d^{3}\sigma \,%
\vec{\sigma}\,\,\Big({\hat{\vec{\pi}}}_{\perp Si}\cdot {\hat{\vec{\pi}}}%
_{\perp Sj}+{\hat{\vec{B}}}_{Si}\cdot {\hat{\vec{B}}}_{Sj}\Big)(\tau ,\vec{%
\sigma})  \label{4.6}
\end{eqnarray}

Then using the results of Eqs.(\ref{c5})-(\ref{c18}) we get%
\begin{eqnarray}
{\vec{\mathcal{K}}}_{(int)} &=&-\sum_{i=1}^{N}\,{\hat{\vec{\eta}}}_{i}(\tau
)\,\Big[\sqrt{m_{i}^{2}\,c^{2}+{\hat{\vec{\kappa}}}_{i}^{2}}+  \nonumber \\
&+&{\frac{{{\hat{\vec{\kappa}}}_{i}}}{{2\,c\,\sqrt{m_{i}^{2}\,c^{2}+{\hat{%
\vec{\kappa}}}_{i}^{2}}}}}\,\cdot \sum_{j\neq i}^{1..N}\,Q_{i}\,Q_{j}\,\Big({%
\frac{1}{2}}\,{\frac{{\partial \,{\hat{\mathcal{K}}}_{ij}({\hat{\vec{\kappa}}%
}_{i},{\hat{\vec{\kappa}}}}_{j},{\hat{\vec{\eta}}_{i}-{\hat{\vec{\eta}}}_{j})%
}}{{\ \partial \,{\hat{\vec{\eta}}}_{i}}}}-2\,\vec{A}_{\perp Sj}({\hat{\vec{%
\kappa}}}_{j},{\hat{\vec{\eta}}}_{i}-{\hat{\vec{\eta}}}_{j})\Big)\Big]-
\nonumber \\
&-&\frac{1}{2}\,\sum_{i=1}^{N}\,\sum_{j\neq i}^{1..N}\,{\frac{{Q_{i}\,Q_{j}}%
}{c}}\,\sqrt{m_{i}^{2}\,c^{2}+{\hat{\vec{\kappa}}}_{i}^{2}}\,{\frac{{%
\partial \,{\hat{\mathcal{K}}}_{ij}({\hat{\vec{\kappa}}}_{i},{\hat{\vec{%
\kappa}}}_{j},{\hat{\vec{\eta}}}_{i}-{\hat{\vec{\eta}}}_{j})}}{{\partial \,{%
\hat{\vec{\kappa}}}_{i}}}}+  \nonumber \\
&+&\sum_{i=1}^{N}\,\sum_{j\not=i}^{1..N}\,\frac{Q_{i}\,Q_{j}}{8\pi \,c}\,%
\frac{{\hat{\vec{\eta}}}_{i}-{\hat{\vec{\eta}}}_{j}}{|{\hat{\vec{\eta}}}_{i}-%
{\hat{\vec{\eta}}}_{j}|}-\sum_{i=1}^{N}\,\sum_{j\not=i}^{1..N}\,\frac{%
Q_{i}\,Q_{j}}{4\pi \,c}\,\int d^{3}\sigma \,\frac{{\hat{\vec{\pi}}}_{\perp
Sj}(\vec{\sigma}-{\hat{\vec{\eta}}}_{j},{\hat{\vec{\kappa}}}_{j})}{|\vec{%
\sigma}-{\hat{\vec{\eta}}}_{i}|}-  \nonumber \\
&-&{\frac{1}{2c}}\,\sum_{i=1}^{N}\,\sum_{j\neq i}^{1..N}\,Q_{i}\,Q_{j}\,\int
d^{3}\sigma \,\vec{\sigma}\,\,\Big[{\hat{\vec{\pi}}}_{\perp Si}(\vec{\sigma}-%
{\hat{\vec{\eta}}}_{i},{\hat{\vec{\kappa}}}_{i})\cdot {\hat{\vec{\pi}}}%
_{\perp Sj}(\vec{\sigma}-{\hat{\vec{\eta}}}_{j},{\hat{\vec{\kappa}}}_{j})+
\nonumber \\
&+&{\hat{\vec{B}}}_{Si}(\vec{\sigma}-{\hat{\vec{\eta}}}_{i},{\hat{\vec{\kappa%
}}}_{i})\cdot {\hat{\vec{B}}}_{Sj}(\vec{\sigma}-{\hat{\vec{\eta}}}_{j},{\hat{%
\vec{\kappa}}}_{j})\Big]-{\frac{1}{2c}}\int d^{3}\sigma \,\vec{\sigma}\,\,%
\Big({\ \vec{\pi}}_{\perp rad}^{2}+{\vec{B}}_{rad}^{2}\Big)(\tau ,\vec{\sigma%
})=  \nonumber \\
&&{}  \nonumber \\
&=&{\vec{\mathcal{K}}}_{matter}+{\vec{\mathcal{K}}}_{rad}=c\,{\vec{K}}%
_{Galilei}+O({\frac{1}{c}})=  \nonumber \\
&&{}  \nonumber \\
&&{}  \nonumber \\
&{\mathring{=}}&-\frac{1}{c}\mathcal{E}_{(int)}\,\vec{R}_{+}=-{\frac{1}{c}}\,%
\Big(\mathcal{E}_{matter}+\mathcal{E}_{rad}\Big)\,{\vec{R}}_{+} =
 \nonumber \\
&{\buildrel {def}\over {=}}&-\frac{1}{c}\mathcal{E}_{(int)}\,
\vec{R}_{+} = - {\frac{1}{c}}\, \Big(\mathcal{E}_{matter} +
\mathcal{E}_{rad}\Big)\, {\vec R}_+ \approx 0.
 \label{4.6a}
\end{eqnarray}

\bigskip

By adding the second class constraints ${\vec A}_{\perp rad}(\tau
,\vec \sigma ) \approx 0$ and ${\vec \pi}_{\perp rad}(\tau ,\vec
\sigma ) \approx 0 $ (implying ${\hat T}_i(\tau ) \approx 0$) we
recover the internal boost of Eqs.(\ref{2.53}) [of Eq.(6.46) of
Ref.\cite{14b}].

\bigskip

Even if all the internal generators are the direct sum of the
generators of the two subsystems, the two subsystems are connected
by the rest-frame conditions and by the vanishing of the internal
boosts, i.e. by the necessity of eliminating the position and
momentum of the internal 3-center of mass.

\subsection{The New Hamilton Equations}

Let us consider the new Hamiltonian (\ref{4.4}) after the canonical
transformation

\begin{eqnarray}
\mathcal{E}_{(int)} &=&M\,c^{2}=  \nonumber \\
&=&c\,\sum_{i=1}^N\,\sqrt{m_{i}^{2}\,c^{2}+{\hat{\vec{\kappa}}}_{i}^{2}(\tau
)} +\sum_{i\not=j}^{1..N}\,{\frac{{Q_{i}\,Q_{j}}}{{4\pi
\,|{\hat{\vec{\eta}}}_{i}(\tau
)-{\hat{\vec{\eta}}}_{j}(\tau )|}}}+V_{DARWIN}+  \nonumber \\
&+&\sum_{i=1}^N\,Q_{i}\,\Big(\int d^{3}\sigma \,\Big[{\vec{\pi}}
_{\perp rad}\cdot \Big({\hat{\vec{\pi}}}_{\perp Si}+({\frac{
{{\hat{\vec{\kappa}}} _{i}}}{\sqrt{m_{i}^{2}\, c^{2} +
{\hat{\vec{\kappa}}}_{i}^{2} }}\cdot }\,{ \frac{{\partial
}}{{\partial \,{\hat{\vec{\eta}}}_{i}}}})\,{\hat{ \vec{A}}}
_{\perp Si}\Big)-  \nonumber \\
&-&{\vec{A}}_{\perp rad}\cdot
\Big({\vec{\partial}}^{2}\,{\hat{\vec{A}}} _{\perp Si}+({\frac{
{{\hat{\vec{\kappa}}}_{i}}}{\sqrt{m_{i}^{2}\,c^{2}+{ \hat{
\vec{\kappa}}}_{i}^{2}}}\cdot }\,\ \ {\frac{{\partial }}{{\partial
\,{ \hat{ \vec{\eta}}}_{i}}}})\, {\hat{\vec{\pi}}}_{\perp
Si}\Big)\Big](\tau ,
\vec{\sigma})-  \nonumber \\
&-&{\frac{{{\hat{\vec{\kappa}}}_{i}\cdot {\vec{A}}_{\perp rad}(\tau
,{\hat{ \vec{\eta}}}_{i}(\tau
))}}{\sqrt{m_{i}^{2}\,c^{2}+{\hat{\vec{\kappa}}}
_{i}^{2}}}}\Big)+{\frac{1}{2}}\,\int d^{3}\sigma \,\Big({\vec{
\pi}}_{\perp
rad}^{2}+{\vec{B}}_{rad}^{2}\Big)(\tau ,\vec{\sigma})=  \nonumber \\
&&  \label{4.7}
\end{eqnarray}

\noindent with the Darwin potential given in Eq.(\ref{4.5}).

\bigskip

The first half of Hamilton equations is (see Eqs.(\ref{3.1}) for the
Poisson brackets)

\begin{eqnarray}
{\frac{{\partial\, A^r_{\perp rad}(\tau ,\vec \sigma )}}{{\partial\,
\tau }}} &{\buildrel \circ \over {=}}& \{ A^r_{\perp rad}(\tau ,\vec
\sigma ),
\mathcal{E}_{int}\} = - \pi^r_{\perp rad}(\tau ,\vec \sigma ) -  \nonumber \\
&-& \sum_{i=1}^N\, Q_i\, \Big[{\hat{\vec{\pi}}}_{\perp Si}(\tau
,\vec \sigma ) + \Big({\frac{
{{\hat{\vec{\kappa}}}_{i}}}{\sqrt{m_{i}^{2}\,c^{2} + {\hat{\vec{
\kappa}}}_{i}^{2}}} \cdot }\, {\frac{{\partial }}{{\partial
\,{\hat{\vec{\eta }}}_{i}}}}\Big)\, {\hat{ \vec{A}}}_{\perp Si}(\tau
,\vec \sigma ) \Big],
 \label{4.9}
\end{eqnarray}

\medskip

\begin{eqnarray}
{\frac{{d\, {\hat \eta}^r_i(\tau )}}{{d\, \tau}}} &{\buildrel \circ
\over {=
}}& \{ {\hat \eta}^r_i(\tau ), \mathcal{E}_{(int)} \} =  \nonumber \\
&=& {\frac{{\ {\hat \kappa}_i^r(\tau )}}{\sqrt{m^2_i\, c^2 + {\hat
{\vec \kappa}}_i^2(\tau )}}} + {\frac{1}{c}}\,{\frac{{\partial\,
V_{DARWIN}}}{{
\partial\, {\hat \kappa}_{ir}}}} +  \nonumber \\
&+& {\frac{{Q_i}}{c}}\, \int d^3\sigma\, \Big({\vec \pi}_{\perp
rad}(\tau ,\vec \sigma )\, \cdot {\frac{{\partial}}{{\partial\,
{\hat \kappa}_{ir}}}} \, \Big[{\hat {\vec \pi}}_{\perp Si}(\tau
,\vec \sigma ) + \Big({\frac{{{
\hat{\vec{\kappa}}}_{i}}}{\sqrt{m_{i}^{2}\,c^{2} +
{\hat{\vec{\kappa}}} _{i}^{2}}} \cdot }\, {\frac{{\partial
}}{{\partial \,{\hat{\vec{\eta}}}_{i}}}
}\Big)\, {\hat {\vec A}}_{\perp Si}(\tau ,\vec \sigma )\Big] -  \nonumber \\
&-& {\vec A}_{\perp rad}(\tau ,\vec \sigma )\, \cdot
{\frac{{\partial}}{{
\partial\, {\hat \kappa}_{ir}}}}\, \Big[{\vec \partial}^2\,\, {\hat {\vec A}}
_{\perp Si}(\tau ,\vec \sigma ) + \Big({\frac{
{{\hat{\vec{\kappa}}}_{i}}}{ \sqrt{m_{i}^{2}\,c^{2} +
{\hat{\vec{\kappa}}}_{i}^{2}}} \cdot }\, {\frac{{
\partial }}{{\partial \,{\hat{\vec{\eta}}}_{i}}}}\Big)\, {\hat {\vec \pi}}
_{\perp Si}(\tau ,\vec \sigma )\Big] \Big) -  \nonumber \\
&-& {\frac{{Q_i}}{c}}\, {\vec A}_{\perp rad}(\tau ,{\hat {\vec
\eta}}_i(\tau ))\, {\frac{{\partial}}{{\partial\, {\hat
\kappa}_{ir}}}}\, {\frac{{{\hat { \vec \kappa}}_i(\tau )}
}{\sqrt{m^2_i\, c^2 + {\hat {\vec \kappa}}_i^2(\tau ) }}}.
  \label{4.10}
\end{eqnarray}

\bigskip

Therefore, like in Eq.(6.2) of Ref.\cite{14b}, we get

\begin{eqnarray}
Q_i\, {\frac{{d\, {\hat \eta}^r_i(\tau )}}{{d\, \tau}}} &{\buildrel
\circ \over {=}}& Q_i\, {\frac{{\ {\hat \kappa}_i^r(\tau
)}}{\sqrt{m^2_i\, c^2 + {
\hat {\vec \kappa}}_i^2(\tau )}}},  \nonumber \\
&&{}  \nonumber \\
&&\Downarrow  \nonumber \\
&&{}  \nonumber \\
Q_{i}\, {\hat{\vec{\pi}}}_{\perp Si} &{\buildrel \circ \over {=}}& -
Q_{i}\, \Big({\frac{
{{\hat{\vec{\kappa}}}_{i}}}{\sqrt{m_{i}^{2}\,c^{2} + {\hat{\vec{
\kappa}}}_{i}^{2}}}\cdot }\,{\frac{{\partial }}{{\partial
\,{\hat{\vec{\eta}} }_{i}}}}\Big)\, {\hat{\vec{A}}}_{\perp Si} = -
Q_{i}\, \frac{\partial {\hat{
\vec{A}}} _{\perp Si}}{\partial \tau},  \nonumber \\
&&{}  \nonumber \\
{\frac{{\partial\, {\vec A}_{\perp rad}(\tau ,\vec \sigma
)}}{{\partial\, \tau}}} &{\buildrel \circ \over {=}}& - {\vec
\pi}_{\perp rad}(\tau ,\vec \sigma ).
  \label{4.11}
\end{eqnarray}
\bigskip

Therefore the new Hamiltonian, i.e. the internal energy, takes the
following simplified form

\begin{eqnarray}
\mathcal{E}_{(int)}
&=&c\,\sum_{i=1}^N\,\sqrt{m_{i}^{2}\,c^{2}+{\hat{\vec{ \kappa
}}}_{i}^{2}(\tau )}+\sum_{i\not=j}^{1..N}\,
{\frac{{Q_{i}\,Q_{j}}}{{4\pi \,|{\ \hat{ \vec{\eta}}}_{i}(\tau
)-{\hat{\vec{\eta}}}_{j}(\tau )|}}}+V_{DARWIN}+\nonumber \\
&+&\sum_{i=1}^N\,{\frac{{Q_{i}}}{c}}\,\Big(\int d^{3}\sigma
\,\Big[-{\vec{A}} _{\perp rad}\cdot
\Big({\vec{\partial}}^{2}\,{\hat{\vec{A}}}_{\perp Si}+({ \frac{
{{\hat{\vec{ \kappa}}}_{i}}}{\sqrt{m_{i}^{2}\,c^{2} +
{\hat{\vec{\kappa} }}_{i}^{2}}}\cdot } \,\ \ {\frac{{\partial
}}{{\partial \,{\hat{\vec{\eta}}} _{i}}}})\,
{\hat{\vec{\pi}}}_{\perp Si}\Big)\Big](\tau ,\vec{\sigma})-
\nonumber \\
&-&{\frac{{{\hat{\vec{\kappa}}}_{i}\cdot {\vec{A}}_{\perp rad}(\tau
,{\hat{ \vec{\eta}}}_{i}(\tau
))}}{\sqrt{m_{i}^{2}\,c^{2}+{\hat{\vec{\kappa}}}
_{i}^{2}}}}\Big)+{\frac{1}{2}}\,\int d^{3}\sigma \,\Big({\vec{
\pi}}_{\perp rad}^{2}+{\vec{B}}_{rad}^{2}\Big)(\tau ,\vec{\sigma}).
  \label{4.12}
\end{eqnarray}

\bigskip

By using Eq.(\ref{b7}) and (\ref{4.11}) we have

\begin{eqnarray}
&&\sum_{i=1}^N\, {\frac{{Q_{i}}}{c}}\,\Big(\int d^{3}\sigma
\,\Big[-{\vec{A}} _{\perp rad}\cdot
\Big({\vec{\partial}}^{2}\,{\hat{\vec{A}}}_{\perp Si}+({ \frac{c\,
{{\hat{\vec{ \kappa}}}_{i}}}{\sqrt{m_{i}^{2}\,c^{2}+{\hat{\vec{
\kappa}}}_{i}^{2}}}\cdot } \,\ \ {\frac{{\partial }}{{\partial
\,{\hat{\vec{ \eta}}}_{i}}}})\, {\hat{\vec{\pi}}}_{\perp
Si}\Big)\Big](\tau ,\vec{\sigma})-
\nonumber \\
&-&{\frac{{{\hat{\vec{\kappa}}}_{i}\cdot {\vec{A}}_{\perp rad}(\tau
,{\hat{ \vec{\eta}}}_{i}(\tau
))}}{\sqrt{m_{i}^{2}\,c^{2}+{\hat{\vec{\kappa}}}
_{i}^{2}}}}\Big) =  \nonumber \\
&=& \sum_{i=1}^N\, {\frac{{Q_i}}{c}} \Big(\int d^3\sigma\, \Big[-
{\vec A}_{\perp rad}(\tau ,\vec \sigma )\, \Big(- \Box\, {\hat {\vec
A}}_{\perp Si}(\tau ,\vec \sigma )\Big)\Big] -
{\frac{{{\hat{\vec{\kappa}}}_{i}\cdot {\vec{A}} _{\perp rad}(\tau
,{\hat{ \vec{\eta}}}_{i}(\tau ))}}{\sqrt{m_{i}^{2}\,c^{2}+{
\hat{\vec{\kappa}}} _{i}^{2}}}}\Big) =  \nonumber \\
&=& \sum_{i=1}^N\, {\frac{{Q_i}}{c}}\, \Big( \int d^3\sigma\,
A^r_{\perp rad}(\tau ,\vec \sigma )\, P^{rs}_{\perp}(\vec \sigma )\,
{\frac{{c\, {\hat \kappa} ^s_i(\tau )}}{\sqrt{m^2_i\, c^2 + {\hat
{\vec \kappa}}_i^2(\tau )}}}\, \delta^3(\vec \sigma - {\vec
\eta}_i(\tau )) - {\frac{{{\hat{\vec{\kappa}}} _{i}\cdot
{\vec{A}}_{\perp rad}(\tau ,{\hat{ \vec{\eta}}}_{i}(\tau ))}}{
\sqrt{m_{i}^{2}\,c^{2}+{\hat{\vec{\kappa}}} _{i}^{2}}}}\Big) =  \nonumber \\
&=& 0.
  \label{4.13}
\end{eqnarray}

\medskip

\noindent so that the final form of the new Hamiltonian and of the
particle velocities are

\begin{eqnarray}
\mathcal{E}_{(int)}
&=&c\,\sum_{i=1}^N\,\sqrt{m_{i}^{2}\,c^{2}+{\hat{\vec{ \kappa
}}}_{i}^{2}(\tau )}+\sum_{i\not=j}^{1..N}\,
{\frac{{Q_{i}\,Q_{j}}}{{4\pi \,|{\ \hat{ \vec{\eta}}}_{i}(\tau
)-{\hat{\vec{\eta}}}_{j}(\tau )|}}}+V_{DARWIN}+
\nonumber \\
&+& {\frac{1}{2}}\,\int d^{3}\sigma \,\Big({\vec{ \pi}}_{\perp
rad}^{2}+{ \vec{B}}_{rad}^{2}\Big)(\tau ,\vec{\sigma}) =
\mathcal{P}_{matter}^{\tau }+
\mathcal{P}_{rad}^{\tau },  \nonumber \\
&&{}  \nonumber \\
{\frac{{d\, {\hat \eta}^r_i(\tau )}}{{d\, \tau}}} &{\buildrel \circ
\over {= }}& {\frac{{\ {\hat \kappa}_i^r(\tau )}}{\sqrt{m^2_i\, c^2
+ {\hat {\vec \kappa}}_i^2(\tau )}}} + {\frac{1}{c}}\,
{\frac{{\partial\, V_{DARWIN}}}{{
\partial\, {\hat \kappa}_{ir}}}}.
  \label{4.14}
\end{eqnarray}

As a consequence, the second half of Hamilton equation for the
radiation field, i.e. for ${\vec{\pi}}_{\perp rad}$, and for the
particle momenta are ( ${\vec B}^2 = \sum_{rs}\, [\partial^r\,
A^s_{\perp} \, \partial^r\, A^s_{\perp} - \partial^r\, A^s_{\perp}\,
\partial^s\, A^r_{\perp}]$)

\begin{eqnarray}
\frac{\partial {\vec{\pi}}_{\perp rad}(\tau ,\vec{\sigma})}{\partial
\tau} &{ \buildrel \circ \over {=}}&\{ {\vec{\pi}}_{\perp rad},
\mathcal{E}_{(int)} \} = - \vec{\partial}^{2}{{\vec{A}}_{\perp
rad}}(\tau ,\vec{\sigma}),\nonumber \\
&&{}  \nonumber \\
\Rightarrow && \Box\, {\vec A}_{\perp rad}(\tau ,\vec \sigma ) {
\buildrel \circ \over {=}} 0,  \nonumber \\
&&{}  \nonumber \\
{\frac{{d\, {\hat {\vec \kappa}}_i(\tau )}}{{d\, \tau}}} &{\buildrel
\circ \over {=}}& - {\frac{1}{c}}\, {\frac{{\partial}}{{\partial\,
{\hat {\vec \eta}}_i}}}\, \Big(
\sum_{i\not=j}\,{\frac{{Q_{i}\,Q_{j}}}{{4\pi \,|{\ \hat{
\vec{\eta}}}_{i}(\tau )-{\hat{\vec{\eta}}}_{j}(\tau )|}}}+V_{DARWIN}
\Big).
 \label{4.15}
\end{eqnarray}

\bigskip

Therefore we get a decoupling of the radiation field from the the
particles, which are mutually interacting not with the Coulomb
potential but with the full Darwin potential. It is a kind of
decoupling like the one assumed to exist to define the
\textit{asymptotic IN} states after having given the asymptotic
conditions.

However the vanishing of the internal boosts {\it reintroduces a
coupling between the particles and the radiation field} at the level
of the reconstruction of the orbits,
\bigskip

Therefore the final form of the new equations of motion is given by
Eqs.(\ref{4.11}), (\ref{4.14}) and (\ref{4.15})

\begin{eqnarray}
{\frac{{d\, {\hat \eta}^r_i(\tau )}}{{d\, \tau}}} &{\buildrel \circ
\over {= }}& {\frac{{\ {\hat \kappa}_i^r(\tau )}}{\sqrt{m^2_i\, c^2
+ {\hat {\vec \kappa}}_i^2(\tau )}}} + {\frac{1}{c}}\,
{\frac{{\partial\, V_{DARWIN}}}{{
\partial\, {\hat \kappa}_{ir}}}},  \nonumber \\
{\frac{{d\, {\hat {\vec \kappa}}_i(\tau )}}{{d\, \tau}}} &{\buildrel
\circ \over {=}}& - {\frac{1}{c}}\, {\frac{{\partial}}{{\partial\,
{\hat {\vec \eta}}_i}}}\, \Big( \sum_{i\not=j}^{1..N}\,
{\frac{{Q_{i}\,Q_{j}}}{{4\pi \,|{\ \hat{ \vec{\eta}}}_{i}(\tau
)-{\hat{\vec{\eta}}}_{j}(\tau )|}}}+V_{DARWIN}\Big),\nonumber \\
&&{}  \nonumber \\
{\frac{{\partial\, {\vec A}_{\perp rad}(\tau ,\vec \sigma
)}}{{\partial\, \tau}}} &{\buildrel \circ \over {=}}& - {\vec
\pi}_{\perp rad}(\tau ,\vec \sigma ),  \nonumber \\
\frac{\partial {\vec{\pi}}_{\perp rad}(\tau ,\vec{\sigma})}{\partial
\tau} &{ \buildrel \circ \over {=}}&\{ {\vec{\pi}}_{\perp rad},
\mathcal{P}^{\tau} = {\frac{{\mathcal{E}_{(int)}}}{c}}\}= -
\vec{\partial}^{2}{{\vec{A}}_{\perp rad}}(\tau ,\vec{\sigma}),  \nonumber \\
&&{}  \nonumber \\
\Rightarrow && \Box\, {\vec A}_{\perp rad}(\tau ,\vec \sigma ) {
\buildrel \circ \over {=}} 0.
  \label{4.16}
\end{eqnarray}

\bigskip

\vfill\eject

\section{Conclusions}

In this paper we have given a complete updated review of the
rest-frame instant form of dynamics for isolated systems in
Minkowski space-time starting from their description in non-inertial
frames done by means of parametrized Minkowski theories. In them the
nature of the isolated system is hidden in their energy-momentum
tensor like in general relativity. We have completely clarified the
problem of the existence of the many extensions of the notion of the
Newtonian center of mass in special relativity. Only the
non-canonical Fokker-Pryce center of inertia $Y^{\mu}(\tau )$ is a
4-vector, so that its world-line describes an inertial observer. The
canonical center of mass ${\tilde x}^{\mu}(\tau )$ and the
non-canonical M$\o$ller center of energy $R^{\mu}(\tau )$ are not
4-vectors. There is a M$\o$ller world-tube of non-covariance,
centered on $Y^{\mu}(\tau )$ and with its M$\o$ller radius
determined by the Poincare' Casimirs, which contains all the
pseudo-world-lines connected with these two non-covariant
quantities. These three collective variables, all tending to the
Newton center of mass in the non-relativistic limit, are determined
by the Poincare' generators, so that they are global non-locally
determinable quantities. This complicated structure is due to the
Lorentz signature of the metric of Minkowski space-time and to the
structure of the Poincare' group, implying that in the instant form
of dynamics the interaction potentials are present also in the
Lorentz boosts and not only in the energy generator.\bigskip

In the rest-frame instant form the instantaneous 3-spaces are the
Wigner hyper-planes orthogonal to the 4-momentum of the isolated
system and the origin of the 3-coordinates is the Fokker-Pryce
center of inertia. The isolated system is described by the
non-covariant canonical center of mass (a decoupled particle
described by six Jacobi data) carrying a pole-dipole structure, i.e.
carrying the invariant mass $M$ (assumed non zero) and the rest spin
${\vec {\bar S}}$ (barycentric angular momentum) of the isolated
system. The external Poincare' generators are built in terms of the
six Jacobi data and of $M$ and ${\vec {\bar S}}$. For each such
system $M$ and ${\vec {\bar S}}$ are functions only of
Wigner-covariant relative canonical variables living in the
instantaneous Wigner 3-space, because the rest-frame conditions
${\vec {\cal P}}_{(int)} \approx 0$ and their gauge fixing ${\vec
{\cal K}}_{(int)} \approx 0$ eliminate the canonical variables of
the internal 3-center of mass. $M$ is the Hamiltonian determining
the evolution of the relative variables.\bigskip

We showed explicitly how to describe massive charged positive-energy
scalar particles, the transverse radiation field and an arbitrary
transverse electro-magnetic field in the rest-frame instant
form.\medskip

Before imposing the rest-frame conditions the basic canonical
variables for describing the particles in the instantaneous Wigner
3-spaces are the Wigner spin-1 3-vectors ${\vec \eta}_i(\tau )$ and
their conjugate 3-momenta ${\vec \kappa}_i(\tau )$. The particle
world-lines are derived covariant quantities, $x^{\mu}_i(\tau ) =
Y^{\mu}(\tau ) + \epsilon_r^{\mu}(\vec h)\, \eta^r_i(\tau )$, which
are non-canonical, $\{ x^{\mu}_i(\tau ), x^{\nu}_j(\tau ) \} \not=
0$. Therefore, they have to be identified with the covariant
non-canonical predictive particle positions of predictive
mechanics.\medskip

Let us remark that also massless positive-energy particles can be
described in this way. There will be the extra condition ${\dot
{\vec \eta}}^2_i(\tau ) = 1$ (so that ${\dot x}_i^2(\tau ) = 0$),
which will induce a constraint on the momenta ${\vec \kappa}_i(\tau
)$. This will be investigated elsewhere, being relevant for the
description of rays of light (see the pseudo-classical photon of
Ref.\cite{34b} and the problem of the spatial localization of
photons in Refs.\cite{bb}).\bigskip

Let us remark that it is possible to extend \cite{cc} the previous
description of isolated systems to non-inertial rest frames, where
the instantaneous 3-spaces are non-Euclidean but only asymptotically
Euclidean: at spatial infinity they are orthogonal to the total
4-momentum of the isolated system.\bigskip

In the second part of the paper we have considered the isolated
system of N positive energy charged scalar particles with mutual
Coulomb interaction plus the electromagnetic field in the radiation
gauge as the classical basis of a relativistic formulation of atomic
physics.\medskip

As shown in Ref.\cite{14b}, in the rest-frame instant form of
dynamics and by using Grassmann-valued electric charges to
regularize the self-energies, we are able to find the explicit
expression of the external and internal Poincare' generators
associated with this isolated system. In the limit $c\,
\rightarrow\, \infty$ we recover the Galilei generators of the
particles plus the $1/c$ corrections connected with the
electro-magnetic field. The same results could be obtained for
spinning particles (with the spin described by Grassmann variables)
by using the results of Ref.\cite{15b}.\bigskip

In Ref.\cite{14b}, relying on the Grassmann regularization, we
evaluated the effective particle-dependent electro-magnetic
potential and fields coming from the Lienard-Wiechert solution in
absence of homogeneous solutions corresponding to incoming radiation
fields. As a consequence, we decided to investigate whether it was
acceptable to put the transverse electro-magnetic potential and
fields equal to the sum of a transverse radiation term plus the
particle-dependent Lienard-Wiechert term.\bigskip

To our surprise, it turned out that at this classical level, with a
fixed number of particles, at every instant (and not only at
asymptotic times $\tau\, \rightarrow\, \pm \infty$) we can define a
canonical transformation sending the isolated system of a transverse
electro-magnetic field plus N charged particles interacting through
the Coulomb potential into a free transverse radiation field plus a
set of N Coulomb-dressed charged particles interacting through a
Coulomb + Darwin potential. Moreover, the internal Poincare'
generators (in particular the interaction-dependent internal mass
and internal Lorentz boosts) become the direct sum of the
corresponding generators of the two non-interacting subsystems.
These subsystems know each other only due to the conditions ${\vec
{\cal P}}_{(int)} \approx 0$, ${\vec {\cal K}}_{(int)} \approx 0$
defining the rest frame and eliminating the  internal 3-center of
mass.\medskip

This shows the limitation of the approximation, often made in atomic
physics, of replacing the electro-magnetic field coupled to the
atoms with a radiation field.
\bigskip

If this canonical transformation will turn out to be unitarily
implementable after a canonical quantization compatible with the
rest-frame instant form of dynamics, we will have an example of
avoidance of the Haag theorem. In QED this theorem says that there
is no unitary transformation sending the asymptotic IN and OUT
radiation field into an interpolating non-radiation field when
charged matter is present. We added some comments on why this
canonical transformation exists: the conditions ${\vec {\cal
P}}_{(int)} \approx 0$, ${\vec {\cal K}}_{(int)} \approx 0$, imply
that the radiation field knows the particles, so that there is some
analogy with the inner reasons allowing the Haag-Ruelle scattering
theory to avoid the Haag theorem.
\bigskip

In the second paper we will investigate how to solve the equations
${\vec {\cal P}}_{(int)} \approx 0$, ${\vec {\cal K}}_{(int)}
\approx 0$ for the elimination of the internal 3-center of mass for
the following isolated systems: a) charged particles with a Coulomb
plus Darwin mutual interaction; b) transverse radiation field; c)
charged particles with a mutual Coulomb interaction plus a
transverse electro-magnetic field. Then we will study the multipolar
expansion of the particle energy-momentum tensor and we will find
the relativistic generalization of the dipole approximation and of
the electric dipole representation used in atomic physics. Then in
the third paper we will define the canonical quantization of the
rest-frame instant form of isolated systems to get a formulation of
relativistic atomic physics, to see whether the previous canonical
transformation is unitarily implementable and to explore the
implications of special relativity for the theory of entanglement.

\vfill\eject

\appendix

\section{Wigner Tetrads for Null Poincar\'e Orbits and the
Transverse Polarization Vectors for the Radiation Field.}

\subsection{Helicity Tetrads}

For $k^2=0$, we have $k^{\mu} = (\omega (\vec k) = |\vec k| =
\sqrt{{\vec k} ^2}; \vec k)$. We have chosen positive energy $k^o >
0$; more in general one should put $|\vec k| \mapsto \eta_s | \vec
k|$ with $\eta_s = \pm 1 = sign\, k^o$. \medskip

By choosing a reference vector ${\buildrel \circ \over {k}}^{\mu} =
\omega_s\, (1;0,0,1)$, where $\omega_s$ is a dimensional parameter,
we can define the standard Wigner helicity boost
${}_HL^{\mu}{}_{\nu}(k,{ \buildrel \circ \over {k}})$ such that
$k^{\mu} = {}_HL^{\mu}{}_{\nu}(k,{ \buildrel \circ \over {k}} ) \,
{\buildrel \circ \over {k}}^{\nu} $. We have \cite{34b}

\begin{eqnarray}
{}_HL^{\mu}{}_{\nu}(k,{\buildrel \circ \over {k}})&=& \left(
\begin{array}{ccc}
{\frac{1}{2}}({\frac{{|\vec k|}}{{\omega_s}}} +
{\frac{{\omega_s}}{{|\vec k|} }}) & 0 & {\frac{1}{2}}({\frac{{|\vec
k|}}{{\omega_s}}} - {\frac{{\omega_s}}{
{|\vec k|}}}) \\
{\frac{1}{2}}({\frac{{|\vec k|}}{{\omega_s}}} -
{\frac{{\omega_s}}{{|\vec k|} }}) {\frac{{k^a}}{{|\vec k|}}} &
\delta^a_b + {\frac{{k^a\, k_b} }{{|\vec k|(|\vec k| + k^3)}}} &
{\frac{1}{2}}({\frac{{|\vec k|} }{{\omega_s}}} + {
\frac{{\omega_s}}{{|\vec k|}}}) {\frac{{k^a}}{{|\vec k|}}} \\
{\frac{1}{2}}({\frac{{|\vec k|}}{{\omega_s}}} -
{\frac{{\omega_s}}{{|\vec k|} }}) {\frac{{k^3}}{{|\vec k|}}} &
{\frac{{k_b}}{{|\vec k|}} } & {\frac{1}{2}}( {\frac{{|\vec
k|}}{{\omega_s}}} + {\frac{{\omega_s}}{{| \vec k|}}}) {\frac{{
k^3}}{{|\vec k|}}}
\end{array}
\right) ,  \nonumber \\
&&{}  \nonumber \\
{}_H\epsilon^{\mu}_A(\vec k)&=&{}_HL^{\mu}{}_{\nu}(k, { \buildrel
\circ \over {k}})\, {\buildrel \circ \over {{\epsilon}}}^{\nu}_A =
{}_HL^{\mu}{}_A(k, {\buildrel \circ \over {k}}),\quad\quad A = (\tau
, r),\nonumber \\
&&{\buildrel \circ \over {{\epsilon}}}^{\mu}_A:\quad (1;0,0,0),\quad
(0;1,0,0),\quad (0;0,1,0),\quad (0;0,0,1),  \nonumber \\
\eta^{\mu\nu}&=&{}_H\epsilon^{\mu}_A(\vec k)\, \eta^{AB}\,
{}_H\epsilon^{\nu}_B(\vec k).
  \label{a1}
\end{eqnarray}

In this way we get a helicity tetrad ${}_H\epsilon^{\mu}_A(\vec k)$
and a null basis

\begin{eqnarray*}
k^{\mu}&=&(|\vec k|; \vec k) = \omega_s\, \Big[ {}_H\epsilon^{\mu}_{
\tau}(\vec k) + {}_H\epsilon^{\mu}_3(\vec k)\Big] ,\quad\quad
when\quad k^2 = 0,  \nonumber \\
{\tilde k}^{\mu}(\vec k)&=&{\frac{1}{{2|\vec k|^2}}}(|\vec k|; -
\vec k) = { \frac{1}{{2\omega_s}}}\, \Big[ {}_H\epsilon^{\mu}_{
\tau} (\vec k) -
{}_H\epsilon^{\mu}_3(\vec k)\Big],  \nonumber \\
{}_H\epsilon^{\mu}_{\lambda}(\vec k)&=& (0; {}_H{\vec \epsilon}%
_{\lambda}(\vec k)) = (0; \delta^a_{\lambda} + {\frac{{k^a\,
k_{\lambda}}}{{ |\vec k|(|\vec k | + k^3)}}},
{\frac{{k_{\lambda}}}{{|\vec
k|}}}),\quad\quad a, \lambda = 1, 2,  \nonumber \\
&&k^2 = {\tilde k}^2(\vec k) = 0,\quad k \cdot {\tilde k}(\vec k) =
1,\nonumber \\
&&k \cdot {}_H\epsilon_{\lambda}(\vec k) = {\tilde k}(\vec k) \cdot
{}_H\epsilon_{\lambda}(\vec k) = 0,  \nonumber \\
&&{}_H\epsilon_{\lambda} (\vec k) \cdot {}_H\epsilon_{{\ \lambda}
^{^{\prime}}}(\vec k) = - {}_H{\vec \epsilon}_{\lambda} ( \vec k)
\cdot {}_H{ \vec \epsilon}_{{\lambda}^{^{\prime}}}(\vec k) = -
\delta _{\lambda {\lambda} ^{^{\prime}}},
 \end{eqnarray*}

\begin{eqnarray*}
 \eta^{\mu\nu}&=&k^{\mu}\, {\tilde k}^{\nu}(\vec k) + k^{\nu}\, {\tilde k}
^{\mu}(\vec k) - \sum_{\lambda = 1}^2\, {}_H\epsilon^{\mu}_{\lambda}
(\vec k)\,\, {}_H\epsilon^{\nu}_{\lambda}(\vec k),  \nonumber \\
&&{}_H\epsilon^{\mu}_{\tau}(\vec k) =
{\frac{{k^{\mu}}}{{2\omega_s}}} +
\omega_s\, {\tilde k}^{\mu}(\vec k),  \nonumber \\
&&{}_H\epsilon^{\mu}_3(\vec k) = {\frac{{k^{\mu}}}{{2\omega_s}}} -
\omega_s\, {\tilde k}_s^{\mu}(\vec k),  \nonumber \\
&&{}  \nonumber \\
&&{}_H\epsilon^a_{\lambda}(- \vec k) = {}_H\epsilon^a_{\lambda}(\vec
k),\qquad {}_H\epsilon^3_{\lambda}(- \vec k) = -
{}_H\epsilon^3_{\lambda}(\vec k),  \nonumber \\
&&\Rightarrow\,\, {}_H{\vec \epsilon}_{\lambda}(- \vec k) \cdot
{}_H{\vec \epsilon}_{\lambda^{^{\prime}}} =
\delta_{\lambda\lambda^{^{\prime}}} - 2\, { \frac{{k_{\lambda}\,
k_{\lambda^{^{\prime}}}}}{{{\vec k}^2}}},
 \end{eqnarray*}

\bea
 &&\Rightarrow\,\, {}_H{\vec \epsilon}_{\lambda}(- \vec k) = \Big(
\delta_{\lambda\lambda^{^{\prime}}} - 2\, {\frac{{k_{\lambda}\,
k_{\lambda^{^{\prime}}}}}{{{\vec k}^2}}}\Big)\, {}_H{\vec \epsilon}
_{\lambda^{^{\prime}}}(\vec k).
  \label{a2}
\end{eqnarray}
\medskip

The quantities ${}_H\epsilon^{\mu}_{\lambda}(\vec k)$ are a possible
set of transverse polarization vectors to be used in Eqs.
(\ref{2.32}) and (\ref{2.33}) for the radiation field in the
radiation gauge.

\subsection{Transformation Properties of Polarization Vectors under Lorentz
Transformations}

We give the transformation properties of the polarization vectors
defined in Eqs.(\ref{a1}) and (\ref{a2}). To this end one needs to
find the Wigner matrix belonging to the little group $E_2$ of
${\buildrel \circ \over {k}} ^{\mu} =\omega_s (1;0,0,1)$. From
Ref.\cite{35b} and from Appendix A of Ref. \cite{34b} we have

\begin{eqnarray}
{}_H\epsilon^{\mu}_A({\vec {(\Lambda\, k)}})&=&\Lambda^{\mu}{}_{
\nu}\, {}_H\epsilon^{\nu}_B(\vec k)\, \mathcal{R}(\vec k, \Lambda
)^B{}_A,\nonumber \\
&&{}  \nonumber \\
\mathcal{R}(\vec k, \Lambda )&=&[{}_HL^{-1}(k, {\buildrel \circ
\over {k}} )]\, \Lambda^{-1}\, [{}_HL({\Lambda\, k}, {\buildrel
\circ \over {k}})]\, \in\, E_2 \subset O(3,1),
  \label{a3}
\end{eqnarray}

\noindent where $\Lambda$ is a Lorentz transformation in O(3,1). If
$\Lambda$ is obtained from the SL(2,C) matrix $\left(
\begin{array}{cc}
\alpha & \beta \\
\gamma & \delta%
\end{array}
\right)$ with $\alpha \delta - \beta \gamma =1$, one obtains the
following parametrization of the Wigner matrix

\begin{eqnarray}
\mathcal{R}(\vec k, \Lambda )&=&  \nonumber \\
&=&\left(
\begin{array}{cccc}
1+{\frac{1}{2}}\mathbf{u}^2 & u^1 & u^2 & -{\frac{1}{2}}\mathbf{u}^2 \\
u^1 cos\, 2\theta +u^2 sin\, 2\theta & cos\, 2\theta & sin\, 2\theta
& -u^1
cos\, 2\theta -u^2 sin\, 2\theta \\
-u^1 sin\, 2\theta +u^2 cos\, 2\theta & -sin\, 2\theta & cos\,
2\theta & u^1
sin\, 2\theta -u^2 cos\, 2\theta \\
{\frac{1}{2}} \mathbf{u}^2 & u^1 & u^2 & 1-{\frac{1}{2}}
\mathbf{u}^2
\end{array}
\right) ,  \nonumber \\
&&\mathbf{u}^2=(u^1)^2+(u^2)^2,  \nonumber \\
&&e^{i\theta}={\frac{{d^{*}}}{{|d|}}},  \nonumber \\
&&u^1 cos\, 2\theta +u^2 sin\, 2\theta +i\Big( u^1 sin\, 2\theta
-u^2 cos\, 2\theta \Big) ={\frac{{\omega_s}}{{|{\vec p}_s|}}}
{\frac{{ac^{*}+bd^{*}}}{{\ |c|^2+|d|^2}}},  \nonumber \\
&&{}  \nonumber \\
&&a = \delta (|\vec k| + k^3) - \gamma (k^1 - i\, k^2),  \nonumber \\
&&b = - \beta (|\vec k| + k^3) + \alpha (k^1 - i\, k^2),  \nonumber \\
&&c = - \gamma (|\vec k| + k^3) - \delta (k^1 + i\, k^2),  \nonumber \\
&&d = \alpha (|\vec k| + k^3) + \beta (k^1 + i\, k^2).
  \label{a4}
\end{eqnarray}

Therefore, we get

\begin{eqnarray}
({\Lambda\, k})^{\mu}&=& \Lambda^{\mu}{}_{\nu} k^{\nu},  \nonumber \\
{}_H\epsilon^{\mu}_{\lambda = 1}({\vec {\Lambda\, k}})
&=&\Lambda^{\mu}{}_{\nu}\, \Big[ cos\, 2\theta \,
{}_H\epsilon^{\nu}_{\lambda = 1}(\vec k) - sin\, 2\theta \,
{}_H\epsilon^{\nu}_{\lambda = 2}(\vec k) + u^1\, k^{\nu} \Big] ,
\nonumber \\
{}_H\epsilon^{\mu}_{\lambda = 2}({\vec {\Lambda\, k}})
&=&\Lambda^{\mu}{}_{\nu}\, \Big[ sin\, 2\theta \,
{}_H\epsilon^{\nu}_{\lambda = 1}(\vec k) + cos\, 2\theta \,
{}_H\epsilon^{\nu}_{\lambda = 2}(\vec k) + u^2\, k^{\nu} \Big] ,
\nonumber \\
{\tilde k}^{\mu}({\vec {\Lambda\, k}})&=& \Lambda^{\mu}{}_{\nu}\,
\Big[ { \tilde k}^{\nu}(\vec k) + {\frac{1}{2}}\, \mathbf{u}^2\,
k^{\nu} +\nonumber \\
&+& {\frac{{u^1\, cos\, 2\theta + u^2\, sin\,
2\theta}}{{\omega_s}}}\, {}_H\epsilon^{\nu}_{\lambda = 1}(\vec k) -
{\frac{{u^1\, sin\, 2\theta - u^2\, cos\, 2\theta}}{{\omega_s}}}\,
{}_H\epsilon^{\nu}_{\lambda = 2}(\vec
k) \Big] ,  \nonumber \\
&&{}  \nonumber \\
{}_H\epsilon^{\mu}_{\tau}({\vec {\Lambda\, k}})&=& {\frac{1}{{\
2\omega_s}}} \, (\Lambda\, k)^{\mu} + \omega_s\, {\tilde
k}^{\mu}({\vec {\Lambda\, k}} ) = \Lambda^{\mu}{} _{\nu}\, \Big[
{}_H\epsilon^{\nu}_{\tau}(\vec k ) + {\frac{
{\omega_s}}{2}}\, \mathbf{u}^2\, k^{\nu} +  \nonumber \\
&+&(u^1\, cos\, 2\theta + u^2\, sin\, 2\theta )\,
{}_H\epsilon^{\nu}_{\lambda = 1}(\vec k) - (u^1\, sin\, 2\theta -
u^2\, cos\, 2\theta )\, {}_H\epsilon ^{\nu}_{\lambda= 2}(\vec
k)\Big] ,  \nonumber \\
{}_H\epsilon^{\mu}_3({\vec {\Lambda\, k}})&=& {\frac{1}{{2\omega_s}}
}\, (\Lambda\, k)^{\mu} - \omega_s\, {\tilde k}^{\mu}({\vec
{\Lambda\, k}} ) = \Lambda^{\mu}{} _{\nu}\, \Big[
{}_H\epsilon^{\nu}_3(\vec k) - {\ \frac{{
\omega_s}}{2}}\, \mathbf{u}^2\, k^{\nu} -  \nonumber \\
&-&(u^1\, cos\, 2\theta + u^2\, sin\, 2\theta )\,
{}_H\epsilon^{\nu}_{\lambda = 1}(\vec k) + (u^1\, sin\, 2\theta -
u^2\, cos\, 2\theta )\, {}_H\epsilon ^{\nu}_{\lambda = 2}(\vec
k)\Big] .
 \label{a5}
\end{eqnarray}

For a circular basis we have

\begin{eqnarray}
{}_H\epsilon^{\mu}_{(\pm )}(\vec k)&=&{\frac{1}{\sqrt{2}}}\, \Big[
{}_H\epsilon^{\mu}_{\lambda = 1}(\vec k)\, \pm\, i\,
{}_H\epsilon^{\mu}_{\lambda = 2}(\vec k)\Big] ,  \nonumber \\
&&{}  \nonumber \\
{}_H\epsilon^{\mu}_{(\pm )}({\vec {\Lambda\, k}})&=&
\Lambda^{\mu}{}_{\nu}\, \Big( e^{\pm\, 2i\, \theta}\,
{}_H\epsilon^{\nu}_{(\pm )}(\vec k) + {\frac{{ u^1\, \pm\, i\,
u^2}}{\sqrt{2}}}\, k^{\nu} \Big).
  \label{a6}
\end{eqnarray}

\vfill\eject

\section{The Lienard-Wiechert Electromagnetic Potentials and Fields.}

\subsection{Properties of the Lienard-Wiechert Fields}

Let us study the properties of the Lienard-Wiechert fields of Eqs.
(\ref{2.50})-(\ref{2.52}) taken from Ref.\cite{14b}. For the vector
potential we have

\begin{eqnarray}
&&{\vec{A}}_{\perp S}(\tau ,\vec{\sigma})\,{\buildrel\circ \over
{{=}}} \,\sum_{i=1}^{N}Q_{i}{\vec{A}}_{\perp
Si}(\vec{\sigma}-{\vec{\eta}}_{i}(\tau
),{\vec{\kappa}}_{i}(\tau )),  \nonumber \\
&&{}  \nonumber \\
&&\vec{A}_{\perp Si}(\vec{\sigma} -
\vec{\eta}_{i},{\vec{\kappa}}_{i}) = { \frac{1 }{4\pi
|\vec{\sigma}-\vec{\eta}_{i}|}}{\frac{1}{{\sqrt{ m_{i}^{2}\,c^{2}+{\
\vec{\kappa}}_{i}^{2}}+\sqrt{m_{i}^{2}\,c^{2}+(\vec{
\kappa}_{i}\cdot {\frac{{\ \vec{\sigma} -
\vec{\eta}_{i}}}{{|\vec{\sigma} -
\vec{\eta}_{i}|}}})^{2}}}}} \times  \nonumber \\
&&\Big[{\vec{\kappa}}_{i}+{\frac{{[\vec{\kappa}_{i}\cdot
(\vec{\sigma} - \vec{ \eta}_{i})]\, (\vec{\sigma} -
\vec{\eta}_{i})}}{{|\vec{\sigma} - \vec{ \eta}
_{i}|^{2}}}}\,{\frac{\sqrt{m_{i}^{2}\,c^{2} +
{{\vec{\kappa}_{i}}^{2}}} }{\sqrt{ m_{i}^{2}\,c^{2} +
(\vec{\kappa}_{i}\cdot {\ \frac{{\vec{\sigma} - \vec{\eta}_{i}
}}{{|\vec{\sigma} - \vec{\eta}_{i}|}}})^{2}}}}\Big]=
\nonumber \\
&=&\Big({\frac{{\vec{\alpha}_{i1}}}{c}}+{\frac{{\vec{\alpha}_{i3}}}{{c^{3}}}}
+\sum_{k=2}^{\infty }\,{\frac{{\vec{\alpha}_{i\,
2k+1}}}{{c^{2k+1}}}}\Big) (\tau ,\vec \sigma ),
  \label{b1}
\end{eqnarray}

\noindent in which

\begin{eqnarray}
{\vec{\alpha}_{i1}}(\tau ,\vec \sigma ) &{=}&\frac{1}{8\pi\,
m_{i}|\vec{ \sigma}-\vec{\eta}_{i}|}({\
\vec{\kappa}}_{i}+{\frac{{[\vec{\kappa}_{i}\cdot
(\vec{\sigma}-\vec{\eta}
_{i})]\,(\vec{\sigma}-\vec{\eta}_{i})}}{{|\vec{
\sigma}-\vec{\eta}_{i}|^{2}}}) },  \nonumber \\
{\vec{\alpha}_{i3}}(\tau ,\vec \sigma ) &{=}&\frac{1}{8\pi\,
m_{i}^{3}|\vec{ \sigma}-\vec{\eta}_{i}|}
[-\frac{1}{4}({\vec{\kappa}}_{i}+{\frac{{[\vec{ \kappa}_{i}\cdot
(\vec{\sigma}-\vec{\eta}_{i})]\,(\vec{\sigma}-\vec{\eta}
_{i})}}{{|\vec{\sigma}-\vec{\eta}
_{i}|^{2}}})({\vec{\kappa}}_{i}^{2}+(\vec{ \kappa}_{i}\cdot
{\frac{{\vec{\sigma}-\vec{\eta}_{i}}}{{|\vec{\sigma}-\vec{
\eta}_{i}|}}})^{2})} +  \nonumber \\
&+&{\frac{{[\vec{\kappa}_{i}\cdot
(\vec{\sigma}-\vec{\eta}_{i})]\,(\vec{
\sigma}-\vec{\eta}_{i})}}{2{|\vec{\sigma}-\vec{\eta}_{i}|^{2}}}({\vec{\kappa}
}_{i}^{2}-(\vec{\kappa}_{i}\cdot
{\frac{{\vec{\sigma}-\vec{\eta}_{i}}}{{|
\vec{\sigma}-\vec{\eta}_{i}|}}})^{2}]}.
  \label{b2}
\end{eqnarray}
\bigskip

From Eqs.(6.2) of Ref.\cite{14b} [by using $Q_i\, {\dot {\vec
\eta}}_i(\tau ) = Q_i\, {\vec \kappa}_i(\tau )/\sqrt{m^2_i\, c^2 +
{\vec \kappa}_i^2(\tau )}$ (see Eqs.(4.5) and after Eq.(5.27) in
Ref.\cite{14b}) and $Q_i\, {\dot {\vec \kappa}}_i(\tau ) {\buildrel
\circ \over {=}} 0$ (see Eqs.(4.5) of Ref.\cite {14b})] it follows
that the electric field is minus the $\tau$-derivative of the vector
potential

\begin{eqnarray}
 \vec{E}_{\perp S}(\tau ,\vec{\sigma}) &=&{\vec{\pi}}_{\perp
S}(\tau ,\vec{ \sigma})= \sum_{i=1}^{N}\,Q_{i}\,{\vec{\pi}}_{\perp
Si}(\vec{
\sigma}-{\vec{\eta}} _{i}(\tau ),{\vec{\kappa}}_{i}(\tau ))=  \nonumber \\
&=&\sum_{i=1}^{N}\,Q_{i}\,{\frac{ {{\vec{\kappa}}_{i}(\tau )\cdot
{\vec{\partial}}_{\sigma }}}{\sqrt{m_{i}^{2}\, c^{2} +
{\vec{\kappa}}_{i}^{2}(\tau )}} }\,{\vec{A}}_{\perp
Si}(\vec{\sigma}-{\vec{\eta}}_{i}(\tau ),{\vec{\kappa}}
_{i}(\tau ))=  \nonumber \\
&=& - \sum_{i=1}^N\, Q_i\, {\frac{ {{\vec{\kappa}}_{i}(\tau )\cdot
{\vec{\partial} }_{{\vec \eta}_i }}}{\sqrt{m_{i}^{2}\,c^{2} +
{\vec{\kappa}}_{i}^{2}(\tau )}} } \,{\vec{A}}_{\perp
Si}(\vec{\sigma}-{\vec{\eta}}_{i}(\tau ),{\vec{\kappa}}
_{i}(\tau ))=  \nonumber \\
&=& - \sum_{i=1}^N\, Q_i\, {\frac{\partial \vec{A}_{\perp S}(\tau
,\vec{\sigma})}{\partial \tau }}{|}_{{\vec \kappa}_i} =
-{\frac{\partial \vec{A}_{\perp
S}(\tau ,\vec{\sigma})}{\partial \tau }}=  \nonumber \\
&=&-\sum_{i=1}^{N}\,Q_{i}\times  \nonumber \\
&&{\frac{1}{{4\pi |\vec{\sigma}-{\vec{\eta}}_{i}(\tau
)|^{2}}}}\,\Big[{\vec{ \kappa}}_{i}(\tau )\,[{\vec{\kappa}}_{i}(\tau
)\cdot {\frac{{\vec{\sigma}-{\ \vec{\eta}}_{i}(\tau
)}}{{|\vec{\sigma}-{\vec{\eta}}_{i}(\tau )|}}}]\,{\frac{
\sqrt{m_{i}^{2}\,c^{2}+{\vec{\kappa}}_{i}^{2}(\tau
)}}{[m_{i}^{2}\,c^{2}+({\ \vec{\kappa}}_{i}(\tau )\cdot
{\frac{{\vec{\sigma}-{\vec{\eta}}_{i}(\tau )}}{
{|\vec{\sigma}-{\vec{\eta}}_{i}(\tau )|}}})^{2}]^{3/2}}}+  \nonumber \\
&+&{\frac{{\vec{\sigma}-{\vec{\eta}}_{i}(\tau
)}}{{|\vec{\sigma}-{\vec{\eta}} _{i}(\tau )|}}}\,\Big({\frac{{\
{\vec{\kappa}}_{i}^{2}(\tau )+({\vec{\kappa}} _{i}(\tau )\cdot
{\frac{{\vec{\sigma}-{\vec{\eta}}_{i}(\tau )}}{{|\vec{ \sigma
}-{\vec{\eta}}_{i}(\tau )|}}})^{2}}}{{{\vec{\kappa}}_{i}^{2}(\tau
)-({ \vec{ \kappa}}_{i}(\tau )\cdot
{\frac{{\vec{\sigma}-{\vec{\eta}}_{i}(\tau )} }{{|
\vec{\sigma}-{\vec{\eta}}_{i}(\tau )|}}})^{2}}}}\,({\frac{\sqrt{
m_{i}^{2}\,c^{2}+{\vec{\kappa}}_{i}^{2}(\tau
)}}{\sqrt{m_{i}^{2}\,c^{2}+({\ \vec{\kappa}}_{i}(\tau )\cdot
{\frac{{\vec{\sigma}-{\vec{\eta}}_{i}(\tau )}}{
{|\vec{\sigma}-{\vec{\eta}}_{i}(\tau )|}}})^{2}}}}-1)+  \nonumber \\
&+&{\frac{{({\vec{\kappa}}_{i}(\tau )\cdot
{\frac{{\vec{\sigma}-{\vec{\eta}} _{i}(\tau
)}}{{|\vec{\sigma}-{\vec{\eta}}_{i}(\tau )|}}})^{2}\,\sqrt{
m_{i}^{2}\,c^{2}+{\vec{\kappa}}_{i}^{2}(\tau
)}}}{{[m_{i}^{2}\,c^{2}+({\vec{ \kappa}}_{i}(\tau )\cdot
{\frac{{\vec{\sigma}-{\vec{\eta}}_{i}(\tau )}}{{|
\vec{\sigma}-{\vec{\eta}}_{i}(\tau )|}}})^{2}\,]^{3/2}}}}\Big)\Big]=
\nonumber \\
&=&\Big({\frac{{\vec{\beta}_{i2}}}{{c^{2}}}}+\sum_{k=2}^{\infty
}\,{\frac{{ \vec{ \beta}_{i\, 2k}}}{{c^{2k}}}}\Big)(\tau ,\vec
\sigma ). \label{b3}
\end{eqnarray}

\noindent with

\begin{eqnarray}
{\vec{\beta}_{i2}}(\tau ,\vec \sigma ) &=& - \sum_{i=1}^N\, Q_i\,
\frac{1}{{ 4\pi m_{i}^{2}|\vec{ \sigma}-{\vec{\eta}}_{i}(\tau
)|^{2}}}\, \Big[\vec{ \kappa}_{i}(\tau )\, \Big({\vec{
\kappa}}_{i}(\tau ) \cdot \frac{{\vec{\sigma }-{\
\vec{\eta}}_{i}(\tau )}} {{| \vec{\sigma}-{\vec{\eta}}_i(\tau
)|}}\Big)
+  \nonumber \\
&+& \frac{{\vec{\sigma}-{\vec{\eta}}_{i}(\tau
)}}{{|\vec{\sigma}-{\vec{\eta}} _{i}(\tau )|}}\, \Big[ \frac{1}{2}\,
\Big(\frac{{\ {\vec{\kappa}} _{i}^{2}(\tau )+({\vec{
\kappa}}_{i}(\tau )\cdot {\frac{{\vec{\sigma}-{\vec{ \eta}}_{i}(\tau
)}}{{| \vec{\sigma}-{\vec{\eta}}_{i}(\tau )|}}})^{2}}}{{{
\vec{\kappa}}_{i}^{2}(\tau )-({\vec{\kappa}}_{i}(\tau )\cdot
\frac{{\vec{ \sigma}-{\vec{\eta}}_{i}(\tau
)}}{{|\vec{\sigma}-{\vec{\eta}}_{i}(\tau )|}} )^2}}\Big)\,
\Big(\vec{\kappa} _{i}^{2}(\tau )-({\vec{\kappa}}_{i}(\tau )\cdot
{\frac{{\vec{\sigma}-{\vec{ \eta}}_{i}(\tau
)}}{{|\vec{\sigma}-{\vec{
\eta}}_{i}(\tau )|}}})^{2}\Big) +  \nonumber \\
&+&\Big({\vec{\kappa}}_{i}(\tau )\cdot
\frac{{\vec{\sigma}-{\vec{\eta}} _{i}(\tau
)}}{{|\vec{\sigma}-{\vec{\eta}}_{i}(\tau )|}}\Big)^{2}\Big].
  \label{b4}
\end{eqnarray}

\bigskip

The magnetic field is

\begin{eqnarray}
\vec{B}_{S}(\tau ,\vec{\sigma})
&=&\sum_{i}^{1..N}\,Q_{i}\,{\vec{B}}_{Si}(
\vec{\sigma}-{\vec{\eta}}_{i}(\tau ),{\vec{\kappa}}_{i}(\tau ))=
\nonumber \\
&=&\sum_{i=1}^{N}\,Q_{i}\,{\frac{1}{{4\pi |\vec{\sigma}-{\vec{\eta}}
_{i}(\tau )|^{2}}}}\, {\frac{{m_{i}^{2}\,c^{2}\,
{\vec{\kappa}}_{i}(\tau )\times
{\frac{{\vec{\sigma}-{\vec{\eta}}_{i}(\tau )}}{{|\vec{\sigma}-{\vec{
\eta}}_{i}(\tau )|}}}}}{{[m_{i}^{2}\,c^{2}+({\vec{\kappa}}_{i}(\tau
)\cdot { \ \frac{{\vec{\sigma}-{\vec{\eta}}_{i}(\tau
)}}{{|\vec{\sigma}-{\vec{\eta}}
_{i}(\tau )|}}})^{2}\,]^{3/2}}}}=  \nonumber \\
&=&\Big({\frac{{\vec{\gamma}_{i1}}}{c}}+{\frac{{\vec{\gamma}_{i3}}}{{c^{3}}}}
+\sum_{k=2}^{\infty }\,{\frac{{\vec{\gamma}_{i\,
2k+1}}}{{c^{2k+1}}}}\Big) (\tau ,\vec \sigma ),
  \label{b5}
\end{eqnarray}

\noindent with

\begin{eqnarray}
{\vec{\gamma}_{i1}}(\tau ,\vec \sigma )
&=&\sum_{i=1}^{N}\,Q_{i}\,{\frac{{{ \vec{\kappa}} _{i}(\tau )\times
{\frac{{\vec{\sigma}-{\vec{\eta}}_{i}(\tau )} }{{|\vec{
\sigma}-{\vec{\eta}}_{i}(\tau )|}}}}}{{4\pi m}_{i}{|\vec{\sigma}-{
\vec{\eta}} _{i}(\tau )|^{2}}}},  \nonumber \\
{\vec{\gamma}_{i3}}(\tau ,\vec \sigma )
&{=-}&\sum_{i=1}^{N}\,Q_{i}\,{\frac{3 {{\vec{\kappa}} _{i}(\tau
)\times {\frac{{\vec{\sigma}-{\vec{\eta}}_{i}(\tau )}}{{|\vec{
\sigma}-{\vec{\eta}}_{i}(\tau )|}}}}}{{8\pi m}_{i}^{3}{|\vec{
\sigma}-{\vec{ \eta}}_{i}(\tau )|^{2}}}({\vec{\kappa}}_{i}(\tau
)\cdot { \frac{{\vec{\sigma}- {\vec{\eta}}_{i}(\tau
)}}{{|\vec{\sigma}-{\vec{\eta}} _{i}(\tau )|}}})^{2}\,}.
  \label{b6}
\end{eqnarray}

\bigskip

From Eqs.(4.17) of Ref.\cite{14b} we have (the source term has
${\dot \eta}^s_i/c$)

\begin{eqnarray}
\Box\, A^r_{\perp S}(\tau ,\vec \sigma ) &=& (\partial^2_{\tau} -
{\vec \partial}^2)\, A^r_{\perp S}(\tau ,\vec \sigma ) = -
\Big({\frac{{\partial\, \pi^r_{\perp S}}}{{\partial\, \tau}}} +
{\vec \partial}^2\,\, A^r_{\perp S}
\Big)(\tau ,\vec \sigma ) =  \nonumber \\
&=& \sum_{i=1}^N\, Q_i\, \Box\, A^r_{\perp Si}(\vec \sigma - {\vec
\eta}_i(\tau );{\vec \kappa}_i(\tau )) {\buildrel \circ \over {=}}  \nonumber \\
&{\buildrel \circ \over {=}}& \sum_{i=1}^N\, Q_i\,
P^{rs}_{\perp}(\vec \sigma )\, {\frac{{\kappa_i^s(\tau
)}}{\sqrt{m^2_i\, c^2 + {\vec \kappa}_i^2(\tau )}}} \, \delta^3(\vec
\sigma - {\vec \eta}_i(\tau ))\, {\buildrel {def}\over {=}}
\, {\vec j}_{\perp}(\tau ,\vec \sigma ).  \nonumber \\
&&{}  \label{b7}
\end{eqnarray}

\subsection{Fourier Transforms of the Lienard-Wiechert Transverse
Electromagnetic Potential, Electric Field and Magnetic Field}

In Appendix A of \cite{14b} we find the following series form for
the Lienard-Wiechert transverse electromagnetic potential

\begin{eqnarray}
\vec{A}_{\perp }(\tau ,\vec{\sigma})
&=&\sum_{i=1}^{N}{\frac{Q_{i}}{4\pi }} \sum_{m=0}^{\infty
}\Big[{\frac{1}{(2m)!}\frac{\vec{\kappa}_{i}}{\sqrt{ m_{i}^{2}\, c^2
+ {\vec{\kappa}_{i}}^{2}}}}({\frac{\vec{\kappa}_{i}}{\sqrt{
m_{i}^{2}\, c^2 + {\vec{\kappa}_{i}}^{2}}}}\cdot
{\vec{\partial}}_{\sigma
})^{2m})\, |\vec{\sigma} -\vec{\eta}_{i}|^{2m-1}-  \nonumber \\
&&-{\frac{1}{(2m+2)!}}{\vec{\partial}}_{\sigma
}({\frac{\vec{\kappa}_{i}}{ \sqrt{m_{i}^{2}\, c^2 +
{\vec{\kappa}_{i}}^{2}}}}\cdot {\vec{\partial}}
_{\sigma })^{2m+1}\,|\vec{\sigma}-\vec{\eta}_{i}|^{2m+1}\Big]:=  \nonumber \\
&:&=\vec{A}_{\perp 1}(\tau ,\vec{\sigma} )+\vec{A}_{\perp 2}(\tau
,\vec{ \sigma} ).
  \label{b8}
\end{eqnarray}

To obtain its Fourier transform we need

\begin{eqnarray}
I_{1} &=&\int d^{3}\sigma e^{i\vec{k}\cdot
\vec{\sigma}}({\frac{\vec{\kappa} _{i}}{\sqrt{m_{i}^{2}\, c^2 +
{\vec{\kappa}_{i}}^{2}}}}\cdot {\vec{\partial}}
_{\sigma })^{2m})\,|\vec{\sigma}-\vec{\eta}_{i}|^{2m-1},  \nonumber \\
I_{2} &=&\int d^{3}\sigma e^{i\vec{k}\cdot
\vec{\sigma}}{\vec{\partial}} _{\sigma
}({\frac{\vec{\kappa}_{i}}{\sqrt{m_{i}^{2}\, c^2 + {\vec{\kappa}_{i}
}^{2}}}} \cdot {\vec{\partial}}_{\sigma
})^{2m+1})\,|\vec{\sigma}-\vec{\eta} _{i}|^{2m+1}.
  \label{b9}
\end{eqnarray}

Clearly each integral converges. Thus

\begin{equation}
I_{i}={{lim}_{\varepsilon \rightarrow 0}}\int_{0}^{\infty }\sigma
^{2}d\sigma \int d\hat{\Omega}_{\sigma }(\ \ )e^{-\varepsilon
|\vec{\sigma}- \vec{\eta}_{i}|}:={lim}_{\varepsilon \rightarrow
0}\int d^{3}\sigma ()e^{-\varepsilon
|\vec{\sigma}-\vec{\eta}_{i}|}={{lim}_{\varepsilon \rightarrow
0}}I_{i}(\varepsilon ).
  \label{b10}
\end{equation}

Change to $\partial /\partial \eta $ from $\partial /\partial \sigma
$, so that we can bring out the derivatives , translate and we
obtain

\begin{eqnarray}
I_{1} &=&(-{\frac{\vec{\kappa}_{i}}{\sqrt{m_{i}^{2}\, c^2 +
{\vec{\kappa}_{i} }^{2}}}} \cdot {\vec{\partial}}_{\eta
_{i}})^{2m}\, e^{i\vec{k}\cdot \vec{ \eta}_{i}}\, \int d^{3}\sigma\,
e^{i\vec{k}\cdot \vec{\sigma}}\,\sigma
^{2m-1}e^{-\varepsilon \sigma } =  \nonumber \\
&=&({\frac{\vec{\kappa}_{i}}{\sqrt{m_{i}^{2}\, c^2 +
{\vec{\kappa}_{i}}^{2}}} }\cdot {\ \ \vec{\partial}}_{\eta
_{i}})^{2m}e^{i\vec{k}\cdot \vec{\eta}_{i}} \frac{4\pi }{k
}\int_{0}^{\infty }d\sigma \sigma ^{2m}(\frac{e^{-\sigma
(\varepsilon - ik)}}{ 2i}+c.c) =  \nonumber \\
&=&({\frac{\vec{\kappa}_{i}}{\sqrt{m_{i}^{2}\, c^2 +
{\vec{\kappa}_{i}}^{2}}} }\cdot {\ \ \vec{\partial}}_{\eta
_{i}})^{2m}e^{i\vec{k}\cdot \vec{\eta}_{i}} \frac{4\pi }{k
}\frac{d^{2m}}{d\varepsilon ^{2m}}\int_{0}^{\infty }d\sigma (
\frac{ e^{-\sigma (\varepsilon -ik)}}{2i}+c.c) =  \nonumber \\
&=&({\frac{\vec{\kappa}_{i}}{\sqrt{m_{i}^{2}\, c^2 +
{\vec{\kappa}_{i}}^{2}}} }\cdot {\ \ \vec{\partial}}_{\eta
_{i}})^{2m}e^{i\vec{k}\cdot \vec{\eta}_{i}} \frac{4\pi }{
k^{2m+2}}(-)^{m}(2m)! = (2m)!\, ({\frac{\vec{\kappa}_{i}}{
\sqrt{m_{i}^{2}\, c^2 + {\ \vec{ \kappa}_{i}}^{2}}}}\cdot
{\vec{k}})^{2m}
\frac{4\pi }{k^{2m+2}}.  \nonumber \\
&&{}  \label{b11}
\end{eqnarray}

\noindent in which we have set $\varepsilon =0$. This integral now
permits us to perform the summation explicitly. \ We find

\begin{eqnarray}
&&\sum_{m=0}^{\infty }\int d^{3}\sigma e^{i\vec{k}\cdot
\vec{\sigma}}{\frac{ 1
}{(2m)!}\frac{\vec{\kappa}_{i}}{\sqrt{m_{i}^{2}\, c^2 +
{\vec{\kappa}_{i}} ^{2}}}}({\ \
\frac{\vec{\kappa}_{i}}{\sqrt{m_{i}^{2}\, c^2 + {\vec{\kappa}
_{i}}^{2}}}}\cdot {\ \vec{ \partial}}_{\sigma
})^{2m})\,|\vec{\sigma}-\vec{
\eta}_{i}|^{2m-1} =  \nonumber \\
&=&\frac{\vec{\kappa}_{i}}{\sqrt{m_{i}^{2}\, c^2 +
{\vec{\kappa}_{i}}^{2}}} \, e^{i\vec{ k }\cdot
\vec{\eta}_{i}}\sum_{m=0}^{\infty }({\frac{\vec{\kappa} _{i}}{\sqrt{
m_{i}^{2}\, c^2 + {\vec{\kappa}_{i}}^{2}}}}\cdot {\vec{k}})^{2m}
\frac{4\pi}{k^{2m+2} } =  \nonumber \\
&=&\frac{4\pi \vec{\kappa}_{i}}{k^{2}\sqrt{m_{i}^{2}\, c^2 +
{\vec{\kappa} _{i}}^{2}}}\, e^{i\vec{k}\cdot \vec{\eta}_{i}}\,
\frac{1}{1 - ({\frac{\vec{ \kappa}_{i}\cdot
\hat{k}}{\sqrt{m_{i}^{2}\, c^2 + {\vec{\kappa}_{i}}^{2}}}} )^{2m}} =
\frac{4\pi \vec{ \kappa}_{i}}{k^{2}}\, e^{i\vec{k}\cdot \vec{\eta}
_{i}}\, \frac{\sqrt{m_{i}^{2}\, c^2 + {\
\vec{\kappa}_{i}}^{2}}}{m_{i}^{2}\, c^2 + {\vec{\kappa}_{i}}^{2} -
\left( \vec{\kappa} _{i}\cdot \hat{k}\right) ^{2}}.
  \label{b12}
\end{eqnarray}

In a similar way we find

\begin{eqnarray}
&&\sum_{m=0}^{\infty }\int d^{3}\sigma e^{i\vec{k}\cdot
\vec{\sigma}}{\frac{ 1 }{(2m+2)!}}{\vec{\partial}}_{\sigma
}({\frac{\vec{\kappa}_{i}}{\sqrt{ m_{i}^{2}\, c^2 +
{\vec{\kappa}_{i}}^{2}}}}\cdot {\vec{\partial}}_{\sigma
})^{2m+1}\,| \vec{\sigma }-\vec{\eta}_{i}|^{2m+1} =  \nonumber \\
&=&\frac{4\pi \vec{k}\vec{\kappa}_{i}\cdot \vec{k}}{k^{4}}\,
e^{i\vec{k} \cdot \vec{\eta}_{i}}\, \frac{\sqrt{m_{i}^{2}\, c^2 +
{\vec{\kappa}_{i}}^{2}} }{m_{i}^{2}\, c^2 + {\
\vec{\kappa}_{i}}^{2}-\left( \vec{\kappa}_{i}\cdot \hat{k}\right)
^{2}}.
  \label{b13}
\end{eqnarray}

Combining we obtain

\begin{eqnarray}
&&{\vec {\tilde A}}_{\perp S}(\tau , \vec k)\, {\buildrel {def}\over
{=}}\, \int d^3\sigma\, e^{- i\, \vec{k} \cdot \vec{\sigma}}\,
\vec{A}_{\perp S}(\tau ,\vec{ \sigma}) = \sum_{i=1}^{N}\,
\frac{Q_{i}\, e^{- i\, \vec{k} \cdot \vec{\eta}_i}}{k^4}\,
\frac{\vec{k}\times (\vec{\kappa}_{i}\times \vec{ k})\,
\sqrt{m_i^2\, c^2 + {\vec{ \kappa}_i}^2}}{m_i^2\, c^2 +
{\vec{\kappa}
_{i}}^{2} - \left( \vec{\kappa} _{i}\cdot \hat{k}\right) ^{2}}.  \nonumber \\
&&{}  \label{b14}
\end{eqnarray}

From this and Eqs.(\ref{2.51}) and (\ref{2.52}) we obtain

\begin{eqnarray}
{\vec {\tilde \pi}}_{\perp S}(\tau , \vec k) &{\buildrel {def}\over
{=}}& { \frac{1}{c}}\, \int d^{3}\sigma e^{- i\, \vec{k} \cdot
\vec{\sigma}}\vec{E}
_{\perp S}(\tau ,\vec{ \sigma}) =  \nonumber \\
&=&-\sum_{i=1}^{N}{\frac{\vec{\kappa}_{i}}{\sqrt{m_{i}^{2}\, c^2 +
{\vec{ \kappa}_{i}}^{2}}}}\cdot {\vec{\partial}}_{\eta _{i}}\int
d^{3}\sigma e^{- i\, \vec{ k}\cdot \vec{\sigma}}\vec{A}_{\perp
}(\tau ,\vec{\sigma}-\vec{\eta}
_{i},\vec{ \kappa}_{i}) =  \nonumber \\
&=&i\, \sum_{i=1}^{N}\frac{Q_{i}\, e^{- i\, \vec{k}\cdot
\vec{\eta}_{i}}}{ k^{4}}\, \frac{ \vec{\kappa}_{i}\cdot
\vec{k}}{\sqrt{m_{i}^{2}\, c^2 + {\vec{ \kappa}_{i}}^{2}}}\, \frac{
\vec{k}\times (\vec{\kappa}_{i}\times \vec{k}) \sqrt{m_{i}^{2}\, c^2
+ {\vec{\kappa} _{i}}^{2}}}{m_{i}^{2}\, c^2 + {\vec{
\kappa}_{i}}^{2} - \left(\vec{\kappa}_{i}\cdot \hat{k}\right) ^{2}},
\nonumber \\
&&{}  \nonumber \\
{\vec {\tilde B}}_S(\tau , \vec k) &{\buildrel {def}\over {=}}& \int
d^{3}\sigma\, e^{- i\, \vec{k}\cdot \vec{\sigma}}\vec{B}_S (\tau
,\vec{\sigma}) =  \nonumber \\
&=&i\, \sum_{i=1}^{N}\, \frac{Q_{i}\, e^{- i\, \vec{k}\cdot
\vec{\eta}_{i}}}{ k^{2}}\frac{ \vec{k}\times
\vec{\kappa}_{i}\sqrt{m_{i}^{2}\, c^2 + {\vec{ \kappa}_{i}}^{2}}}{
m_{i}^{2}\, c^2 + {\vec{\kappa}_{i}}^{2} - \left( \vec{
\kappa}_{i}\cdot \hat{k}\right) ^{2}}.
 \label{b15}
\end{eqnarray}

\subsection{The $1/c$ Expansion of $\mathcal{E}_{(int)}$ after the Canonical
Transformation (\protect\ref{3.6}).}

By using Eqs. (\ref{b1}) - (\ref{b4}) the non-relativistic limit of
the Darwin potential (\ref{4.5}) is

\begin{eqnarray}
V_{Darwin}({\hat {\vec \eta}}_i(\tau ) - {\hat {\vec \eta}}_j(\tau
); {\hat { \vec \kappa}}_i(\tau )) &\rightarrow_{c \rightarrow
\infty}& {\frac{1}{{c^2}} }\, \sum_{i\not= j}^{1..N}\, Q_i\, Q_j\,
\Big({\frac{{{\hat {\vec \kappa}}_i(\tau )} }{{m_i}}} \cdot {\vec
\alpha}_{j1}(\tau , {\hat {\vec \eta}}_i(\tau )) +
\nonumber \\
&+& {\frac{1}{2}}\, \int d^3\sigma\, \big[{\vec \gamma}_{i1} \cdot
{\vec \gamma}_{j1}\Big](\tau , \vec \sigma ) +  \nonumber \\
&+& {\frac{1}{{c^4}}}\, \sum_{i\not= j}^{1..N}\, Q_i\, Q_j\,
\Big({\frac{{{\hat { \vec \kappa}}_i(\tau )}}{{m_i}}} \cdot
\Big[{\vec \alpha}_{j3} - {\frac{{{\ \hat {\vec \kappa}}_i^2(\tau
)}}{{2 m_i^2}}}\, {\vec \alpha}_{j1}\Big](\tau
, {\hat {\vec \eta}}_i(\tau )) +  \nonumber \\
&+& \int d^3\sigma\, \Big[{\frac{1}{2}}\, \Big({\vec \beta}_{i2}
\cdot {\vec \beta}_{j2} + {\vec \gamma}_{i1} \cdot {\vec
\gamma}_{j3} + {\vec \gamma}
_{i3} \cdot {\vec \gamma}_{j1}\Big) +  \nonumber \\
&+& \Big({\frac{{{\hat {\vec \kappa}}_i(\tau )}}{{m_i}}}\,
{\frac{{\partial} }{{\partial\, {\hat {\vec \eta}}_i}}}\Big)\,
\Big({\vec \alpha}_{i1} \cdot { \ \vec \beta}_{j2} - {\vec
\beta}_{i2} \cdot {\vec \alpha}_{j1}\Big) \Big] (\tau ,\vec \sigma )
\Big).
  \label{b18}
\end{eqnarray}
\bigskip

Therefore the non-relativistic limit of $\mathcal{E}_{(int)}$, given
by Eq.( \ref{4.4}), is

\bea
 \mathcal{E}_{(int)}\rightarrow _{c\rightarrow \infty }
&&(\sum_{i=1}^N\, m_{i})\, c^{2} + \sum_{i=1}^N\,
{\frac{{{\hat{\vec{\kappa}}} _{i}^{2}(\tau )}}{{2m_{i}}}} +
\sum_{i\not=j}^{1..N}\, {\frac{{Q_{i}\,Q_{j}}}{{4\pi \,| {\
\hat{\vec{\eta}}}_{i}(\tau ) -
{\hat{\vec{\eta}}}_{j}(\tau )|}}} +\nonumber \\
 &+& {\frac{1}{2\,}}\, \int d^{3}\sigma \,\Big({\vec{\pi}}_{\perp
rad}^{2}+{\vec{B}}_{rad}^{2} \Big)(\tau ,\vec{\sigma})+  \nonumber \\
&+&{\frac{1}{{c^{2}}}}\,\Big(-\sum_{i}\,{\frac{{{\hat{\vec{\kappa}}}
_{i}^{4}(\tau )}}{{8m_{i}^{3}}}} + \sum_{i\not=j}^{1..N}\,\,
\Big[Q_{i}\,Q_{j}{\frac{{{ \ \hat{\vec{\kappa}}}_{i}(\tau
)}}{{m_{i}}}}\cdot {\vec{\alpha}}_{j1}(\tau ,{
\hat{\vec{\eta}}}_{i}(\tau )) + \nonumber \\
 &+& {\frac{1}{2}}\,\int
d^{3}\sigma \,\Big({\vec{ \gamma}}_{i1}\cdot
{\vec{\gamma}}_{j1}\Big)(\tau ,\vec{\sigma})\Big]\Big)+
\nonumber \\
&+&O(c^{-4}).
  \label{b19}
\end{eqnarray}

\vfill\eject

\section{The Internal Poincare' Generators after the Canonical Transformation}

\subsection{The New Internal Momentum and Angular Momentum.}

To obtain the instant form of the 3-momentum (\ref{4.1}) we used the
following results [remember from Eqs.(\ref{2.50}) and (\ref{2.51})
that in the new canonical basis the fields of the Lienard-Wiechert
solution depend upon $\vec \sigma - {\hat {\vec \eta}}_i(\tau )$ and
${\hat {\vec \kappa}}_i(\tau )$]

\begin{eqnarray}
&&\sum_{i\not=j}^{1..N}\,Q_{i}\,Q_{j}\,\int d^{3}\sigma
\,\Big({\hat{\vec{\pi}}} _{\perp Si}\times
{\hat{\vec{B}}}_{Sj}\Big)^{r}(\tau ,\vec{\sigma})=
\nonumber \\
&=&\sum_{i\not=j}^{1..N}\,Q_{i}\,Q_{j}\,\int d^{3}\sigma
\Big({\hat{\pi}}_{\perp Si}^{s}\frac{{\partial \,}}{{\partial \sigma
}^{r}}{\hat{A}}_{\perp Sj}^{s} \Big)(\tau ,\vec{\sigma}),
  \label{c1}
\end{eqnarray}

\noindent where the last line was obtained by integration by parts
and using the transversality condition; then we get

\begin{eqnarray}
&&\sum_{i\not=j}^{1..N}\,Q_{i}\,Q_{j}\,\int d^{3}\sigma
\Big({\hat{\pi}}_{\perp Si}^{s}\frac{{\partial \,}}{{\partial \sigma
}^{r}}{\hat{A}}_{\perp Sj}^{s}
\Big)(\tau ,\vec{\sigma})=  \nonumber \\
&=&\frac{1}{2}\sum_{i\not=j}^{1..N}\,Q_{i}\,Q_{j}\,\int d^{3}\sigma
\Big(-\frac{{\ \partial \,}}{{\partial \hat{\eta}}_{i}^{r}}
({\hat{\pi}}_{\perp Sj}^{s}{\hat{ A}}_{\perp
Si}^{s})+\frac{{\partial \,}}{{\partial \hat{\eta}}_{i}^{r}}({\
\hat{A}}_{\perp Sj}^{s}{\hat{\pi}}_{\perp Si}^{s})\Big)(\tau
,\vec{\sigma})\nonumber \\
&=&\frac{1}{2}\,\sum_{i\not=j}^{1..N}\,Q_{i}\,Q_{j}\,\frac{{\partial
\,}}{{\ \partial \hat{\eta}}_{i}^{r}}\,
{\hat{\mathcal{K}}}_{ji}=-\frac{1}{2}
\,\sum_{i\not=j}^{1..N}\,Q_{i}\,Q_{j}\,\frac{{\partial
\,}}{{\partial \hat{\eta}} _{i}^{r}}\,{\hat{\mathcal{K}}}_{ij}.
  \label{c2}
\end{eqnarray}
\bigskip

Moreover we have

\begin{eqnarray}
&&\int d^{3}\sigma \,\Big({\vec{\pi}}_{\perp rad}\times
{\hat{\vec{B}}}_{Si}+ {\hat{\vec{\pi}}}_{\perp Si}\times
{\vec{B}}_{rad}\Big)^{r}(\tau ,\vec{\sigma })=  \nonumber \\
&=&\frac{{\partial \,}}{{\partial \hat{\eta}}_{i}^{r}}\int
d^{3}\sigma \, \Big(-\pi _{\perp rad}^{s}\hat{A}_{\perp
Si}^{s}+\hat{\pi}_{\perp Si}^{s}
\hat{A}_{\perp rad}^{s}\Big)^{r}(\tau ,\vec{\sigma})=  \nonumber \\
&=&\frac{{\partial \,}}{{\partial \hat{\eta}}_{i}^{r}}\int
d^{3}\sigma \, \Big(-\vec{\pi}_{\perp rad}\cdot
{\hat{\vec{A}}}_{\perp Si}+\vec{A}_{\perp rad}\cdot
{\hat{\vec{\pi}}}_{\perp Si}\Big)^{r}(\tau ,\vec{\sigma})=-{\frac{{
\partial \,{\hat{T}}_{i}(\tau )}}{{\partial \,{\hat{\vec{\eta}}}_{i}}}}.
  \label{c3}
\end{eqnarray}

\bigskip

The instant form of the internal angular momentum has been obtained
by using Eqs.(6.41)-(6.45) of Ref.\cite{14b}, based on the explicit
knowledge of the Lienard Wiechert fields (\ref{2.50}) and
(\ref{2.51}), since they imply

\begin{eqnarray}
&&\int d^{3}\sigma \Big[\vec{\sigma}\times \Big({\vec{\pi}}_{\perp
rad}\times {\hat{\vec{B}}}_{Si}+{\hat{\vec{\pi}}}_{\perp Si}\times
{\vec{B}}_{rad}\Big)(\tau ,\vec{\sigma})\Big]  \nonumber \\
&=&-({\hat{\vec{\eta}}}_{i}\times {\frac{{\partial }}{{\partial
\,{\hat{\vec{ \eta}}}_{i}}}}+{\hat{\vec{\kappa}}}_{i}\times
{\frac{{\partial }}{{\partial \,{\ \ \hat{\kappa}}_{i}}}})\,\int
d^{3}\sigma \,\Big({\hat{\vec{A}}}_{\perp Si}\cdot
{\vec{\pi}}_{\perp rad}-\vec{A}_{\perp rad}\cdot {\ \hat{\vec{\pi}}}
_{\perp Si}\Big)=  \nonumber \\
&=&-({\hat{\vec{\eta}}}_{i}\times {\frac{{\partial }}{{\partial
\,{\hat{\vec{ \eta}}}_{i}}}}+{\hat{\vec{\kappa}}}_{i}\times
{\frac{{\partial }}{{\ \partial \,{\hat{\kappa}}_{i}}}})\,
{\hat{T}}_{i}(\tau ),  \nonumber \\
&&{}  \nonumber \\
\sum_{i\neq j}^{1..N} &&Q_{i}\,Q_{j}\,\int d^{3}\sigma
\,\vec{\sigma}\times \Big({\ \hat{\vec{\pi}}}_{\perp Si}\times
{\hat{\vec{B}}}_{Sj}\Big)(\tau ,\vec{\sigma })=  \nonumber \\
&=&-\frac{1}{2}\,\sum_{i\neq j}^{1..N}\,Q_{i}\,Q_{j}\,
\Big({\hat{\vec{\eta}}} _{i}\times \frac{{\partial \,}}{{\partial
{\hat{\vec{\eta}}}_{i}}}+{{\hat{ \vec{\kappa}}}_{i}}\times
\frac{{\partial \,}}{{\partial \,{\hat{\vec{\kappa}
}}_{i}}}\Big)\,{\hat{\mathcal{K}}}_{ij}.
  \label{c4}
\end{eqnarray}

\subsection{The Internal Boosts.}

By using  Eqs.(\ref{3.4}) the internal boost in Eq.(\ref {4.6})
below can be shown to assume the following form

\begin{eqnarray*}
\mathcal{\vec{K}}_{(int)}
&=&-\sum_{i=1}^{N}\,{\hat{\vec{\eta}}}_{i}(\tau
)\, \Big[\sqrt{m_{i}^{2}\,c^{2}+{\hat{\vec{\kappa}}}_{i}^{2}}+ \\
&+&{\frac{{{\hat{\vec{\kappa}}}_{i}}}{{2\, c\,
\sqrt{m_{i}^{2}\,c^{2}+{\hat{ \vec{ \kappa}}}_{i}^{2}}}}}\,\cdot
\sum_{j\neq i}^{1..N}\,Q_{i}\,Q_{j}\,\Big({\frac{ 1}{2}}
\,{\frac{{\partial \,{\hat{\mathcal{K}}}_{ij}({\hat{\vec{\kappa}}}
_{i},{\hat{\vec{\kappa}}}}_{j},{\hat{\vec{\eta}}_{i}-{\hat{\vec{\eta}}}
_{j}) }}{{\ \partial \,{\hat{\vec{\eta}}}_{i}}}}- 2\,\vec{A}_{\perp
Sj}({\hat{\vec{\kappa}}}_{j},{\hat{\vec{\eta}}}_{i}-{\hat{\vec{\eta}}}_{j})
\Big)\Big]- \\
&-&\frac{1}{2}\,\sum_{i=1}^{N}\,\sum_{j\neq i}^{1..N}\,
{\frac{{Q_{i}\,Q_{j}}}{c}} \, \sqrt{
m_{i}^{2}\,c^{2}+{\hat{\vec{\kappa}}}_{i}^{2}}\,{\frac{{\partial \,
{\hat{\mathcal{K}}}_{ij}({\hat{\vec{\kappa}}}_{i},{\hat{\vec{\kappa}}}
_{j},{\ \hat{ \vec{\eta}}}_{i}-{\hat{\vec{\eta}}}_{j})}}{{\partial
\,{\hat{\vec{\kappa}}}_{i}}}}+ \\
&+&{\frac{1}{c}}\,\sum_{i=1}^{N}\,\sum_{j\not=i}^{1..N}\,\frac{Q_{i}\,Q_{j}}{8\pi}
\,\frac{{\hat{\vec{\eta}}}_{i}-{\hat{\vec{\eta}}}_{j}}{|{\hat{\vec{\eta}}}
_{i}-{\hat{\vec{\eta}}}_{j}|} -
\sum_{i=1}^{N}\,\sum_{j\not=i}^{1..N}\,\frac{ Q_{i}\,Q_{j}}{4\pi\, c
}\,\int d^{3}\sigma \,\frac{{\hat{\vec{\pi}}}_{\perp Sj}(
\vec{\sigma}-{\hat{\vec{\eta}}}_{j},{\hat{\vec{\kappa}}}_{j})}{|\vec{
\sigma} -{\hat{\vec{\eta}}}_{i}|}- \\
&-&{\frac{1}{2c}}\,\sum_{i=1}^{N}\,\sum_{j\neq
i}^{1..N}\,Q_{i}\,Q_{j}\,\int d^{3}\sigma
\,\vec{\sigma}\,\,\Big[\vec{\pi}_{\perp Si}(\vec{\sigma}-{\hat{
\vec{\eta}}}_{i},{\hat{\vec{\kappa}}}_{i})\cdot {\vec{\pi}}_{\perp
Sj}(\vec{\sigma} - {\hat{\vec{\eta}}}_{j},{\hat{\vec{\kappa}}}_{j})+ \\
&+&\vec{B}_{Si}(\vec{\sigma}-{\hat{\vec{\eta}}}_{i},{\hat{\vec{\kappa}}}
_{i})\cdot {\vec{B}}_{Sj}(\vec{\sigma} -
{\hat{\vec{\eta}}}_{j},{\hat{\vec{ \kappa}}}_{j})\Big] -
{\frac{1}{2c}}\int d^{3}\sigma \,\vec{\sigma}\,\,\Big({ \ \
\vec{\pi}}_{\perp rad}^{2}+{\vec{B}}_{rad}^{2}\Big)(\tau
,\vec{\sigma})-
\end{eqnarray*}

\begin{eqnarray}
&+&{\frac{1}{c}}\, \sum_{i=1}^N\,{\hat{\vec{\eta}}}_{i}(\tau
)\,Q_{i}\,{\frac{{{ \ \hat{\vec{\kappa} }}_{i}(\tau )\cdot
{\vec{A}}_{\perp rad}(\tau ,{\hat{ \vec{ \eta}}}_{i}(\tau ))
}}{\sqrt{m_{i}^{2}\,c^{2}+{\hat{\vec{\kappa}}}
_{i}^{2}(\tau )}}}+  \nonumber \\
&+&{\frac{1}{c}}\,\sum_{i=1}^N\,Q_{i}\,\Big[\sqrt{m_{i}^{2}\,c^{2}+{\hat{\vec{
\kappa}}}_{i}^{2}(\tau )}\,{\frac{{\partial }}{{\partial
\,{\hat{\vec{\kappa} }}_{i}}}}-{\hat{\vec{\eta}}}_{i}(\tau
)\,{\frac{{{\hat{\vec{\kappa}}} _{i}(\tau
)}}{\sqrt{m_{i}^{2}\,c^{2}+{\hat{\vec{\kappa}}}_{i}(\tau )}}}\cdot
{\frac{{\partial }}{{\partial \,{\hat{\vec{\eta}}}_{i}}}}\Big]  \nonumber \\
&&\times \int d^{3}\sigma \,\Big({\vec{\pi}}_{\perp rad}\cdot
{\hat{\vec{A}}} _{\perp Si}-{\vec{A}}_{\perp rad}\cdot
{\hat{\vec{\pi}}}_{\perp Si}\Big) (\tau
,\vec{\sigma})-{\frac{1}{c}}\,\sum_{i=1}^N\,Q_{i}\,\int d^{3}\sigma
\,{\ \frac{{{\vec{\pi}}_{\perp rad}(\tau ,\vec{\sigma})}}{{4\pi
\,|\vec{\sigma}-{\ \ \hat{\vec{\eta}}}_{i}(\tau )|}}}-  \nonumber \\
&-&{\frac{1}{c}}\,\sum_{i=1}^N\,Q_{i}\,\int d^{3}\sigma
\,\vec{\sigma}\,\,\Big({ \ \ \vec{\pi}}_{\perp rad}\cdot
{\hat{\vec{\pi}}}_{\perp Si}+{\vec{B}}
_{rad}\cdot {\hat{\vec{B}}}_{Si}\Big)(\tau ,\vec{\sigma}).
  \label{c5}
\end{eqnarray}
\bigskip

The final four lines of the long expression on the right hand side
contain terms that involve both radiation and Lienard-Wiechert
fields. \ We have seen that in the final form of the Hamiltonian
given in Eq.(\ref{4.14}) that all such terms cancel. \ We
investigate whether such cancelations occur here as well. We first
point out that Eq.(\ref{4.11}) implies

\begin{eqnarray}
&+&{\frac{1}{c}}\,\sum_{i=1}^N\,Q_{i}\,\Big[-{\hat{\vec{\eta}}}_{i}(\tau
)\,{\ \frac{{{\hat{\vec{\kappa}}}_{i}(\tau
)}}{\sqrt{m_{i}^{2}\,c^{2}+{\hat{\vec{ \kappa}}}_{i}(\tau )}}}\cdot
{\frac{{\partial }}{{\partial \,{\hat{\vec{\eta}
}}_{i}}}}\Big]  \nonumber \\
&&\times \int d^{3}\sigma \,\Big({\vec{\pi}}_{\perp rad}\cdot
{\hat{\vec{A}}} _{\perp Si}-{\vec{A}}_{\perp rad}\cdot
{\hat{\vec{\pi}}}_{\perp Si}\Big)
(\tau ,\vec{\sigma})  \nonumber \\
&=&{\frac{1}{c}}\,\sum_{i=1}^N\,Q_{i}\,\Big[{\hat{\vec{\eta}}}_{i}(\tau
)\,\int d^{3}\sigma \,\Big({\vec{\pi}}_{\perp rad}\cdot
{\hat{\vec{\pi}}}_{\perp Si}- {\vec{A}}_{\perp rad}\cdot
\frac{\partial ^{2}{\hat{\vec{A}}}_{\perp
Si} }{ \partial \tau ^{2}}\Big)(\tau ,\vec{\sigma})\Big]  \nonumber \\
&=&{\frac{1}{c}}\,\sum_{i=1}^N\,Q_{i}\,\Big[{\hat{\vec{\eta}}}_{i}(\tau
)\,\int d^{3}\sigma \,\Big({\vec{\pi}}_{\perp rad}\cdot
{\hat{\vec{\pi}}}_{\perp Si}- {\vec{A}}_{\perp rad}\cdot (\Box
\,{\hat{\vec{A}}}_{\perp Si}+\vec{
\partial} ^{2}{\hat{\vec{A}}}_{\perp Si}\Big)(\tau ,\vec{\sigma})\Big]
\nonumber \\
&=&{\frac{1}{c}}\,\sum_{i=1}^N\,Q_{i}\,\Big[{\hat{\vec{\eta}}}_{i}(\tau
)\,\int d^{3}\sigma \,\Big({\vec{\pi}}_{\perp rad}\cdot
{\hat{\vec{\pi}}}_{\perp Si}- {\vec{A}}_{\perp rad}\cdot
\vec{\partial}^{2}{\hat{\vec{A}}}_{\perp Si}
\Big) (\tau ,\vec{\sigma})\Big]  \nonumber \\
&&-\sum_{i=1}^N\,{\hat{\vec{\eta}}}_{i}(\tau
)\,Q_{i}\,{\frac{{{\hat{\vec{\kappa} }}_{i}(\tau )\cdot
{\vec{A}}_{\perp rad}(\tau ,{\hat{\vec{\eta}}}_{i}(\tau ))
}}{c\sqrt{m_{i}^{2}\,c^{2}+{\hat{\vec{\kappa}}}_{i}^{2}(\tau )}}},
  \label{c6}
\end{eqnarray}

\noindent and this last term cancels the seventh line on the right
hand side of Eq.(\ref{c5}). \ Furthermore the second part of the
tenth line

\begin{eqnarray}
&&-{\frac{1}{c}}\,\sum_{i=1}^N\,Q_{i}\,\int d^{3}\sigma
\,\,\vec{\sigma}\,\, \Big( {\vec{B}}_{rad}\cdot
{\hat{\vec{B}}}_{Si}\Big)(\tau ,\vec{\sigma})
\nonumber \\
&=&-{\frac{1}{c}}\,\sum_{i=1}^N\,Q_{i}\,\int d^{3}\sigma
\Big((\vec{\partial} _{\sigma }{A}_{\perp rad}^{r}){\hat{A}}_{\perp
Si}^{r}-\,\,\vec{\sigma}(\vec{
\partial}_{\sigma }^{2}{\vec{A}}_{\perp rad})\cdot {\hat{\vec{A}}}_{\perp
Si} \Big)(\tau ,\vec{\sigma}).
  \label{c7}
\end{eqnarray}

Taking into account the canceling from Eqs. (\ref{c6}) and
(\ref{c7}), the final four lines of the long expression on the right
hand side of Eq.( \ref {c5}) thus reduce to

\begin{eqnarray}
&+&{\frac{1}{c}}\,\sum_{i=1}^N\,Q_{i}\,\Big[\sqrt{m_{i}^{2}\,c^{2}+{\hat{\vec{
\kappa}}}_{i}^{2}(\tau )}\,{\frac{{\partial }}{{\partial
\,{\hat{\vec{\kappa} }}_{i}}}}\Big]  \nonumber \\
&&\times \int d^{3}\sigma \,\Big({\vec{\pi}}_{\perp rad}\cdot
{\hat{\vec{A}}} _{\perp Si}-{\vec{A}}_{\perp rad}\cdot
{\hat{\vec{\pi}}}_{\perp Si}\Big)(\tau ,\vec{\sigma})  \nonumber \\
&&-{\frac{1}{c}}\,\sum_{i=1}^N\,Q_{i}\,\int d^{3}\sigma
\Big({\frac{{{\vec{\pi}} _{\perp rad}(\tau ,\vec{\sigma})}}{{4\pi
\,|\vec{\sigma}-{\hat{\vec{\eta}}} _{i}(\tau
)|}}}+(\vec{\partial}_{\sigma }{A}_{\perp rad}^{r}){\hat{A}}
_{\perp Si}^{r}\Big)(\tau ,\vec{\sigma})  \nonumber \\
&+&{\frac{1}{c}}\,\sum_{i=1}^N\,Q_{i}\,\int d^{3}\sigma
\,({\hat{\vec{\eta}}} _{i}(\tau
)-\,\vec{\sigma}\,)\,\Big({\vec{\pi}}_{\perp rad}\cdot {\hat{\vec{
\pi}}}_{\perp Si}-(\vec{\partial}_{\sigma }^{2}{\vec{A}}_{\perp
rad})\cdot { \ \ \hat{\vec{A}}}_{\perp Si}\Big)(\tau ,\vec{\sigma}).
  \label{c8}
\end{eqnarray}

To further simplify this expression we make use of the Fourier
transforms of the potentials and fields given in Appendix B. \ In
particular the first part of the last integral in Eq.(\ref{c8})
becomes

\begin{eqnarray}
&&-{\frac{1}{c}}\,\sum_{i=1}^N\,Q_{i}\,\int d^{3}\sigma
\,\,\vec{\sigma}\,\Big({ \ \vec{\pi}}_{\perp rad}\cdot
{\hat{\vec{\pi}}}_{\perp Si}\Big)(\tau ,\vec{
\sigma})=  \nonumber \\
&=&-{\frac{1}{c}}\,\sum_{i=1}^N\,Q_{i}\int d^{3}\tilde{k}\omega
(\vec{k} )\,\sum_{\lambda =1,2}\,{\vec{\epsilon}}_{\lambda
}(\vec{k})\{\,\Big[ ia_{\lambda }(\vec{k})\int d^{3}\sigma
\,\vec{\sigma}\,e^{-i\,[\omega (\vec{k })\,\tau -\vec{k}\cdot
\vec{\sigma}]}\cdot {\hat{\vec{\pi}}}_{\perp
Si}+c.c.\}=  \nonumber \\
&=&{\frac{1}{c}}\,\sum_{i=1}^N\,Q_{i}\int d^{3}\tilde{k}\omega
(\vec{k} )\,\sum_{\lambda =1,2}\,{\epsilon }_{\lambda
}^{r}(\vec{k})\{\,\Big[ ia_{\lambda }(\vec{k})\,e^{-i\,\omega
(\vec{k})\,\tau }i\frac{\partial }{\partial \vec{k}}(-i)
\frac{e^{i\vec{k} \cdot \vec{\eta}_{i}}}{k^{4}}\frac{
\vec{\kappa}_{i}\cdot \vec{k}[\vec{k}\times (\vec{\kappa}_{i}\times
\vec{k} )]_{r}}{m_{i}^{2}\,c^{2}+{\vec{\kappa}_{i}}^{2}-\left(
\vec{\kappa}_{i}\cdot \hat{k}\right) ^{2}}+c.c\}=  \nonumber \\
&=&{\frac{1}{c}}\,\sum_{i=1}^N\,Q_{i}\int d^{3}\tilde{k}\omega
(\vec{k} )\,\sum_{\lambda =1,2}\,{\epsilon }_{\lambda
}^{r}(\vec{k})\{\,\Big[ ia_{\lambda }(\vec{k})\,e^{-i\,[\omega
(\vec{k})\,\tau -\vec{k}\cdot \vec{ \eta}_{i}]}\frac{\partial
}{\partial \vec{k}}\frac{1}{k^{4}}\frac{\vec{\kappa }_{i}\cdot
\vec{k}[\vec{k}\times (\vec{\kappa}_{i}\times \vec{k})]_{r}}{
m_{i}^{2}\,c^{2}+{\vec{\kappa}_{i}}^{2}-\left( \vec{\kappa}_{i}\cdot
\hat{k}\right) ^{2}}+c.c\}-  \nonumber \\
&-&{\frac{1}{c}}\,\sum_{i=1}^N\,Q_{i}\,\int d^{3}\sigma
\,{\hat{\vec{\eta}}} _{i}(\tau )[\,{\vec{\pi}}_{\perp rad}\cdot
{\hat{\vec{\pi}}}_{\perp Si}](\tau ,\vec{\sigma}),
  \label{c9}
\end{eqnarray}

\noindent while the second part becomes

\begin{eqnarray}
&&{\frac{1}{c}}\,\sum_{i=1}^N\,Q_{i}\,\int d^{3}\sigma
\,\,\vec{\sigma}\,\,\Big( ( \vec{\partial}_{\sigma
}^{2}{\vec{A}}_{\perp rad})\cdot {\hat{\vec{A}}}
_{\perp Si}\Big)(\tau ,\vec{\sigma}) =  \nonumber \\
&=&{\frac{1}{c}}\,\sum_{i=1}^N\,Q_{i}\,\int d^{3}\sigma
{\hat{\vec{\eta}}} _{i}(\tau )\,\,\,\Big((\vec{\partial}_{\sigma
}^{2}{\vec{A}}_{\perp rad})\cdot {\hat{\vec{A}}}_{\perp
Si}\Big)(\tau ,\vec{\sigma}) +  \nonumber \\
&+&{\frac{1}{c}}\,\sum_{i=1}^N\,Q_{i}\,\int
d^{3}\tilde{k}\,\sum_{\lambda =1,2}{ \ \ \epsilon }_{\lambda
}^{r}(\vec{k})\Big(ia_{\lambda }(\vec{k} )\,e^{-i\,[\omega
(\vec{k})\,\tau -\vec{k}\cdot \vec{\eta}_{i}]}\vec{k}^{2}
\frac{\partial }{\partial
\vec{k}}\frac{1}{k^{4}}\frac{[\vec{k}\times (\vec{ \kappa}_{i}\times
\vec{k})]_{r}\sqrt{m_{i}^{2}\, c^2 + {\vec{\kappa}_{i}} ^{2} }}{
m_{i}^{2}\, c^2 + {\vec{\kappa}_{i}}^{2} - \left( \vec{\kappa}
_{i}\cdot \hat{k}\right) ^{2}}+c.c\Big).  \nonumber \\
&&{}
  \label{c10}
\end{eqnarray}

Substituting Eqs.(\ref{c9}) and (\ref{c10}) into Eq.(\ref{c8}) gives

\begin{eqnarray}
&+&{\frac{1}{c}}\,\sum_{i=1}^N\,Q_{i}\,\Big[\sqrt{m_{i}^{2}\,c^{2}+{\hat{\vec{
\kappa}}}_{i}^{2}(\tau )}\,{\frac{{\partial }}{{\partial
\,{\hat{\vec{\kappa}}}_{i}}}}\Big]  \nonumber \\
&&\times \int d^{3}\sigma \,\Big({\vec{\pi}}_{\perp rad}\cdot
{\hat{\vec{A}}} _{\perp Si}-{\vec{A}}_{\perp rad}\cdot
{\hat{\vec{\pi}}}_{\perp Si}\Big)(\tau ,\vec{\sigma})-  \nonumber \\
&-&{\frac{1}{c}}\,\sum_{i=1}^N\,Q_{i}\,\int d^{3}\sigma
\Big({\frac{{{\vec{\pi}} _{\perp rad}(\tau ,\vec{\sigma})}}{{4\pi
\,|\vec{\sigma}-{\hat{\vec{\eta}}} _{i}(\tau
)|}}}+(\vec{\partial}_{\sigma }{A}_{\perp rad}^{r}){\hat{A}}
_{\perp Si}^{r}\Big)(\tau ,\vec{\sigma})+  \nonumber \\
&+&{\frac{1}{c}}\,\sum_{i=1}^N\,Q_{i}\,\int
d^{3}\tilde{k}\,\sum_{\lambda =1,2}{ \ \epsilon }_{\lambda
}^{r}(\vec{k})\{ia_{\lambda }(\vec{k})\,e^{-i\,[\omega
( \vec{k})\,\tau -\vec{k}\cdot \vec{\eta}_{i}]}  \nonumber \\
&&\times \lbrack \vec{k}^{2}\frac{\partial }{\partial
\vec{k}}\frac{1}{k^{4}} \frac{[\vec{k}\times (\vec{\kappa}_{i}\times
\vec{k})]_{r}\sqrt{m_{i}^{2}\,c^{2}+{\vec{\kappa}_{i}}^{2}}}{m_{i}^{2}\,c^{2}+{\vec{\kappa}_{i}}
^{2}-\left( \vec{\kappa}_{i}\cdot \hat{k}\right) ^{2}}+\omega
(\vec{k})\frac{\partial }{\partial
\vec{k}}\frac{1}{k^{4}}\frac{\vec{\kappa}_{i}\cdot \vec{ k
}[\vec{k}\times (\vec{\kappa}_{i}\times
\vec{k})]_{r}}{m_{i}^{2}\,c^{2}+{\ \vec{\kappa}_{i}}^{2}-\left(
\vec{\kappa}_{i}\cdot \hat{k}\right) ^{2}}]+c.c\}.  \nonumber \\
&&{}
  \label{c11}
\end{eqnarray}
Including all Fourier transforms the above becomes

\begin{eqnarray*}
&+&{\frac{1}{c}}\,\sum_{i=1}^N\,Q_{i}\,\int
d^{3}\tilde{k}\sum_{\lambda =1,2} \Big[i\, a_{\lambda
}(\vec{k})\,e^{-i\,[\omega (\vec{k})\,\tau
- \vec{k}\cdot \vec{\eta}_{i}]}  \nonumber \\
&&\Big(\omega (\vec{k})\sqrt{m_{i}^{2}\, c^{2} +
{\hat{\vec{\kappa}}} _{i}^{2}(\tau )} \,\frac{{\partial }}{{\partial
\,{\hat{\vec{\kappa}}}_{i}}} \frac{1}{k^{4}}
\frac{{\vec{\epsilon}}_{\lambda }(\vec{k})\cdot \lbrack \vec{
k}\times (\vec{ \kappa}_{i}\times \vec{k})]\sqrt{m_{i}^{2}\, c^2 +
{\vec{ \kappa}_{i}}^{2}}}{ m_{i}^{2}\, c^2 + {\vec{\kappa}_{i}}^{2}
- \left( \vec{\kappa}_{i}\cdot \hat{k}\right)^{2}} +  \nonumber \\
&+&\sqrt{m_{i}^{2}\,c^{2}+{\hat{\vec{\kappa}}}_{i}^{2}(\tau
)}\,\frac{{\ \partial }}{{\partial
\,{\hat{\vec{\kappa}}}_{i}}}\frac{1}{k^{4}}\frac{{\vec{
\epsilon}}_{\lambda }(\vec{k})\cdot \lbrack \vec{k}\times
(\vec{\kappa} _{i}\times \vec{k})]\vec{\kappa}_{i}\cdot
\vec{k}}{m_{i}^{2}\, c^2 + {\vec{ \kappa} _{i}}^{2}-\left(
\vec{\kappa}_{i}\cdot \hat{k}\right) ^{2}}-\frac{
\omega ( \vec{k})}{k^{2}}{\vec{\epsilon}}_{\lambda }(\vec{k}) +  \nonumber \\
&+&\vec{k}\, \frac{1}{k^{4}}
\frac{{\vec{\epsilon}}_{\lambda}(\vec{k}) \cdot \lbrack
\vec{k}\times (\vec{\kappa}_{i}\times \vec{k})]\sqrt{m_{i}^{2}\, c^2
+ {\vec{ \kappa}_{i}}^{2}}}{m_{i}^{2}\, c^2 +
{\vec{\kappa}_{i}}^{2}-\left( \vec{\kappa}_{i}\cdot \hat{k}\right)
^{2}} +
 \end{eqnarray*}

\bea
 &+&{\epsilon }_{\lambda }^{r}(\vec{k})\Big[\vec{k}^{2}\frac{\partial
}{\partial \vec{k}}\frac{1}{k^{4}}\frac{[\vec{k}\times
(\vec{\kappa}_{i}\times \vec{k} )]_{r}\sqrt{m_{i}^{2}\, c^2 +
{\vec{\kappa}_{i}}^{2}}}{m_{i}^{2}\, c^2 + {\vec{\kappa}_{i}} ^{2} -
\left( \vec{\kappa}_{i}\cdot \hat{k}\right)
^{2}}+  \nonumber \\
 &+&\omega(\vec{k})\frac{ \partial }{\partial
\vec{k}}\frac{1}{k^{4}}\frac{ \vec{\kappa}_{i}\cdot \vec{k
}[\vec{k}\times (\vec{\kappa}_{i}\times \vec{k} )]_{r}}{m_{i}^{2}\,
c^2 + {\vec{\kappa }_{i}}^{2}-\left( \vec{\kappa}
_{i}\cdot \hat{k}\right)^{2}}\Big]\Big)+c.c\Big].  \nonumber \\
 &&{}  \label{c12}
\end{eqnarray}

Expanding the triplet products in the above and using the
transversality yields for Eq.(\ref{c8})

\begin{eqnarray}
&+&{\frac{1}{c}}\,\sum_{i=1}^N\,Q_{i}\,\int
d^{3}\tilde{k}\sum_{\lambda =1,2} \Big[ia_{\lambda
}(\vec{k})\,e^{-i\,[\omega (\vec{k})\,\tau
-\vec{k}\cdot \vec{\eta}_{i}]}  \nonumber \\
&&\Big(\omega (\vec{k})\sqrt{m_{i}^{2}\,c^{2}+{\hat{\vec{\kappa}}}
_{i}^{2}(\tau )} \,\frac{{\partial }}{{\partial
\,{\hat{\vec{\kappa}}}_{i}}} \frac{1}{k^{2}}
\frac{{\vec{\epsilon}}_{\lambda }(\vec{k})\cdot \vec{\kappa}
_{i}\sqrt{ m_{i}^{2}\, c^2 + {\vec{\kappa}_{i}}^{2}}}{m_{i}^{2}\,
c^2 + {\ \vec{\kappa}_{i}}^{2} - \left(\vec{\kappa}_{i}\cdot
\hat{k}\right) ^{2}} +\nonumber \\
&+&\sqrt{m_i^2\, c^2 + {\hat{\vec{\kappa}}}_i^2(\tau )}\,\frac{{\
\partial }}{{\partial \,{\hat{\vec{\kappa}}}_{i}}}\,
\frac{1}{k^{2}}\, \frac{ {\vec{ \epsilon}}_{\lambda }(\vec{k})\cdot
\vec{\kappa}_{i}\vec{\kappa} _{i}\cdot \vec{k}}{m_{i}^{2}\, c^2 +
{\vec{\kappa}_{i}}^{2} - \left( \vec{ \kappa}_{i}\cdot \hat{k
}\right) ^{2}} - \frac{\omega (\vec{k})}{k^{2}}{\vec{
\epsilon}}_{\lambda }( \vec{k}) +  \nonumber \\
&+&\vec{k}\, \frac{1}{k^{2}}\,
\frac{{\vec{\epsilon}}_{\lambda}(\vec{k} ) \cdot
\vec{\kappa}\sqrt{m_{i}^{2}\, c^2 + {\vec{\kappa}_{i}}^{2}}}{
m_{i}^{2}\, c^2 + {\vec{\kappa} _{i}}^{2} - \left(
\vec{\kappa}_{i}\cdot \hat{k}\right) ^{2}} +  \nonumber \\
&+&{\epsilon }_{\lambda }^{r}(\vec{k})\Big[\vec{k}^{2}\frac{\partial
}{\partial \vec{k}}\frac{1}{k^{4}}\frac{[\vec{k}\times
(\vec{\kappa}_{i}\times \vec{k} )]_{r}\sqrt{m_{i}^{2}\, c^2 +
{\vec{\kappa}_{i}}^{2}}}{m_{i}^{2}\, c^2 + {\vec{\kappa}_{i}}
^{2}-\left( \vec{\kappa}_{i}\cdot \hat{k}\right) ^{2}} + \omega
(\vec{k})\frac{\partial }{\partial \vec{k}}\, \frac{1}{k^{4}} \,
\frac{\vec{\kappa}_{i}\cdot \vec{k }[\vec{k}\times (\vec{\kappa}
_{i}\times \vec{k})]_{r}}{m_{i}^{2}\, c^2 + {\vec{\kappa }_{i}}^{2}
- \left(\vec{\kappa}_{i}\cdot \hat{k}\right)^{2}}\Big]\Big)+c.c\Big].
 \nonumber \\
&&{}
  \label{c13}
\end{eqnarray}

Using

\begin{equation}
\frac{\partial }{\partial \vec{k}}\,
\frac{1}{m_{i}^{2}\,c^{2}+{\vec{\kappa}_{i} }^{2}-\left(
\vec{\kappa}_{i}\cdot \hat{k}\right) ^{2}}=\frac{2(\vec{\kappa}
_{i}\cdot \vec{k}\vec{\kappa}_{i}-\left( \vec{\kappa}_{i}\cdot
\hat{k} \right) ^{2}\vec{k})}{\left(
m_{i}^{2}\,c^{2}+{\vec{\kappa}_{i}}^{2}-\left( \vec{\kappa}_{i}\cdot
\hat{k}\right) ^{2}\right) ^{2}},
  \label{c14}
\end{equation}
performing, the indicated operations and combining like terms yields

\begin{eqnarray*}
 &=&{\frac{1}{c}}\,\sum_{i=1}^N\,Q_{i}\,\int d^{3}\tilde{k}\sum_{\lambda
=1,2} \Big[ia_{\lambda }(\vec{k})\,e^{-i\,[\omega (\vec{k})\,\tau
-\vec{k}\cdot\vec{\eta}_{i}]}  \nonumber \\
 &&\Big({\vec{\epsilon}}_{\lambda }(\vec{k})\,\Big[\frac{\omega
(\vec{k})}{k^{2}}\frac{(m_{i}^{2}\,c^{2}+{\vec{\kappa}_{i}}^{2})}{m_{i}^{2}\,c^{2}+{\
\vec{\kappa}_{i}}^{2}-\left( \vec{\kappa}_{i}\cdot \hat{k}\right)
^{2}}- \frac{\omega (\vec{k})}{k^{2}}-\frac{(\vec{\kappa}_{i}\cdot
\vec{k} )^{2}\omega (\vec{k})}{\vec{k}^{4}\left(
m_{i}^{2}\,c^{2}+{\vec{\kappa}_{i}}
^{2}-\left( \vec{\kappa}_{i}\cdot \hat{k}\right) ^{2}\right) }+  \nonumber \\
 &+&\frac{\sqrt{m_{i}^{2}\,c^{2}+{\hat{\vec{\kappa}}}_{i}^{2}(\tau
)}}{k^{2}} \,\frac{\vec{\kappa}_{i}\cdot
\vec{k}}{m_{i}^{2}\,c^{2}+{\vec{\kappa}_{i}} ^{2}-\left(
\vec{\kappa}_{i}\cdot \hat{k}\right) ^{2}}-\frac{\vec{\kappa}
_{i}\cdot \vec{k}}{\vec{k}^{2}\left(
m_{i}^{2}\,c^{2}+{\vec{\kappa}_{i}} ^{2}-\left(
\vec{\kappa}_{i}\cdot \hat{k}\right) ^{2}\right) }\,\sqrt{
m_{i}^{2}\,c^{2}+{\vec{\kappa}_{i}}^{2}}\Big]+
 \end{eqnarray*}

\bea
 &+&{\vec{\epsilon}}_{\lambda }(\vec{k})\cdot
\vec{\kappa}_{i}\,\vec{\kappa} _{i}\,\Big[\frac{\omega
(\vec{k})}{k^{2}}\,\Big(\frac{{1}}{m_{i}^{2}\,c^{2}+{
\vec{\kappa}_{i}}^{2}-\left( \vec{\kappa}_{i}\cdot \hat{k}\right)
^{2}}- \frac{2(m_{i}^{2}\,c^{2}+{\ \vec{\kappa}_{i}}^{2})}{\left(
m_{i}^{2}\,c^{2}+{ \vec{\kappa}_{i}}^{2}-\left(
\vec{\kappa}_{i}\cdot \hat{k}\right)
^{2}\right) ^{2}}+  \nonumber \\
&+&\frac{2\left( \vec{\kappa}_{i}\cdot \vec{k}\right)
^{2}}{k^{2}\left( m_{i}^{2}\,c^{2}+{\vec{\kappa}_{i}}^{2}-\left(
\vec{\kappa}_{i}\cdot \hat{k} \right) ^{2}\right)
^{2}}+\frac{1}{m_{i}^{2}\,c^{2}+{\vec{\kappa}_{i}}
^{2}-\left( \vec{\kappa}_{i}\cdot \hat{k}\right) ^{2}}\Big)-  \nonumber \\
&-&\frac{\sqrt{m_{i}^{2}\,c^{2}+{\hat{\vec{\kappa}}}_{i}^{2}(\tau
)}}{k^{2}} \,\frac{2\vec{\kappa}_{i}\cdot \vec{k}}{\left(
m_{i}^{2}\,c^{2}+{\vec{\kappa} _{i}}^{2}-\left(
\vec{\kappa}_{i}\cdot \hat{k}\right) ^{2}\right) ^{2}}+
\frac{2\sqrt{m_{i}^{2}\,c^{2}+{\vec{\kappa}_{i}}^{2}}\vec{\kappa}_{i}\cdot
\vec{k}}{k^{2}\left( m_{i}^{2}\,c^{2}+{\vec{\kappa}_{i}}^{2}-\left(
\vec{\kappa}_{i}\cdot \hat{k}\right) ^{2}\right) ^{2}}\Big]+  \nonumber \\
&+&{\vec{\epsilon}}_{\lambda }(\vec{k})\cdot
\vec{\kappa}_{i}\,\vec{k}\,\Big[ \frac{\omega
(\vec{k})}{k^{2}}\,\Big(\frac{2(m_{i}^{2}\,c^{2}+{\vec{\kappa}
_{i}}^{2})(\vec{\kappa}_{i}\cdot \vec{k})}{k^{2}\left(
m_{i}^{2}\,c^{2}+{ \vec{\kappa}_{i}}^{2}-\left(
\vec{\kappa}_{i}\cdot \hat{k}\right) ^{2}\right)
^{2}}-\frac{4}{k^{2}}\frac{\vec{\kappa}_{i}\cdot \vec{k}}{
m_{i}^{2}\,c^{2}+{\vec{\kappa}_{i}}^{2}-\left( \vec{\kappa}_{i}\cdot
\hat{k}\right) ^{2}}-  \nonumber \\
&-&\frac{2\left( \vec{\kappa}_{i}\cdot \hat{k}\right)
^{2}}{k^{2}\left( m_{i}^{2}\,c^{2}+{\vec{\kappa}_{i}}^{2}-\left(
\vec{\kappa}_{i}\cdot \hat{k} \right) ^{2}\right)
^{2}}+\frac{2\vec{\kappa}_{i}\cdot \vec{k}}{\vec{k} ^{2}\left(
m_{i}^{2}\,c^{2}+{\vec{\kappa}_{i}}^{2}-\left( \vec{\kappa}
_{i}\cdot \hat{k}\right) ^{2}\right) }\Big)\Big]+  \nonumber \\
&+&\frac{\sqrt{m_{i}^{2}\,c^{2}+{\hat{\vec{\kappa}}}_{i}^{2}(\tau
)}}{k^{2}}\,\Big(\frac{{1}}{m_{i}^{2}\,c^{2}+{\vec{\kappa}_{i}}^{2}-\left(
\vec{\kappa} _{i}\cdot \hat{k}\right)
^{2}}+\frac{2\vec{\kappa}_{i}\cdot \vec{k}(\vec{ \kappa}_{i}\cdot
\vec{k})}{k^{2}\left( m_{i}^{2}\,c^{2}+{\vec{\kappa}_{i}}
^{2}-\left( \vec{\kappa}_{i}\cdot \hat{k}\right) ^{2}\right) ^{2}}+
\nonumber \\
&+&\frac{1}{m_{i}^{2}\,c^{2}+{\vec{\kappa}_{i}}^{2}-\left(
\vec{\kappa} _{i}\cdot \hat{k}\right)
^{2}}-\frac{4}{m_{i}^{2}\,c^{2}+{\vec{\kappa}_{i}} ^{2}-\left(
\vec{\kappa}_{i}\cdot \hat{k}\right) ^{2}}-\frac{2\left( \vec{
\kappa}_{i}\cdot \vec{k}\right) ^{2}}{k^{2}\left(
m_{i}^{2}\,c^{2}+{\ \vec{ \kappa}_{i}}^{2}-\left(
\vec{\kappa}_{i}\cdot \hat{k}\right) ^{2}\right) ^{2}
}=  \nonumber \\
&+&\frac{2}{\left( m_{i}^{2}\,c^{2}+{\vec{\kappa}_{i}}^{2}-\left(
\vec{\kappa }_{i}\cdot \hat{k}\right) ^{2}\right)
}\Big)\,\,\Big)+c.c\Big],
  \label{c15}
\end{eqnarray}

\noindent or

\begin{eqnarray*}
&=&\frac{1}{c}\,\sum_{i=1}^N\,Q_{i}\,\int
d^{3}\tilde{k}\,\sum_{\lambda =1,2}\, \Big[ia_{\lambda
}(\vec{k})\,e^{-i\,[\omega (\vec{k})\,\tau
-\vec{k}\cdot \vec{\eta}_{i}]}  \nonumber \\
&&\Big({\vec{\epsilon}}_{\lambda
}(\vec{k})[0+0]\,+{\vec{\epsilon}}_{\lambda }(\vec{k})\cdot
\vec{\kappa}_{i}\,\vec{\kappa}_{i}\,\Big[\frac{\omega (\vec{k
})}{k^{2}}(0)-\frac{\sqrt{m_{i}^{2}\,c^{2}+{\hat{\vec{\kappa}}}_{i}^{2}(\tau
)}}{k^{2}}0\Big]+  \nonumber \\
&+&\frac{{\vec{\epsilon}}_{\lambda }(\vec{k})\cdot
\vec{\kappa}_{i}\hat{k}}{ \left(
m_{i}^{2}\,c^{2}+{\vec{\kappa}_{i}}^{2}-\left( \vec{\kappa}_{i}\cdot
\hat{k}\right) ^{2}\right) }\,\frac{\omega
(\vec{k})}{k^{2}}\,\Big[\Big(
\frac{2(m_{i}^{2}\,c^{2}+{\vec{\kappa}_{i}}^{2})}{\left(
m_{i}^{2}\,c^{2}+{\ \vec{\kappa}_{i}}^{2}-\left(
\vec{\kappa}_{i}\cdot \hat{k}\right) ^{2}\right) }-
 \end{eqnarray*}

  \bea
&-&\frac{2\left( \vec{\kappa}_{i}\cdot \hat{k}\right) ^{2}}{\left(
m_{i}^{2}\,c^{2}+{\vec{\kappa}_{i}}^{2}-\left( \vec{\kappa}_{i}\cdot
\hat{k} \right) ^{2}\right)
})\Big]+\sqrt{m_{i}^{2}\,c^{2}+{\hat{\vec{\kappa}}} _{i}^{2}(\tau
)}(0)\Big)+c.c\Big]=0.
 \label{c16}
\end{eqnarray}
Thus all four lines of the long expression on the right hand side of
Eq.( \ref{c5}) cancel among one another

\bigskip Hence our boost reduces to

\begin{eqnarray}
\mathcal{\vec{K}}_{(int)}
&=&-\sum_{i=1}^{N}\,{\hat{\vec{\eta}}}_{i}(\tau
)\, \Big[\sqrt{m_{i}^{2}\,c^{2}+{\hat{\vec{\kappa}}}_{i}^{2}}+  \nonumber \\
&+&{\frac{{{\hat{\vec{\kappa}}}_{i}}}{{2\, c\,
\sqrt{m_{i}^{2}\,c^{2}+{\hat{ \vec{ \kappa}}}_{i}^{2}}}}}\,\cdot
\sum_{j\neq i}^{1..N}\,Q_{i}\,Q_{j}\,\Big({\frac{ 1}{2}}
\,{\frac{{\partial \,{\hat{\mathcal{K}}}_{ij}({\hat{\vec{\kappa}}}
_{i},{\hat{\vec{\kappa}}}}_{j},{\hat{\vec{\eta}}_{i}-{\hat{\vec{\eta}}}
_{j}) }}{{\ \partial \,{\hat{\vec{\eta}}}_{i}}}}- 2\,\vec{A}_{\perp
Sj}({\hat{\vec{\kappa}}}_{j},{\hat{\vec{\eta}}}_{i}-{\hat{\vec{\eta}}}_{j})\Big)
\Big]-  \nonumber \\
&-&\frac{1}{2}\,\sum_{i=1}^{N}\,\sum_{j\neq i}^{1..N}\,
{\frac{{Q_{i}\,Q_{j}}}{c}} \, \sqrt{
m_{i}^{2}\,c^{2}+{\hat{\vec{\kappa}}}_{i}^{2}}\,{\frac{{\partial \,
 {\ \hat{\mathcal{K}}}_{ij}({\hat{\vec{\kappa}}}_{i},{\hat{\vec{\kappa}}}_{j},
{\ \hat{ \vec{\eta}}}_{i}-{\hat{\vec{\eta}}}_{j})}}{{\partial
\,{\hat{\vec{\kappa}}}_{i}}}}+  \nonumber \\
&+&{\frac{1}{c}}\,\sum_{i=1}^{N}\,\sum_{j\not=i}^{1..N}\,\frac{Q_{i}\,Q_{j}}{8\pi
}\,\frac{{\hat{\vec{\eta}}}_{i}-{\hat{\vec{\eta}}}_{j}}{|{\hat{\vec{\eta}}}
_{i}-{\hat{\vec{\eta}}}_{j}|}-\sum_{i=1}^{N}\,\sum_{j\not=i}^{1..N}\,\frac{
Q_{i}\,Q_{j}}{4\pi\, c }\,\int d^{3}\sigma
\,\frac{{\hat{\vec{\pi}}}_{\perp Sj}(
\vec{\sigma}-{\hat{\vec{\eta}}}_{j},{\hat{\vec{\kappa}}}_{j})}{|\vec{
\sigma} -{\hat{\vec{\eta}}}_{i}|}-  \nonumber \\
&-&{\frac{1}{2c}}\,\sum_{i=1}^{N}\,\sum_{j\neq
i}^{1..N}\,Q_{i}\,Q_{j}\,\int d^{3}\sigma
\,\vec{\sigma}\,\,\Big[\vec{\pi}_{\perp Si}(\vec{\sigma}-{\hat{
\vec{\eta}}}_{i},{\hat{\vec{\kappa}}}_{i})\cdot {\vec{\pi}}_{\perp
Sj}(\vec{\sigma}-{\hat{\vec{\eta}}}_{j},{\hat{\vec{\kappa}}}_{j})+  \nonumber \\
&+&\vec{B}_{Si}(\vec{\sigma}-{\hat{\vec{\eta}}}_{i},{\hat{\vec{\kappa}}}
_{i})\cdot {\vec{B}}_{Sj}(\vec{\sigma} -
{\hat{\vec{\eta}}}_{j},{\hat{\vec{
\kappa}}}_{j})\Big]-{\frac{1}{2c}}\int d^{3}\sigma
\,\vec{\sigma}\,\,\Big({\ \vec{\pi}}_{\perp
rad}^{2}+{\vec{B}}_{rad}^{2}\Big)(\tau ,\vec{\sigma}),
  \label{c17}
\end{eqnarray}

\noindent which consists only of the decoupled radiation fields
portion at the end plus particle and Lienard-Wiechert potential
induced terms.

\subsection{Semi-relativistic Expansions of $\mathcal{E}_{(int)}$ and $
\mathcal{\vec{K}}_{(int)}$ after the Canonical Transformation}

By using Eqs.(\ref{b1}) - (\ref{b4}) we get ${\hat
{\mathcal{K}}}_{12} = O(c^{-3})$, so that all the terms after the
first in Eq.(\ref{4.6}) are of order $O(c^{-2})$, $O(c^{-3})$,
$O(c^{-3})$, $O(c^{-5})$, respectively. Therefore we get

\begin{eqnarray}
\mathcal{\vec{K}}_{(int)} &=&- \sum_{i=1}^2\,
{\hat{\vec{\eta}}}_{i}(\tau )\, \Big(m_{i}\, c +
\frac{{\hat{\vec{\kappa}}}_{i}^{2}}{2m_{i}c}\Big) +
O(c^{-2}) -  \nonumber \\
&-& {\frac{1}{2c}}\int d^{3}\sigma \,\vec{\sigma}\,\,\Big({\vec{
\pi}}
_{\perp rad}^{2}+{\vec{B}}_{rad}^{2}\Big)(\tau ,\vec{\sigma}) =  \nonumber \\
&&{}  \nonumber \\
&=&{\vec{\mathcal{K}}}_{matter}+{\vec{\mathcal{K}}}_{rad}=c\,{\vec{K}}
_{Galilei}+O({\frac{1}{c}}) = - c\, \sum_{i=1}^2\, m_i\, {\hat {\vec \eta}}_i
\approx 0.  \nonumber \\
&&{}
  \label{c18}
\end{eqnarray}

\vfill\eject

\section{Dimensions}

Let $[t]$, $[l]$ and $[m]$ be the units of time, length and mass.
Then for the proper time we have $[\tau ] = [x^o = c\, t] = [l]$.

\bigskip

For the 4-momentum $P^{\mu} = (P^o = {E\over c}; \vec P)$ with $Mc =
\sqrt{P^2}$ we have the following dimensions $[M] = [m]$, $[E = c\,
P^0] = [m\, l^2\, t^{-2}]$ , $[\vec P] = [m\, l\, t^{-1}]$.
\medskip

For the Lorentz generators $J^i = \epsilon^{ijk}\, J^{jk}$, $K^r =
J^{or}$ we have $[J^{\mu\nu}] = [\vec J] = [\vec K] = [m\, l^2\,
t^{-1}]$. For the non-relativistic Galilei boosts we have $[{\vec
K}^{^{\prime}}] = [\vec K/c] = [m\, l]$.

\bigskip

For the particles we have $[{\vec \eta}_i] = [l]$ , $[{\vec
\kappa}_i] = [m\, l\, t^{-1}]$, $[{\frac{{d\, {\hat \eta} ^r_i(\tau
)}}{{d\, \tau}}}] = [0]$ a-dimensional, $[{\frac{{d\, {\hat {\vec
\kappa}}_i(\tau )}}{{d\, \tau}}}] = [m\, t^{-1}]$. For the
energy-momentum tensor we have $[T^{AB}] = [m\, l^{-2}\, t^{-1}]$.

\bigskip

For the electro-magnetism we adopt the Heaviside-Lorentz system of
units, so that the Coulomb potential is ${{Q_1\, Q_2}\over {4\pi\,
|{\vec \eta}_1(\tau ) - {\vec \eta}_2(\tau )|}}$ and $\epsilon_o =
\mu_o = 1$ for the electric permittivity and magnetic permeability
of the vacuum (this implies $\vec D = \vec E$ and $\vec B = \vec
H$).\medskip

Therefore the dimensions of the electric charge and of the
electro-magnetic potentials and fields are $[Q_i] = [m^{1/2}\,
l^{3/2}\, t^{-1}]$, $[{\vec A}_{\perp}] = [Q\, l^{-1}] = [m^{1/2}\,
l^{1/2}\, t^{-1}]$, $[{\vec \pi}_{\perp} = {\vec E }_{\perp}] =
[\vec B]  = [{\vec A}_{\perp}\, l^{-1}] = [m^{1/2}\, l^{-1/2}\,
t^{-1}]$. Consistently we have $[Q^2\, l^{-1}] = [Q\, {\vec
A}_{\perp}] = [m\, c^2]$.\medskip

In Eqs.(\ref{2.32}) we have $[k] = [l^{-1}]$, $[{\vec A}_{\perp}] =
[m^{1/2}\, l^{1/2}\, t^{-1}]$, $[a_{\lambda}] = [m^{1/2}\, l^{5/2}\,
t^{-1}]$, while for the helicity in Eqs.(\ref{2.34}) we have $[h] =
[\vec j] = [m\, l^2\, t^{-1}]$.\medskip

For the Lienard-Wiechert electro-magnetic potentials and fields of
Eqs. (\ref{2.49})-(\ref{2.53}) we have $ [{\vec A}_{\perp Si}] =
[l^{-1}]$, $[{\vec \pi}_{\perp Si} = {\vec E}_{\perp Si}] = [{\vec
B}_{Si}] = [l^{-2}]$, $[V_{Darwin}] = [m\, l^2\, t^{-2}]$. \medskip

For the functionals (\ref{3.4}) and (\ref{3.5}) we have $[T_i] =
[m^{1/2}\, l^{3/2}\, t^{-1}] = [Q_i]$, $[\mathcal{K} _{ij}] = [0]$
a-dimensional.

\vfill\eject


\begin{thebibliography}{99}

\bibitem{1b} C.Cohen-Tannoudji, J. Dupont-Roc and G.Grynberg, \textit{
Photons and Atoms. Introduction to Quantum Electrodynamics} (Wiley,
New York, 1989).

\bibitem{2b} C.Cohen-Tannoudji, J. Dupont-Roc and G.Grynberg, \textit{
Atom-Photon Interactions. Basic Processes and Applications} (Wiley,
New York, 1992).

\bibitem{3b} W.P.Schleich, \textit{Quantum Optics in Phase Space} (Wiley-VCH,
Berlin, 2001).


\bibitem{4b} M. LeBellac and J.M.Levy-Leblond, {\it Galilean
Electromagnetism}, Nuovo Cimento {\bf 14B}, 217 (1973).

\bibitem{5b}L.Cacciapuoti and C.Salomon, {\it ACES: Mission Concept and
Scientific Objective}, 28/03/2007, ESA document, Estec
($ACES_Science_v1_printout.doc$).\hfill\break
 L.Cacciapuoti, N.Dimarcq and C.Salomon, {\it Atomic Clock Ensemble
in Space: Scientific Objectives and Mission Status}.\hfill\break
 See also the talks at the {\it SIGRAV Graduate School on Experimental
Gravitation in Space}(Firenze, September 25-27, 2006)
(http://www.fi.infn.it/GGI-grav-space/$egs\_s$.html); at the
Workshop {\it Advances in Precision Tests and Experimental
Gravitation in Space} (Firenze, September 28/30, 2006)
(http://www.fi.infn.it/GGI-grav-space/$egs\_w$.html); at the
Workshop "Theoretical Aspects of the ACES Mission" (Firenze, April
29-30, 2008) (ftp://cacciapuoti:In73rn0@ftp.estec.esa.int/ ); at the
Workshop on "ACES and Future GNSS-based Earth Observation and
Navigation" (Muenchen, May 26-27, 2008)
(http://www.iapg.bv.tum.de/12735--~aces~programme.html).



\bibitem{6b}A.Peres and D.R.Terno, {\it Quantum Information and
Relativity Theory}, Rev.Mod.Phys. {\bf 76}, 93
(2004)(quant-ph/0212023).\hfill\break
 D.R.Terno, {\it Introduction
to Relativistic Quantum Information} (quant-ph/0508049).



\bibitem{7b}L.Lusanna, {\it The Chrono-Geometrical Structure of Special and
General Relativity: A Re-Visitation of Canonical Geometrodynamics},
lectures at 42nd Karpacz Winter School of Theoretical Physics:
Current Mathematical Topics in Gravitation and Cosmology, Ladek,
Poland, 6-11 Feb 2006, Int.J.Geom.Methods in Mod.Phys. {\bf 4}, 79
(2007). (gr-qc/0604120).


\bibitem{8b}L. Lusanna, {\it The N- and 1-Time Classical Descriptions of N-Body
Relativistic Kinematics and the Electromagnetic Interaction}, Int.
J. Mod. Phys. {\bf A12}, 645 (1997).\hfill\break
 D.Alba, L.Lusanna and M.Pauri, \textit{New Directions in
Non-Relativistic and Relativistic Rotational and Multipole
Kinematics for N-Body and Continuous Systems} (2005), in
\textit{Atomic and Molecular Clusters: New Research}, ed.Y.L.Ping
(Nova Science, New York, 2006) (hep-th/0505005).\hfill\break
 D.Alba, L.Lusanna and M.Pauri, \textit{Centers of Mass and Rotational
Kinematics for the Relativistic N-Body Problem in the Rest-Frame
Instant Form}, J.Math.Phys. \textbf{43}, 1677-1727 (2002)
(hep-th/0102087).\hfill\break
 D.Alba, L.Lusanna and M.Pauri,
\textit{ Multipolar Expansions for Closed and Open Systems of
Relativistic Particles} , J. Math.Phys. \textbf{46}, 062505, 1-36
(2004) (hep-th/0402181).

\bibitem{9b}D. Alba and L.Lusanna, {\it Simultaneity, Radar 4-Coordinates
and the 3+1 Point of View about Accelerated Observers in Special
Relativity} (2003) (gr-qc/0311058); {\it Generalized Radar
4-Coordinates and Equal-Time Cauchy Surfaces for Arbitrary
Accelerated Observers} (2005), Int.J.Mod.Phys. {\bf D16}, 1149
(2007) (gr-qc/0501090).

\bibitem{10b} D.Alba, H.W.Crater and L.Lusanna, \textit{Hamiltonian
Relativistic Two-Body Problem: Center of Mass and Orbit
Reconstruction}, J.Phys. {\bf A40}, 9585 (2007) (gr-qc/0610200).






\bibitem{11b}C.M$\o$ller, {\it Sur la dinamique des syste'mes ayant
un moment angulaire interne}, Ann.Inst.H.Poincare' {\bf 11}, 251
(1949).\hfill\break
 C.M\"oller, \textit{The Theory of Relativity} (Oxford Univ.Press,
Oxford, 1957).


\bibitem{12b}E.Schmutzer and J.Plebansi, {\it  Quantum Mechanics in
Noninertial Frames of Reference}, Fortschr.Phys. {\bf 25}, 37
(1978).

\bibitem{13b}M.Pauri and G.Prosperi, {\it Canonical Realizations of
Lie Symmetry Groups}, {\bf 7}, 366 (1966); {\it Canonical
Realizations of the Rotation Group }, J.Math.Phys. {\bf 8}, 2256
(1967); {\it Canonical Realizations of the Galilei Group},
J.Math.Phys. {\bf 9}, 1146 (1968); \textit{Canonical Realizations of
the Poincare' group: I. General Theory}, J.Math.Phys. \textbf{16},
1503 (1975); \textit{Canonical Realizations of the Poincare' Group:
II. Space-Time Description of Two Particles Interacting at a
Distance, Newtonian-like Equations of Motion and Approximately
Relativistic Lagrangian Formulation}, J.Math.Phys. \textbf{17}, 1468
(1976).

\bibitem{14b} H.W.Crater and L.Lusanna, \textit{The Rest-Frame Darwin
Potential from the Lienard-Wiechert Solution in the Radiation
Gauge}, Ann.Phys. (N.Y.) \textbf{289}, 87 (2001).

\bibitem{15b} D.Alba, H.W.Crater and L.Lusanna, \textit{The Semiclassical
Relativistic Darwin Potential for Spinning Particles in the Rest
Frame Instant Form: Two-Body Bound States with Spin 1/2
Constituents}, Int.J.Mod.Phys. \textbf{A16}, 3365-3478 (2001)
(hep-th/0103109).\hfill\break
 F.Bigazzi and L.Lusanna, {\it Spinning Particles on Spacelike
 Hypersurfaces and their Rest Frame Description}, Int.J.Mod.Phys.
 {\bf A14}, 1429 (1999) (hep-th/9807052).

\bibitem{16b}N.N.Bogoliubov and D.V. Shirkov,{\it Introduction To The Theory
of Quantized Fields}, (Wiley, New York, 1980).


\bibitem{a}J.Earman and D.Fraser, {\it Haag's Theorem and its Implications
for the Foundations of Quantum Field Theory}, Erkenntnis {\bf 64},
305 (2006) (philsci-archive.pitt.edu/archive/00002673/).

\bibitem{17b} G.Longhi and L.Lusanna, \textit{
Bound-State Solutions, Invariant Scalar Products and Conserved
Currents for a Class of Two-Body Relativistic Systems}, Phys.Rev.
\textbf{D34}, 3707 (1986).



\bibitem{18b} H.Leutwyler and J.Stern, {\it Relativistic Dynamics on a Null Plane},
Ann.Phys. (N.Y.) \textbf{112}, 94 (1978).



\bibitem{19b}G.Veneziano, {\it Quantum Strings and the Constants of
Nature}, in {\it The Challenging Questions}, ed. A.Zichichi, the
Subnuclear Series n. 27 (Plenum Press, NewYork, 1990).

\bibitem{20b}H.Epstein, V.Glaser and A.Jaffe, {\it Nonpositivity of the
Energy Density in Quantized Field Theories}, Nuovo Cimento {\bf 36},
1016 (1965).



\bibitem{21b}L.Lusanna, {\it   Gauge Fixings, Evolution Generators and World-line
Conditions in Relativistic Classical Mechanics with Constraints},
Nuovo Cimento {\bf 65B}, 135 (1981).

\bibitem{22b} D.G.Currie, T.F.Jordan and E.C.G.Sudarshan, \textit{Relativistic
Invariance and Hamiltonian Theories of Interacting Particles},
Rev.Mod.Phys. \textbf{35}, 350 (1965(.\hfill\break H.Leutwyler,
\textit{A no Interaction Theorem in Classical Relativistic
Hamiltonian Particle Mechanics}, Nuovo Cimento \textbf{37}, 556
(1965).\hfill\break S.Chelkowski, J.Nietendel and R.Suchanek,
\textit{The No-Interaction Theorem in Relativistic Particle
Mechanics}, Acta Phys.Pol. \textbf{B11}, 809 (1980).

\bibitem{23b}G.Longhi, L.Lusanna and G.Longhi, {\it On the
Many-Time Formulation of Classical Particle Dynamics}, J.Math.Phys.
{\bf 30}, 1893 (1989).

\bibitem{24b} I. T. Todorov, \textit{Dynamics of Relativistic Point Particles
as a Problem with Constraints} , Dubna Joint Institute for Nuclear
Research No. E2-10175, 1976; \textit{On the Quantification of a
Mechanical System with Second Class Constraints}, Ann. Inst. H.
Poincare' \textbf{A28}, 207 (1978).\hfill\break
 A. Komar, \textit{Constraint Formalism of Classical Mechanics},
Phys. Rev. \textbf{D18}, 1881 and \textit{Interacting Relativistic
Particles}, Phys.Rev. \textbf{D18}, 1887 (1978).

\bibitem{25b} Ph.Droz Vincent, \textit{Is Interaction Possible without
Heredity?}, Lett.Nuovo Cimento \textbf{1}, 839 (1969);
\textit{Relativistic
Systems of Interacting Particles}, Phys.Scr. \textbf{2}, 129 (1970); \textit{%
\ \ Hamiltonian Systems of Relativistic Particles}, Rep. Math. Phys.,\textbf{%
8} ,79 (1975); \textit{Two-Body Relativistic Systems}, Ann.Inst.H.Poincar%
\'{e} \textbf{27}, 407 (1977) and \textit{N-Body Relativistic
Systems}, \textbf{32A }, 377 (1980); \textit{Action at a Distance
and Relativistic Wave Equations for Spinless Quarks}, Phys.Rev.
\textbf{D19}, 702 (1979).

\bibitem{26b}N.Mukunda and E.C.G.Sudarshan, {\it  Form Of Relativistic
Dynamics With World Lines}, Phys.Rev. {\bf D23}, 2210
(1981).\hfill\break
 A.Kihlberg, R.Marnelius and N.Mukunda, {\it  Relativistic Potential Models
 As Systems With Constraints And Their Interpretation}, Phys.Rev. {\bf
 D23}, 2201 (1981).\hfill\break
 J.N.Goldberg, {\it Relativistically Interacting Particles and
 World-lines}, Syracuse Univ. preprint (1980).



\bibitem{27b} M. Kalb and P. Van Alstine, \textit{Invariant Singular Actions
for the Relativistic Two-Body Problem: a Hamiltonian Formulation},
Yale Reports, C00-3075-146 (1976),C00-3075-156 (1976).\hfill\break
 D.Dominici, J.Gomis and G.Longhi, \textit{A Lagrangian for Two
Interacting Relativistic Particles}, Nuovo Cimento \textbf{B48}, 152
(1978); \textit{A Lagrangian For Two Interacting Relativistic
Particles: Canonical Formulation}, Nuovo Cimento \textbf{A48}, 257
(1978); \textit{A Possible Approach To The Two-Body Relativistic
Problem}, Nuovo Cimento \textbf{A56}, 263 (1980).\hfill\break
 J.Gomis, J.A.Lobo and J.M.Pons, \textit{A Singular Lagrangian Model
For Two Interacting Relativistic Particles}, Ann.Inst.H.Poincare'
\textbf{A35}, 17 (1981).\hfill\break
 R.Giachetti and E.Sorace, \textit{Canonical Theory of Relativistic
 Interactions}, Nuovo Cimento \textbf{A43}, 281 (1978); \textit{Relativistic
 Two-Body Interactions: A Hamiltonian Formulation}, Nuovo Cimento
\textbf{A56}, 263 (1980).\hfill\break
 L.Lusanna, \textit{A Model for N Classical Relativistic Particles},
 Nuovo Cimento \textbf{A64}, 65 (1981).

\bibitem{28b} F.Rohrlich, {\it Many-Body Forces and the Cluster
Decomposition}, Phys.Rev. {\bf D23}, 1305 (1981).\hfill\break
 L.P.Horwitz and F.Rohrlich, {\it Constraint Relativistic Quantum
 Dynamics}, Phys.Rev. {\bf D24}, 1928 (1981).\hfill\break
 H.Sazdjian, {\it  Separable Interactions In Classical Relativistic
 Hamiltonian Mechanics}, Lett.Math.Phys. {\bf 5}, 319 (1981); {\it
 Position Variables in Classical Relativistic Hamiltonian
 Mechanics}, Nucl.Phys. {\bf B161}, 469 (1979).\hfill\break
 S.N.Sokolov, {\it Theory of Relativistic Direct Interaction},
 Serpukhov report IHEP, OTF 78-125 (1978).

 \bibitem{29b}D.G.Currie, {\it Poincaré-Invariant Equations of
 Motion for Classical Particles}, Phys.Rev. {\bf 142}, 817
 (1966).\hfill\break
 R.H.Hill, {\it Inatantaneous Action-at-a-Distance in Classical
 Relativistic Mechanic}, J.Math.Phys. {\bf 8}, 201 (1967).

\bibitem{30b} L.Bel, {\it Mecanica Relativista Predictiva}, courso
impartido en el Departamento di Fisica Teorica de la Univrsidad
Autonoma de Barcelona, UAB FT-34 (1977); {\it Dynamique des systèmes
de N particules ponctuelles en relativité restreinte},
Ann.Inst.Henry Poincare' {\bf 12}, 307 (1970); {\it Predictive
Relativistic Mechanics.}, Ann.Inst.Henry Poincare' {\bf 14}, 189
(1971).\hfill\break
 L.Bel and F.X.Fustero, {\it  Predictive Relativistic Mechanics of n
 Particle Systems.}, Ann.Inst.Henry Poincare' {\bf 25},
 411 (1976).\hfill\break
 L.Bel and J.Martin, {\it  Hamiltonians and Conservative Systems},
 Ann.Inst.Henry Poincare' {\bf 22}, 173
 (1975) and {\it Predictive Relativistic Mechanics of Systems of
 N Particles with Spin}, {\bf 33}, 409 (1980).





\bibitem{31b}D. Alba, L. Lusanna and M. Pauri, {\it Dynamical
Body Frames, Orientation-Shape Variables and Canonical Spin Bases
for the Nonrelativistic N-Body Problem }, J. Math. Phys. {\bf 43},
373 (2002) (hep-th/0011014).

\bibitem{32b}M.Pauri and G.Prosperi, {\it Canonical Realizations of
the Galilei Group}, J.Math.Phys. {\bf 9}, 1146 (1968).\hfill\break
 R.DePietri, L.Lusanna and M.Pauri, {\it Gauging Kinematical and
Internal Symmetry Groups for Extended Systems: the Galilean One-Time
and Two-Times Harmonic Oscillators}, Class.Quantum Grav. {\bf 11},
1417 (1996).




\bibitem{33b} L.Assenza and G.Longhi, \textit{Collective and Relative
Variables for Massless Fields}, Int.J.Mod.Phys. \textbf{A15},
4575-4601 (2000).


\bibitem{b}D.Alba and L.Lusanna, {\it The Lienard-Wiechert Potential
of Charged Scalar Particles and their Relation to Scalar
Electrodynamics in the Rest-Frame Instant Form}, Int.J.Mod.Phys.
{\bf A13}, 2791 (1998).

\bibitem{34b} A.Barducci and L.Lusanna, {\it The Photon in Pseudo-Classical Mechanics},
Nuovo Cimento \textbf{77A}, 39 (1983).



\bibitem{bb}
  N.H.Lindner, A.Peres and D.R.Terno, {\it  Wigner's Little Group
 and Berry's Phase for Massless Particles}, J.Phys. {\bf
 A36}, L449 (2003) (hep-th/0304017).\hfill\break
 J.A.Brooke and F.E.Schroeck, {\it Localization of the Photon on
 Phase Space}, J.Math.Phys. {\bf 37}, 5958 (1996).\hfill\break
 O.Keller, {\it On the Theory of Spatial Photon Localization},
 Phys.Rep. {\bf 411}, 1 (2005).

\bibitem{cc}D.Alba and L.Lusanna, {\it Charged Particles and the
Electro-Magnetic Field in Non-Inertial Frames}, in preparation.


\bibitem{35b} S.Weinberg, {\it Feynman Rules for Any Spin. II. Massless Particles},
Phys.Rev. \textbf{134}, B882 (1964); {\it Feynman Rules for Any Spin
}, Phys. Rev. \textbf{133}, B1318.








\end{thebibliography}
\end{document}